\documentclass[twocolumn]{aastex631}

\usepackage{indentfirst}
\usepackage{graphicx}
\usepackage{subfigure}

\usepackage{amsmath}
\usepackage{multirow}

\usepackage{longtable}
\usepackage{threeparttable}
\begin{document}

\title{Luminosity function of quasars at $1.0<z<3.5$ from SDSS and DESI}

\author[0000-0003-3292-8631]{Gaocheng Yin}
\affiliation{Department of Astronomy, School of Physics, Peking University, Beijing 100871, People’s Republic of China}
\affiliation{Kavli Institute for Astronomy and Astrophysics, Peking University, Beijing 100871, People’s Republic of China}

\author[0000-0003-4176-6486]{Linhua Jiang}
\affiliation{Department of Astronomy, School of Physics, Peking University, Beijing 100871, People’s Republic of China}
\affiliation{Kavli Institute for Astronomy and Astrophysics, Peking University, Beijing 100871, People’s Republic of China}

\author[0000-0003-0230-6436]{Zhiwei Pan}
\affiliation{Department of Astronomy, School of Physics, Peking University, Beijing 100871, People’s Republic of China}
\affiliation{Kavli Institute for Astronomy and Astrophysics, Peking University, Beijing 100871, People’s Republic of China}

\author[0000-0002-4279-4182]{Paul Martini}
\affiliation{Center for Cosmology and AstroParticle Physics, The Ohio State University, 191 West Woodruff Avenue, Columbus, OH 43210, USA}
\affiliation{Department of Astronomy, The Ohio State University, 4055 McPherson Laboratory, 140 W 18th Avenue, Columbus, OH 43210, USA}
\affiliation{The Ohio State University, Columbus, 43210 OH, USA}

\author[0000-0001-9457-0589]{Wei-Jian Guo}
\affiliation{Key Laboratory of Optical Astronomy, National Astronomical Observatories, Chinese Academy of Sciences, Beijing 100012, People’s Republic of China}

\author[0000-0002-3983-6484]{Siwei Zou}
\affiliation{Chinese Academy of Sciences South America Center for Astronomy, National Astronomical Observatories, CAS, Beijing 100101, People’s Republic of China}
\affiliation{Department of Astronomy, Tsinghua University, Beijing 100084, People’s Republic of China}

\author[0000-0002-1234-552X]{Shengxiu Sun}
\affiliation{Department of Astronomy, School of Physics, Peking University, Beijing 100871, People’s Republic of China}
\affiliation{Kavli Institute for Astronomy and Astrophysics, Peking University, Beijing 100871, People’s Republic of China}

\author[0000-0002-5854-7426]{Swayamtrupta Panda}
\affiliation{International Gemini Observatory/NSF NOIRLab, Casilla 603, La Serena, Chile}

\author[0000-0003-2923-1585]{Abhijeet Anand}
\affiliation{Lawrence Berkeley National Laboratory, 1 Cyclotron Road, Berkeley, CA 94720, USA}

\author{Benjamin Alan Weaver}
\affiliation{NSF NOIRLab, 950 N. Cherry Ave., Tucson, AZ 85719, USA}

\author[0000-0002-1125-7384]{Aaron Meisner}
\affiliation{NSF NOIRLab, 950 N. Cherry Ave., Tucson, AZ 85719, USA}

\author[0000-0002-2169-0595]{Andrei Cuceu}
\affiliation{Lawrence Berkeley National Laboratory, 1 Cyclotron Road, Berkeley, CA 94720, USA}

\author[0000-0002-4928-4003]{Arjun Dey}
\affiliation{NSF NOIRLab, 950 N. Cherry Ave., Tucson, AZ 85719, USA}

\author[0000-0002-1769-1640]{Axel de la Macorra}
\affiliation{Instituto de F\'{\i}sica, Universidad Nacional Aut\'{o}noma de M\'{e}xico,  Circuito de la Investigaci\'{o}n Cient\'{\i}fica, Ciudad Universitaria, Cd. de M\'{e}xico  C.~P.~04510,  M\'{e}xico}

\author{Christophe Magneville}
\affiliation{IRFU, CEA, Universit\'{e} Paris-Saclay, F-91191 Gif-sur-Yvette, France}

\author{David Brooks}
\affiliation{Department of Physics \& Astronomy, University College London, Gower Street, London, WC1E 6BT, UK}

\author[0000-0002-8828-5463]{David Kirkby}
\affiliation{Department of Physics and Astronomy, University of California, Irvine, 92697, USA}

\author{David Schlegel}
\affiliation{Lawrence Berkeley National Laboratory, 1 Cyclotron Road, Berkeley, CA 94720, USA}

\author{David Sprayberry}
\affiliation{NSF NOIRLab, 950 N. Cherry Ave., Tucson, AZ 85719, USA}

\author[0000-0001-9712-0006]{Davide Bianchi}
\affiliation{Dipartimento di Fisica ``Aldo Pontremoli'', Universit\`a degli Studi di Milano, Via Celoria 16, I-20133 Milano, Italy}
\affiliation{INAF-Osservatorio Astronomico di Brera, Via Brera 28, 20122 Milano, Italy}

\author[0000-0003-0201-5241]{Dick Joyce}
\affiliation{NSF NOIRLab, 950 N. Cherry Ave., Tucson, AZ 85719, USA}

\author[0000-0001-9632-0815]{Enrique Gaztañaga}
\affiliation{Institut d'Estudis Espacials de Catalunya (IEEC), c/ Esteve Terradas 1, Edifici RDIT, Campus PMT-UPC, 08860 Castelldefels, Spain}
\affiliation{Institute of Cosmology and Gravitation, University of Portsmouth, Dennis Sciama Building, Portsmouth, PO1 3FX, UK}
\affiliation{Institute of Space Sciences, ICE-CSIC, Campus UAB, Carrer de Can Magrans s/n, 08913 Bellaterra, Barcelona, Spain}

\author[0000-0002-9646-8198]{Eusebio Sanchez}
\affiliation{CIEMAT, Avenida Complutense 40, E-28040 Madrid, Spain}

\author[0000-0001-7316-4573]{Francisco Javier Castander}
\affiliation{Institut d'Estudis Espacials de Catalunya (IEEC), c/ Esteve Terradas 1, Edifici RDIT, Campus PMT-UPC, 08860 Castelldefels, Spain}
\affiliation{Institute of Space Sciences, ICE-CSIC, Campus UAB, Carrer de Can Magrans s/n, 08913 Bellaterra, Barcelona, Spain}

\author[0000-0001-7145-8674]{Francisco Prada}
\affiliation{Instituto de Astrof\'{i}sica de Andaluc\'{i}a (CSIC), Glorieta de la Astronom\'{i}a, s/n, E-18008 Granada, Spain}

\author{Gaston Gutierrez}
\affiliation{Fermi National Accelerator Laboratory, PO Box 500, Batavia, IL 60510, USA}

\author{Graziano Rossi}
\affiliation{Department of Physics and Astronomy, Sejong University, 209 Neungdong-ro, Gwangjin-gu, Seoul 05006, Republic of Korea}

\author[0000-0003-1704-0781]{Gregory Tarlé}
\affiliation{University of Michigan, 500 S. State Street, Ann Arbor, MI 48109, USA}

\author[0000-0002-9136-9609]{Hiram K. Herrera-Alcantar}
\affiliation{IRFU, CEA, Universit\'{e} Paris-Saclay, F-91191 Gif-sur-Yvette, France}
\affiliation{Institut d'Astrophysique de Paris. 98 bis boulevard Arago. 75014 Paris, France}

\author[0000-0002-6684-3997]{Hu Zou}
\affiliation{National Astronomical Observatories, Chinese Academy of Sciences, A20 Datun Road, Chaoyang District, Beijing, 100101, P.~R.~China}

\author[0000-0001-6979-0125]{Ignasi Pérez-Ràfols}
\affiliation{Departament de F\'isica, EEBE, Universitat Polit\`ecnica de Catalunya, c/Eduard Maristany 10, 08930 Barcelona, Spain}

\author[0000-0002-2890-3725]{Jaime E. Forero-Romero}
\affiliation{Departamento de F\'isica, Universidad de los Andes, Cra. 1 No. 18A-10, Edificio Ip, CP 111711, Bogot\'a, Colombia}
\affiliation{Observatorio Astron\'omico, Universidad de los Andes, Cra. 1 No. 18A-10, Edificio H, CP 111711 Bogot\'a, Colombia}

\author{Jessica Nicole Aguilar}
\affiliation{Lawrence Berkeley National Laboratory, 1 Cyclotron Road, Berkeley, CA 94720, USA}

\author[0000-0002-2733-4559]{John Moustakas}
\affiliation{Department of Physics and Astronomy, Siena University, 515 Loudon Road, Loudonville, NY 12211, USA}

\author[0000-0002-3461-0320]{Joseph Harry Silber}
\affiliation{Lawrence Berkeley National Laboratory, 1 Cyclotron Road, Berkeley, CA 94720, USA}

\author[0000-0002-6550-2023]{Klaus Honscheid}
\affiliation{Center for Cosmology and AstroParticle Physics, The Ohio State University, 191 West Woodruff Avenue, Columbus, OH 43210, USA}
\affiliation{Department of Physics, The Ohio State University, 191 West Woodruff Avenue, Columbus, OH 43210, USA}
\affiliation{The Ohio State University, Columbus, 43210 OH, USA}

\author[0000-0001-7178-8868]{Laurent Le Guillou}
\affiliation{Sorbonne Universit\'{e}, CNRS/IN2P3, Laboratoire de Physique Nucl\'{e}aire et de Hautes Energies (LPNHE), FR-75005 Paris, France}

\author[0000-0003-4962-8934]{Marc Manera}
\affiliation{Departament de F\'{i}sica, Serra H\'{u}nter, Universitat Aut\`{o}noma de Barcelona, 08193 Bellaterra (Barcelona), Spain}
\affiliation{Institut de F\'{i}sica d’Altes Energies (IFAE), The Barcelona Institute of Science and Technology, Edifici Cn, Campus UAB, 08193, Bellaterra (Barcelona), Spain}

\author[0000-0003-1838-8528]{Martin Landriau}
\affiliation{Lawrence Berkeley National Laboratory, 1 Cyclotron Road, Berkeley, CA 94720, USA}

\author{Michael Schubnell}
\affiliation{Department of Physics, University of Michigan, 450 Church Street, Ann Arbor, MI 48109, USA}
\affiliation{University of Michigan, 500 S. State Street, Ann Arbor, MI 48109, USA}

\author[0000-0002-6024-466X]{Mustapha Ishak}
\affiliation{Department of Physics, The University of Texas at Dallas, 800 W. Campbell Rd., Richardson, TX 75080, USA}

\author[0000-0003-3188-784X]{Nathalie Palanque-Delabrouille}
\affiliation{IRFU, CEA, Universit\'{e} Paris-Saclay, F-91191 Gif-sur-Yvette, France}
\affiliation{Lawrence Berkeley National Laboratory, 1 Cyclotron Road, Berkeley, CA 94720, USA}

\author{Peter Doel}
\affiliation{Department of Physics \& Astronomy, University College London, Gower Street, London, WC1E 6BT, UK}

\author{Ramon Miquel}
\affiliation{Instituci\'{o} Catalana de Recerca i Estudis Avan\c{c}ats, Passeig de Llu\'{\i}s Companys, 23, 08010 Barcelona, Spain}
\affiliation{Institut de F\'{i}sica d’Altes Energies (IFAE), The Barcelona Institute of Science and Technology, Edifici Cn, Campus UAB, 08193, Bellaterra (Barcelona), Spain}

\author{Robert Kehoe}
\affiliation{Department of Physics, Southern Methodist University, 3215 Daniel Avenue, Dallas, TX 75275, USA}

\author[0000-0003-3142-233X]{Satya Gontcho A Gontcho}
\affiliation{Lawrence Berkeley National Laboratory, 1 Cyclotron Road, Berkeley, CA 94720, USA}
\affiliation{University of Virginia, Department of Astronomy, Charlottesville, VA 22904, USA}

\author[0000-0001-9070-3102]{Seshadri Nadathur}
\affiliation{Institute of Cosmology and Gravitation, University of Portsmouth, Dennis Sciama Building, Portsmouth, PO1 3FX, UK}

\author[0000-0003-4992-7854]{Simone Ferraro}
\affiliation{Lawrence Berkeley National Laboratory, 1 Cyclotron Road, Berkeley, CA 94720, USA}
\affiliation{University of California, Berkeley, 110 Sproul Hall \#5800 Berkeley, CA 94720, USA}

\author[0000-0002-0000-2394]{Stephanie Juneau}
\affiliation{NSF NOIRLab, 950 N. Cherry Ave., Tucson, AZ 85719, USA}

\author[0000-0001-6098-7247]{Steven Ahlen}
\affiliation{Department of Physics, Boston University, 590 Commonwealth Avenue, Boston, MA 02215 USA}

\author[0000-0003-3510-7134]{Theodore Kisner}
\affiliation{Lawrence Berkeley National Laboratory, 1 Cyclotron Road, Berkeley, CA 94720, USA}

\author{Todd Claybaugh}
\affiliation{Lawrence Berkeley National Laboratory, 1 Cyclotron Road, Berkeley, CA 94720, USA}

\author[0000-0002-0644-5727]{Will Percival}
\affiliation{Department of Physics and Astronomy, University of Waterloo, 200 University Ave W, Waterloo, ON N2L 3G1, Canada}
\affiliation{Perimeter Institute for Theoretical Physics, 31 Caroline St. North, Waterloo, ON N2L 2Y5, Canada}
\affiliation{Waterloo Centre for Astrophysics, University of Waterloo, 200 University Ave W, Waterloo, ON N2L 3G1, Canada}

\author[0000-0001-6356-7424]{Anthony Kremin}
\affiliation{Lawrence Berkeley National Laboratory, 1 Cyclotron Road, Berkeley, CA 94720, USA}

\author[0000-0002-6731-9329]{Claire Lamman}
\affiliation{The Ohio State University, Columbus, 43210 OH, USA}

\author{Claire Poppett}
\affiliation{Lawrence Berkeley National Laboratory, 1 Cyclotron Road, Berkeley, CA 94720, USA}
\affiliation{Space Sciences Laboratory, University of California, Berkeley, 7 Gauss Way, Berkeley, CA  94720, USA}
\affiliation{University of California, Berkeley, 110 Sproul Hall \#5800 Berkeley, CA 94720, USA}

\author[0000-0001-5381-4372]{Rongpu Zhou}
\affiliation{Lawrence Berkeley National Laboratory, 1 Cyclotron Road, Berkeley, CA 94720, USA}

\begin{abstract}

We present a study of the evolution of type 1 quasars at $1.0<z<3.5$, covering the peak epoch of quasar activity. The quasar evolution has been extensively explored by a variety of previous works and the derived quasar luminosity functions (QLFs) are not well consistent with each other, presumably due to the complexities introduced by different quasar selection techniques and associated completeness corrections. We use a new strategy to construct QLFs based on a library of all known quasars. We focus on a wide region of $\sim$1700 deg$^2$ and a deep field of $\sim$265 deg$^2$ that have rich spectroscopic data primarily from SDSS and DESI. We then apply traditional color cuts in the rest-frame UV/optical to select quasar candidates and use the quasar library to identify them. Our final sample consists of 62,426 quasars at $1.0<z<3.5$, with a high completeness ($\sim$96\%) and a high purity ($\sim$93\%) in the color selection. Simple color cuts can potentially minimize selection biases for the study of quasar evolution. We derive binned QLFs and characterize them using a double power-law model. Sample incompleteness and contamination are considered as part of the uncertainties in the calculation. Compared to previous results, our QLFs are slightly higher at the faint end, and also higher at the bright end at $2.5<z<3.5$. The QLFs suggest that the quasar evolution at $1.0 < z < 2.5$ can be well described by the pure luminosity evolution model, while at $2.5 < z < 3.5$, it can be described by either the pure luminosity evolution or the pure density evolution model.

\end{abstract}


\section{Introduction}

Quasars are among the brightest astrophysical objects and provide an important probe for studying the distant universe. 
They are powered by active supermassive black holes (SMBHs), and thus trace their formation and evolution \citep[e.g.,][]{2000ApJ...533..631K, 2004A&A...424..793W, 2005Natur.433..604D, 2005MNRAS.361..776S, 2013ARA&A..51..511K} and shape the evolution of their host galaxies \citep[e.g.,][]{2003MNRAS.346.1055K, 2011ApJ...739...56D, 2021MNRAS.507....1V, 2024A&A...689A.219D, 2025NatAs...9.1541O}. Due to their extreme luminosity, quasars are often used as background sources to study the state of the intergalactic medium (IGM) \citep[e.g.,][]{2006AJ....132..117F, 2010ApJ...714..834C, 2015MNRAS.447..499M, 2022NatAs...6..850J, 2023ApJ...949L..42M}. Large samples of quasars are also used to study the large-scale structure of the universe \citep[e.g.,][]{2013AJ....145...10D, 2014JCAP...05..027F, 2015A&A...574A..59D, 2021PhRvD.103h3533A}. Therefore, quasars play an important role in our understanding of SMBHs, galaxy evolution, IGM, and large-scale structure of the universe.

In the past three decades, large sky surveys have produced more than one million quasars with spectra. The 2dF Project was initiated in 1990 \citep{1998ASPC..152...71L}. Based on a sample of over 6000 quasars identified from the initial observations of the 2dF QSO Redshift Survey (2QZ), \cite{2000MNRAS.317.1014B} reported the quasar luminosity function (QLF) and its redshift evolution. \cite{2004MNRAS.349.1397C} presented the final catalogue of the 2QZ that contains 23,338 quasars. 
The Sloan Digital Sky Survey (SDSS) quasar survey is one of the most widely used quasar surveys. Its initial goal was to obtain spectra for 100,000 quasars \citep{2000AJ....120.1579Y}. As SDSS continued to operate, the number of observed quasar spectra increased significantly. The Baryon Oscillation Spectroscopic Survey (BOSS) aimed to observe more than 150,000 quasars with $g < 22$ mag \citep{2013AJ....145...10D}. Moreover, the SDSS-IV quasar catalog from the SDSS Data Release 16 contains 750,414 quasars \citep{2020ApJS..250....8L}.
The Large Sky Area Multi-Object Fiber Spectroscopic Telescope (LAMOST) can obtain up to 4000 spectra in a single exposure, reaching a limiting magnitude of $r \approx 19$ mag at a resolution of $R = 1800$. The LAMOST Data Releases 4 and 5 quasar catalog contains a total of 19,253 quasars, primarily in the redshift range of $0 < z < 4$ \citep{2019ApJS..240....6Y}.
The Dark Energy Spectroscopic Instrument (DESI) offers a substantial advance over previous wide-area spectroscopic surveys in quasar detection and redshift coverage \citep{2016arXiv161100037D, 2022AJ....164..207D, desicollaboration2025desidr2resultsii, 2024AJ....168...95M, 2024AJ....168..245P, 2024AJ....168..124A, 2025arXiv250501596G}. Recently, the DESI survey is producing a large number of quasars \citep{2023ApJ...944..107C}. Its quasar target selection was based on a random forest algorithm and color selection that identifies quasars in a magnitude range of $16.5 < r < 23.0$ mag. This selection process enables DESI to target over $200 \deg^{-2}$ quasars, including $60 \deg^{-2}$ quasars at $z > 2.1$.

There are many methods to select quasars. The color selection method has been widely used \citep[e.g.,][]{2002AJ....123.2945R, 2006AJ....131.2766R, 2013AJ....145...10D, 2015ApJS..221...27M, 2016ApJ...819...24W, 2022ApJ...928..172P, 2023ApJ...944..107C}. This is because quasars with power-law spectra and stars with blackbody spectra have different distributions in the color space. Quasars are variable objects with different variability timescales ranging from hours to years \citep[e.g.,][]{2011ApJ...728...26M, 2016A&A...587A..41P}, so the method based on variability is efficient to select quasars. Quasars can also be selected by their strong X-ray and radio emission \citep[e.g.,][]{2005A&A...441..417H, 2009AJ....138.1925M, 2011ApJ...736...57Z, 2014MNRAS.445.3557V, 2015MNRAS.453.1946G, 2015ApJ...804..118B}. With Gaia's advancements in measuring proper motions with greater accuracy, the proper motion method has been applied to quasar selection in recent years \citep[e.g.,][]{2020A&A...644A..17H, 2021ApJS..254....6F, 2023arXiv230617749S}. Recently, machine-learning techniques have become a powerful tool for quasar selection. Supervised classifiers, such as support vector machines (SVMs) and gradient-boosted decision trees (e.g., XGBoost), have been widely used in this field. These methods have been applied to time-domain variability features \citep{2011ApJ...735...68K} and optical--infrared colors from large-area surveys \citep{2019MNRAS.485.4539J, 2026ApJS..282...38Z}. They have also been extended to transfer-learning frameworks, which can help address data-set shifts in dense stellar regions such as the Galactic plane \citep{2021ApJS..254....6F}. Although many methods have been developed to select quasars, no single method is complete. Each method has its own selection biases and can miss certain types of quasars. For example, not all quasars show strong variability over short timescales. Color selection can also be incomplete in redshift ranges where quasar colors overlap with the stellar locus.

At present, more than one million quasars have been identified \citep{2020ApJS..250....8L, 2023ApJ...944..107C}, with an increasing number of quasar candidates discovered, such as CatNorth \citep{2024ApJS..271...54F}. Different selection methods were used to find these quasars. This largely increases the selection efficiency, but makes it complicated to derive QLFs because of the mix of different selections. It consequently leads to discrepancies in the QLF measurements. For example, in the redshift range of $1.0<z<2.0$, the faint end of the QLFs from \cite{2016A&A...587A..41P} are significantly higher than the results of \cite{2013ApJ...773...14R} and \cite{2018AJ....155..110Y}. 

In our work, we make use of the SDSS DR16 \citep{2020ApJS..249....3A} and DESI Year 3 quasars as an existing library, regardless of how they were selected. We then apply a straightforward color selection method to select quasars and estimate sample completeness using this library. We introduce the data and target selection in Section \ref{sec: Data and target selection}. In Section \ref{sec: Results}, we present our completeness corrections, model fitting, and derive QLFs. In Section \ref{sec: Discussion}, we compare our results with previous studies and discuss the evolution of the QLFs. We summarize the paper in Section \ref{sec: Summary}. For SDSS, we use the point-spread function (PSF) magnitudes in the AB system. For the UKIRT Infrared Deep Sky Survey (UKIDSS) and Visible and Infrared Survey Telescope for Astronomy (VISTA), we use aperture magnitudes in the Vega system. The Wide-field Infrared Survey Explorer (WISE) PSF magnitudes are also expressed in the Vega system. We adopt a $\Lambda$-dominated flat cosmology with $H_0=70 \mathrm{\ km \ s^{-1} \ Mpc^{-1}}$, $\Omega_{M} = 0.3$, and $\Omega_{\Lambda} = 0.7$.

\section{Data and target selection}
\label{sec: Data and target selection}

We study the quasar evolution of type 1 quasars at $1.0 < z < 3.5$. There are two main reasons for selecting this intermediate redshift range. The first one is that the density peak of luminous quasars occurs between $z=2$ and $z=3$. The other reason is that the colors of quasars at $2 < z < 3$ are similar to those of type A and type F stars, making it difficult to select these quasars \citep[e.g.,][]{1999AJ....117.2528F, 2002AJ....123.2945R, 2006AJ....131.2766R}. Therefore, our goal is to improve the QLF measurement at this redshift range.

\subsection{Data}

We select type 1 quasars using their optical-infrared colors. From a morphological perspective, both quasars and stars appear as point sources in ground-based images, making it challenging to distinguish quasars based solely on morphology. To efficiently select quasars and achieve higher completeness, we use the SDSS optical data, UKIDSS and VISTA near-IR data, and WISE $W1$- and $W2$- band mid-IR data. These bands cover a large wavelength range for an efficient selection of quasars from point-like sources.

Our targets are first selected from the SDSS imaging data. We exclude objects with the SDSS processing flags `BRIGHT', `EDGE', `SATUR', and `BLENDED'. The extinction data are obtained from SDSS DR16, and we correct the optical magnitudes for Galactic extinction before applying the color selection. We identify point-like objects using the SDSS photometric morphological classification. In the SDSS imaging pipeline, sources are classified as point-like (type = 6) when the difference between the PSF magnitude and the cmodel magnitude satisfies $\mathrm{psfMag} - \mathrm{cmodelMag} < 0.145$ mag (\href{https://www.sdss4.org/dr16/algorithms/classify/\#photo_class}{SDSS documentation}). In the following analysis, the color selection is applied only to these point-like sources.

The near-IR imaging data are from UKIDSS and the VISTA Hemisphere Survey (VHS). UKIDSS is a near-IR survey carried out with the Wide Field Camera (WFCAM) on UKIRT \citep{2007MNRAS.379.1599L}. We use their photometric data from the UKIDSS Large Area Survey (LAS). VHS is a near-IR survey designed to reach a depth 30 times fainter than the Two Micron All Sky Survey in at least two wavebands $J$ and $Ks$ \citep{2013Msngr.154...35M}. When performing cross-matching between SDSS and UKIDSS/VISTA, the matching radius is $1 {\arcsec} $. 

WISE performed an all-sky survey in the 3.4, 4.6, 12, and 22 $\mathrm{\mu m}$ bands \citep{2010AJ....140.1868W}. We use the unWISE Catalog and the matching radius is $1\farcs5$ \citep{2014AJ....147..108L, 2016AJ....151...36L}. The unWISE Catalog offers two key advantages over the existing WISE catalog (AllWISE): it is built on substantially deeper imaging data and provides enhanced modeling for densely crowded regions \citep{2019ApJS..240...30S}.

In addition to the photometric data mentioned above, we also use spectral data from SDSS and DESI. The SDSS DR16 SpecObj table provides redshifts and spectral classifications of galaxies, stars, and quasars, including 1,362,512 quasar spectra \citep{2020ApJS..249....3A}. For DESI, we use the spectral data from a version of the Year 3 dataset (DESI Jura dataset), which contains information for 3,372,691 quasar spectra.

To achieve a high sample completeness, we target sky regions where a large fraction of the candidate quasars have already been spectroscopically confirmed by SDSS and DESI. We define two distinct sky zones. The first is a deep region that covers $\mathrm{20^h40^m < R.A. < 3^h40^m, -1\fdg26 < decl. < 1\fdg26}$. It is located in SDSS Stripe 82 and its area is 264.6 $\mathrm{deg}^2$. The second region is a wide region that encompasses two areas: $\mathrm{22^h40^m < R.A. < 2^h0^m, 1\fdg26 < decl. < 10\fdg00}$ and $\mathrm{8^h50^m < R.A. < 15^h50^m, -2\fdg00 < decl. < 10\fdg00}$. The total area is 1689.1 $\mathrm{deg}^2$.

\subsection{Color selection}

\begin{figure*}[htbp]
    \centering
    \subfigure[Deep region]{
        \includegraphics[width=0.48\textwidth]{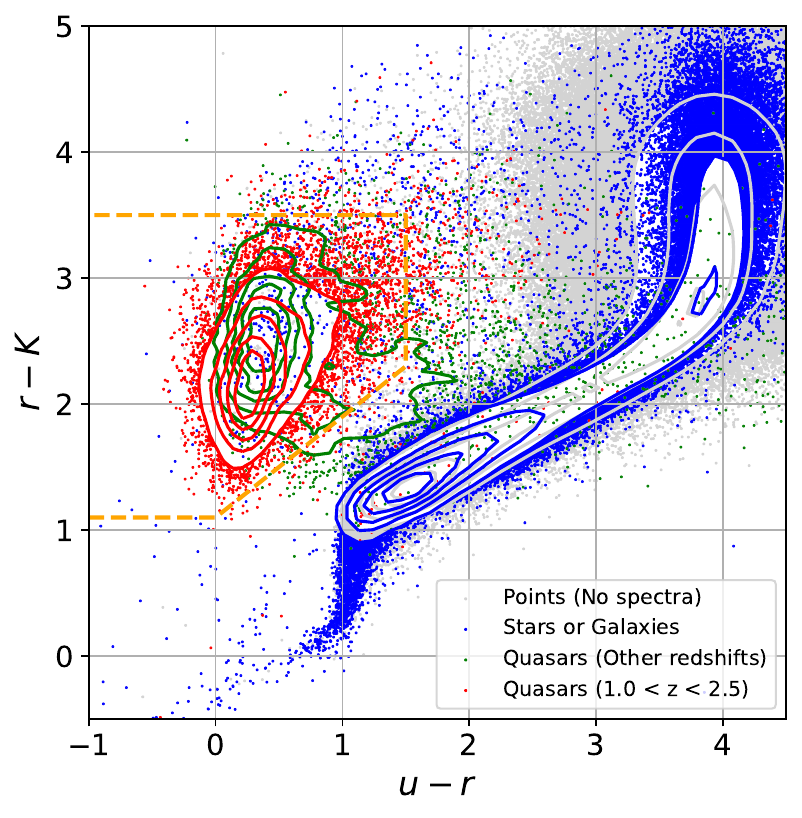}
        \label{fig:urK_diagram}
    }
    \hfill
    \subfigure[Wide region]{
        \includegraphics[width=0.48\textwidth]{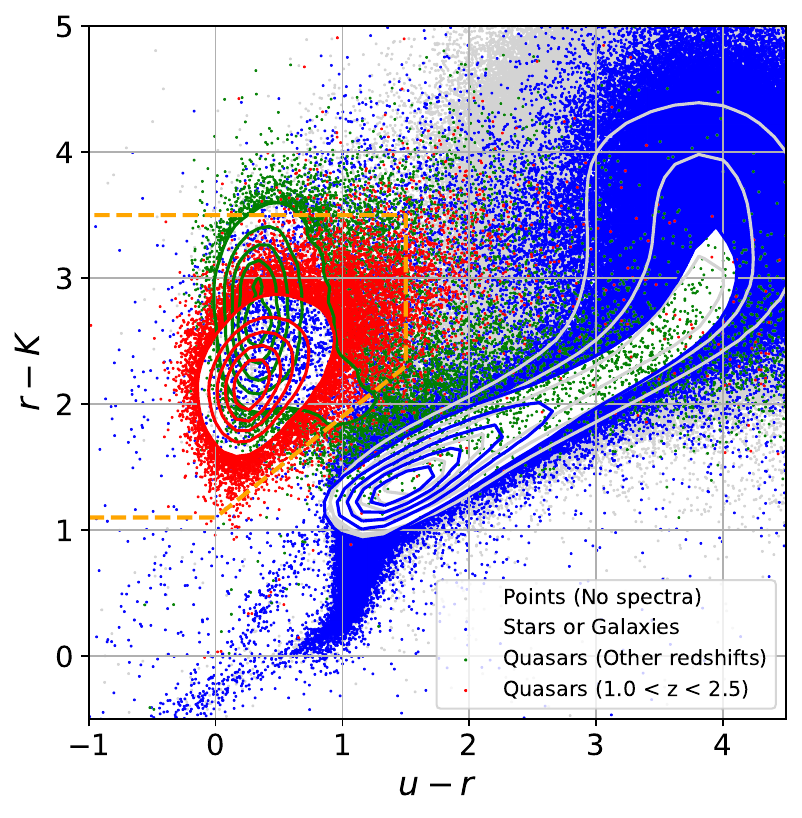}
        \label{fig:urK_diagram_OtherRegion}
    }

    \subfigure[Deep region]{
        \includegraphics[width=0.48\textwidth]{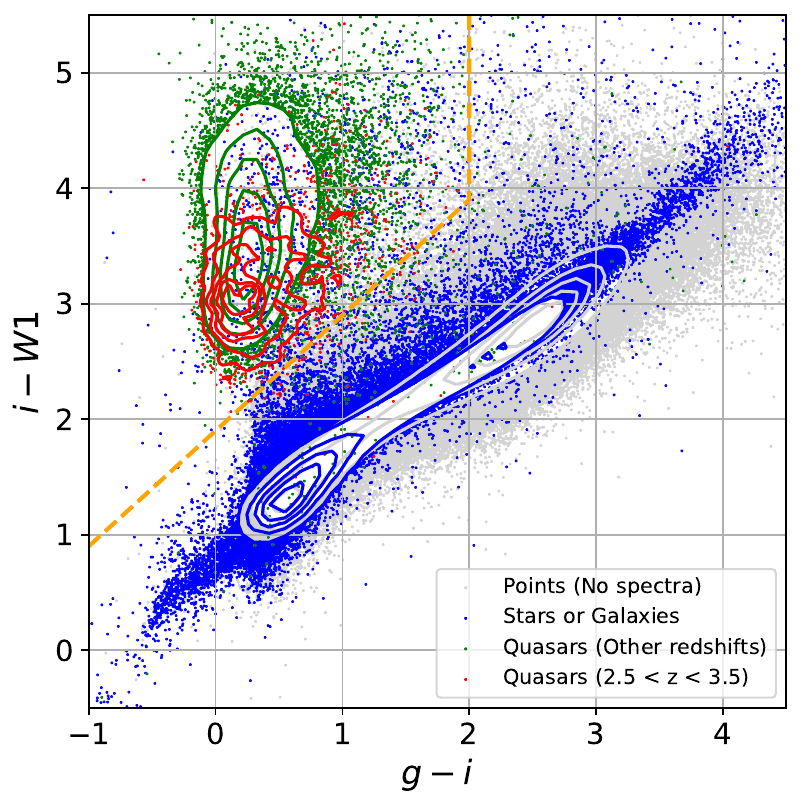}
        \label{fig:giW1_diagram}
    }
    \hfill
    \subfigure[Wide region]{
        \includegraphics[width=0.48\textwidth]{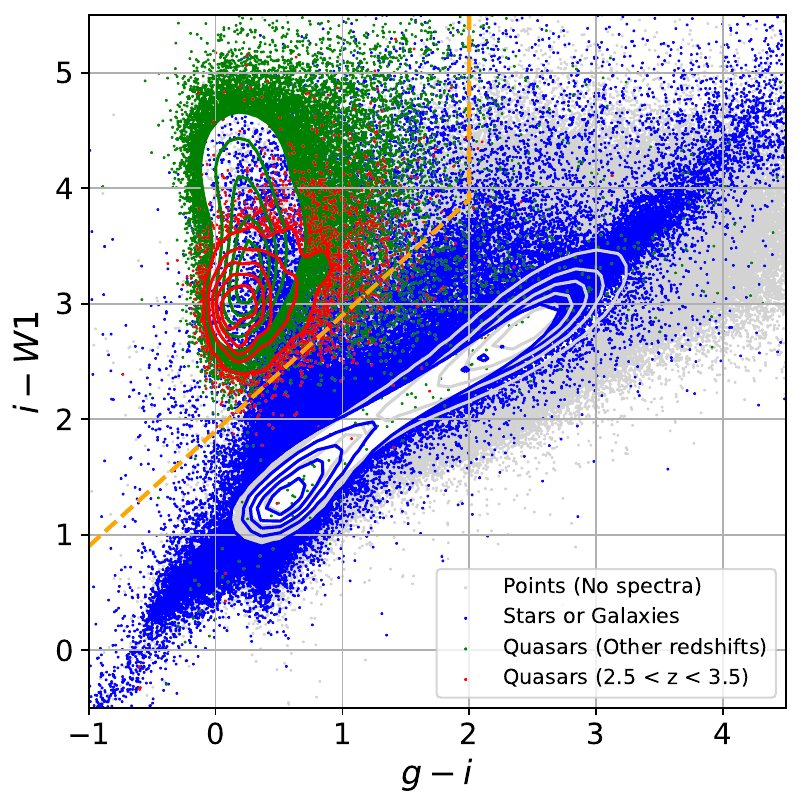}
        \label{fig:giW1_diagram_OtherRegion}
    }

    \caption{Our color selection criteria. Panels (a) and (b) show the selection of quasars at $1.0 < z < 2.5$ in the deep region and wide region, respectively. The magnitude range of the objects is $15.0 < r < 21.5$ mag in (a) and $15.0 < r < 20.5$ mag in (b). The light grey dots represent point sources without spectra. Quasars at $1.0 < z < 2.5$ are marked by the red dots, and other quasars are represented by the green dots. Stars and galaxies are indicated by the blue dots. The orange dashed line denotes our color selection criteria. 
Panels (c) and (d) show the selection of quasars at $2.5 < z < 3.5$ in the deep region and wide region, respectively. The magnitude range of the objects is $15.0 < i < 21.5$ mag in (c) and $15.0 < i < 20.5$ mag in (d). The red dots indicate quasars at $2.5 < z < 3.5$, and the orange dashed line represents our color selection criteria. We adopt the AB magnitude system for SDSS photometry, and the Vega system for VISTA, UKIDSS, and WISE data. Zero-point corrections are not applied in this work. \label{fig: urKgiW1_diagram_all}}
        
\end{figure*}

To ensure a uniform quasar selection over a wide redshift range, we develop a selection method that balances sample completeness with selection efficiency and construct two sets of color–color selection criteria, optimized separately for low-redshift($1.0 < z < 2.5$) and high-redshift ($2.5 < z < 3.5$) quasars. We examine the distributions of stars and quasars in different color-color diagrams and find that in the redshift range of $1.0 < z < 2.5$, the $u-r$ vs. $r-K$ color-color diagram (Figure \ref{fig:urK_diagram}, \ref{fig:urK_diagram_OtherRegion}) is effective to separate quasars from stars. This is because quasars at $1.0 < z < 2.5$ exhibit a significantly bluer $u-r$ color compared with late-type stars, while their $r-K$ color appears slightly redder than that of early-type stars. Therefore, we apply the following color selection criteria:
\[
\left\{
\begin{array}{ll}
u - r < 1.5, \\
1.1 < r - K < 3.5, \\
r - K > 0.8 (u - r) + 1.1.
\end{array}
\right.
\]
The magnitude range for target selection is $15.0 < r < 21.5$ mag in the deep region, and is $15.0 < r < 20.5$ mag in the wide region.

We summarize sample completeness and purity in Table \ref{tab: Completeness and purity for quasars}. We define color-selection completeness as the ratio of color-selected quasars to all quasars within a redshift range. We define purity as the ratio of color-selected quasars to all color-selected sources with spectra (including quasars, stars, and galaxies). We also define spectral completeness as the ratio of color-selected sources with spectral observations to all color-selected sources. In the deep region, we find 11,611 quasars at $1.0 < z < 2.5$. Similarly, in the wide region we find 42,119 quasars in the same redshift range. The color-selection completeness of the sample is 95\% (97\%) in the deep (wide) region, in all magnitude ranges that we consider. Only a small number of quasars at $1.0 < z < 2.5$ are excluded by our selection method, and this is mainly due to two factors. First, the $u$-band magnitude errors for some quasars are large, which results in large color errors. Second, there are a small number of very red quasars that cannot be identified by our current color selection method. The purity of the sample is 97\% (95\%) in the deep (wide) region, in all magnitude ranges that we consider.  Some contaminants are white dwarfs that typically exhibit a very blue color. Other contaminants include stars with large photometric errors. When we computed sample completeness and purity in the next section, their associated uncertainties are estimated in a straightforward way that the fraction of missed quasars or the fraction of contaminants correspond to a $2 \sigma$ uncertainty. For example, if a completeness is 96\%, then its associated $2 \sigma$ uncertainty is 4\%, or $1 \sigma$ uncertainty is 2\%.

In the redshift range of $2.5<z<3.5$, quasars appear much fainter in the rest-frame UV continuum due to the IGM absorption. This results in the quasar's $u-r$ color being very similar to that of low-temperature stars. We find that the $g-i$ and $i-W1$ color-color diagram (Figure \ref{fig:giW1_diagram}, \ref{fig:giW1_diagram_OtherRegion}) works well in distinguishing quasars from stars. This is because quasars in this redshift range have a notably bluer $g-i$ color than that of late-type stars, while their $i-W1$ color is slightly redder than that of early-type stars. Therefore, we apply the following color selection criteria:
\[
\left\{
\begin{array}{ll}
g - i < 2.0, \\
i - W1 < (g - i) + 1.9.
\end{array}
\right.
\]
The magnitude range for target selection is $15.0 < i < 21.5$ mag in the deep region, and is $15.0 < i < 20.5$ mag in the wide region.

With the above selection criteria, we find 1768 quasars in the deep region and 6928 quasars in the wide region. Table \ref{tab: Completeness and purity for quasars} shows that we maintain a reasonable balance between completeness and purity. 
Our selection only excludes a small number of real quasars, mainly due to large $W1$-band magnitude errors for some quasars. The color-based selection inevitably includes certain contaminants, such as intrinsically blue white dwarfs and galaxies that are occasionally misclassified as point sources. Additionally, some stars have positional offsets between their WISE coordinates and SDSS coordinates, which results in cross-matching errors. Currently, our cross-matching radius with the UnWISE Catalog is $1\farcs5$. Although reducing it to $1 {\arcsec}$ could help mitigate this interference, it may also prevent some sources from successfully matching with the WISE data.


\begin{table*}
\renewcommand{\arraystretch}{1.2}
\centering
\caption{Completeness and Purity for the Quasar Selection}
\begin{tabular}{ccccccc}
\hline
\hline
Regions & Morphology & Cross-matching & Redshift & Color & Purity & Spectral \\
\hline
\multirow{2}{*}{Deep region} 
& \multirow{2}{*}{$86.9 \pm 6.5 \%$} 
& \multirow{2}{*}{$71.7 \pm 14.1 \%$} 
& $1.0 < z < 2.5$ & $95.2 \pm 2.4 \%$ & $96.8 \pm 1.6 \%$ & $98.4 \pm 0.8 \%$ \\
&  &  & $2.5 < z < 3.5$ & $96.4 \pm 1.8 \%$ & $93.3 \pm 3.3 \%$ & $91.0 \pm 4.5 \%$ \\
\hline
\multirow{2}{*}{Wide region} 
& \multirow{2}{*}{$99.2 \pm 0.4 \%$} 
& \multirow{2}{*}{$85.7 \pm 7.1 \%$} 
& $1.0 < z < 2.5$ & $97.0 \pm 1.5 \%$ & $95.4 \pm 2.3 \%$ & $96.4 \pm 1.8 \%$ \\
&  &  & $2.5 < z < 3.5$ & $96.8 \pm 1.6 \%$ & $91.4 \pm 4.3 \%$ & $87.1 \pm 6.4 \%$ \\
\hline
\multicolumn{7}{l}{\textup{Note: Morphology, cross-matching, color, and spectral refer to morphology completeness,}} \\
\multicolumn{7}{l}{\textup{cross-matching completeness, color completeness, and spectral completeness, respectively.}} \\
\end{tabular}
\label{tab: Completeness and purity for quasars}
\end{table*}

\begin{figure}
    \centering
    \subfigure[$1.0 < z < 2.5$]{
        \includegraphics[width=0.48\textwidth]{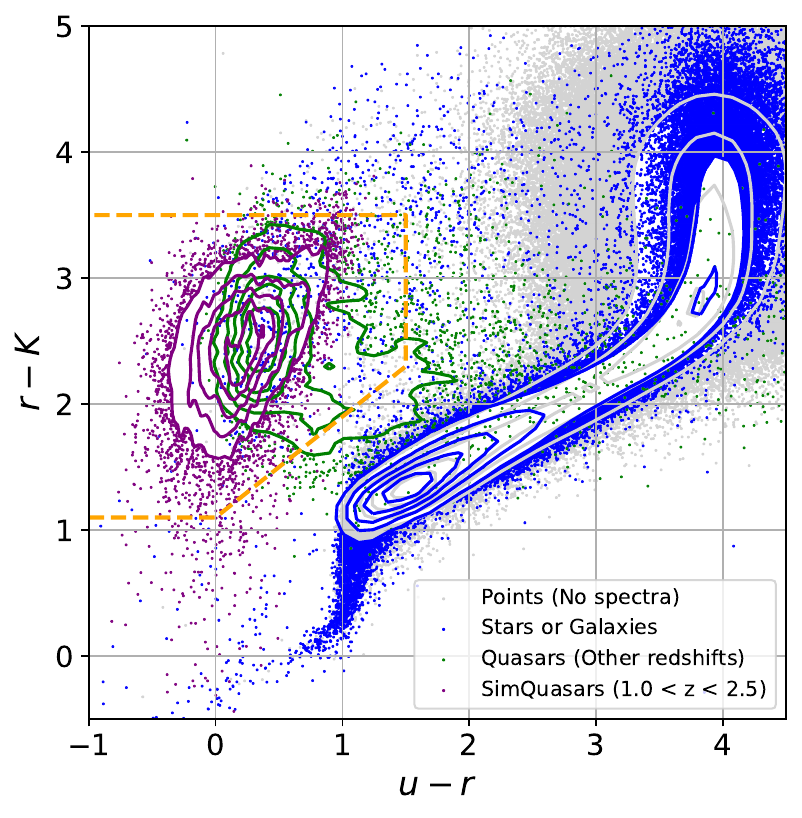}
        \label{fig:urK_simulation}
    }
    \hfill
    \subfigure[$2.5 < z < 3.5$]{
        \includegraphics[width=0.48\textwidth]{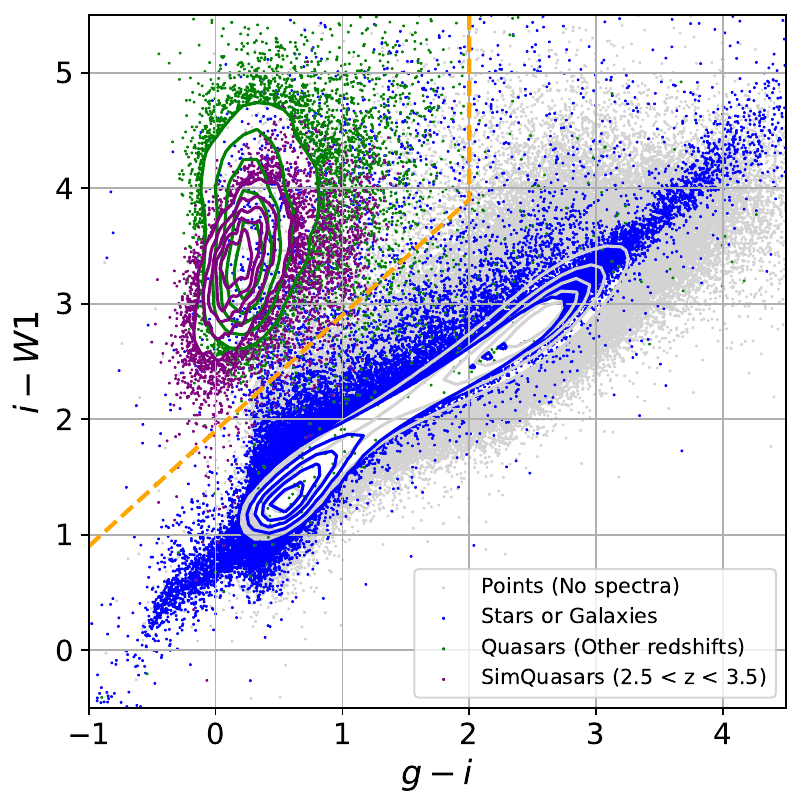}
        \label{fig:giW1_simulation}
    }
    \caption{Color-color diagrams of our simulated quasars. Panel (a) displays the redshift range of $1.0 < z < 2.5$ and covers a magnitude range of $15.0 < r < 21.5$. Panel (b) shows the redshift range of $2.5 < z < 3.5$ and covers a magnitude range of $15.0 < i < 21.5$. The purple dots represent the simulated quasars. For the convenience of comparison, we also plot the true distributions of point-like stars and galaxies with the blue dots. The orange dashed line represents our color selection criteria.}
    \label{fig:simulation_all}
\end{figure}

\subsection{Simulation}

To further test the effectiveness of our color selection method, we performed a simulation using the \href{https://simqso.readthedocs.io/en/latest/#}{simqso} package \citep{2021ascl.soft06008M}. Simqso is a collection of tools designed to generate simulated quasar spectra and photometric data. The quasar evolution model we use for simulation is the pure luminosity evolution (PLE) model at $1.0 < z < 2.5$ and the luminosity evolution density evolution (LEDE) model at $2.5 < z < 3.5$ of BOSS DR9 \citep{2013ApJ...773...14R}. The power-law slopes follow a Gaussian distribution, and the continuum is modeled as a series of broken power laws. We use the default values for these parameters. For each quasar, we simulate a spectral range from 3000-300,000 \AA. We set the spectral resolution as $R = 1000$. We simulate about 100,000 quasars for each of the two redshift ranges.

We use the SDSS, UKIDSS, and WISE photometric systems to calculate object magnitudes. Magnitude errors are also added so that these errors follow the magnitude-error distributions of real objects in each band. 
We randomly draw 10,000 quasars from the simulated samples and plot them in the color-color diagram (Figure \ref{fig:simulation_all}). For comparison, we also plot real stars, galaxies, and point sources without spectral data from the deep region. It shows that our color selection method is robust. For the simulated quasars, our color selection completeness is $97.8 \%$ for $1.0 < z < 2.5$ and $98.9 \%$ for $2.5 < z < 3.5$. Our color selection method effectively recovers the vast majority of quasars.

\section{Results}
\label{sec: Results}

\subsection{Completeness corrections}

In this subsection, we applied four major completeness corrections—morphology, cross-matching, color selection, and spectroscopic coverage—to derive the QLFs. The first incompleteness comes from morphology. In order to quickly exclude galaxies, we only selected point sources (SDSS type = 6) when selecting quasars. Distant quasars with strong detections are typically point sources. However, very faint point sources may be misclassified by the SDSS photometric pipeline as extended sources. We cross-match the quasar catalog with the SDSS photometric data to obtain the morphological classification of quasars and calculate the proportion of point-source quasars among all quasars. The calculated completeness is $99.2\% \pm 0.4\%$ in the wide region, and is $86.9\% \pm 6.5\%$ in the deep region.

The second incompleteness is from the cross-matching procedure. To obtain multi-band photometric data, we cross-matched SDSS photometry data with the UKIDSS catalog (VISTA catalog) and the UnWISE catalog in the wide region (deep region). However, some sources are faint in the infrared bands, resulting in a lack of corresponding UKIDSS, VISTA, or WISE data, which leads to missing matches in the cross-matching process. In the wide region, our average completeness in this procedure is $85.7\% \pm 7.1\%$, and in the deep region, the average completeness is $71.7\% \pm 14.1\%$. In addition, fainter quasars have lower completeness. The uncertainties of the completeness are calculated using the same approach as we did earlier. All these uncertainties will be propagated to the final calculation of QLFs.

The third incompleteness comes from the color selection method. Figure \ref{fig: urKgiW1_diagram_all} shows that a small fraction of quasars are outside of our selection criteria. We calculate the completeness based on this fraction. In the wide region, our completeness of the color selection is over 96\%, and in the deep region, our completeness of the color selection is about 95\%.

The final sample incompleteness comes from the incompleteness of spectral observations. Not all sources have spectral data. We assume that sources without spectra have the same fraction of quasars as those with spectra. In both wide and deep regions, our spectral completeness is over 87\%. 

\begin{figure}
    \centering
    \subfigure[Deep region]{
        \includegraphics[width=0.48\textwidth]{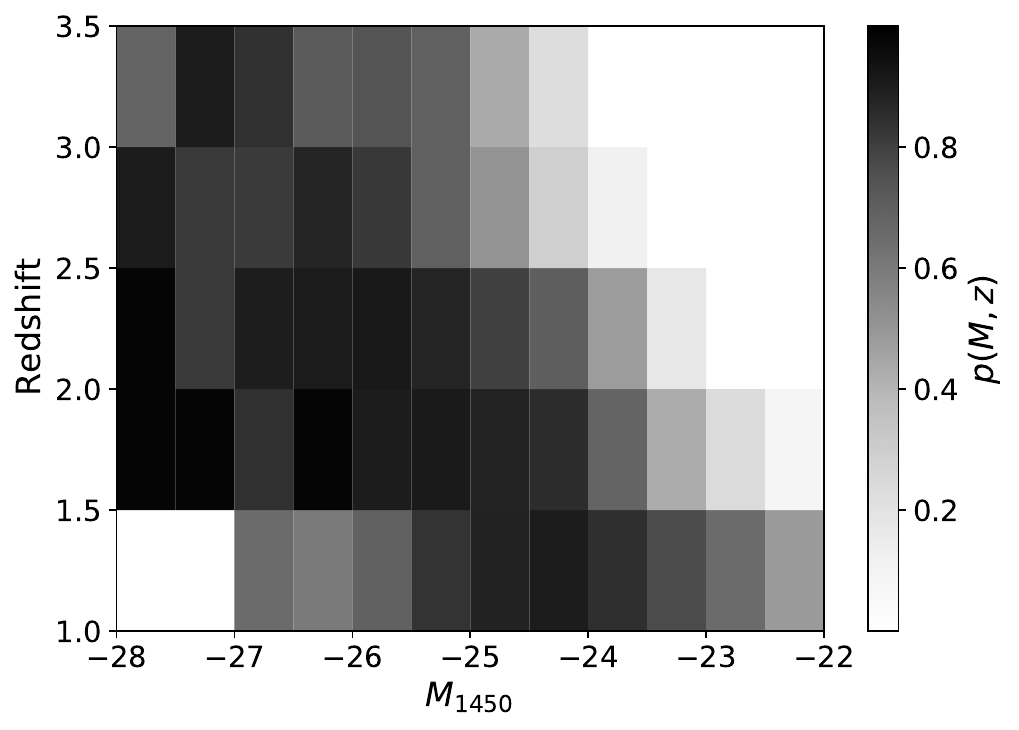}
        \label{fig:Selection_function_Stripe82Region}
    }
    \hfill
    \subfigure[Wide region]{
        \includegraphics[width=0.48\textwidth]{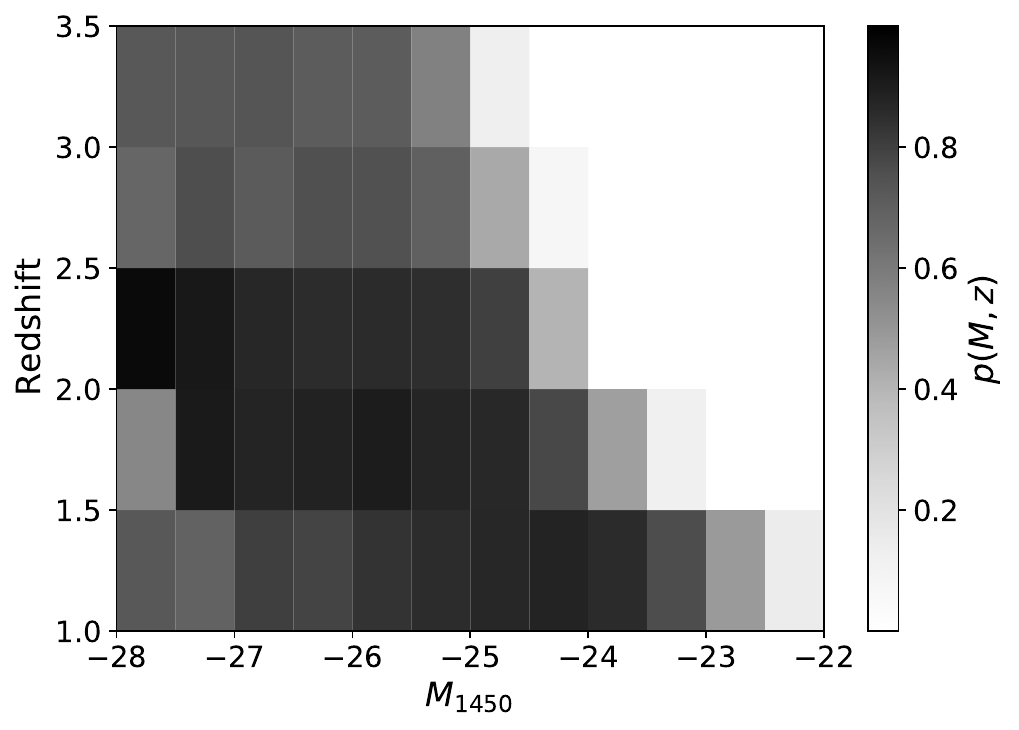}
        \label{fig:Selection_function_OtherRegion}
    }
    \caption{Distribution of the quasar selection function $p(M,z)$. Panel (a) shows the distribution of $p(M,z)$ in the deep region. A magnitude limit of $15.0 < r < 21.5\,\mathrm{mag}$ is adopted for $1.0<z<2.5$, while a magnitude limit of $15.0 < i < 21.5\,\mathrm{mag}$ is adopted for $2.5<z<3.5$. Panel (b) shows the distribution of $p(M,z)$ in the wide region. A magnitude limit of $15.0 < r < 20.5\,\mathrm{mag}$ is adopted for $1.0<z<2.5$, while a magnitude limit of $15.0 < i < 20.5\,\mathrm{mag}$ is adopted for $2.5<z<3.5$.}

    \label{fig:Selection_function}
\end{figure}

By combining these completeness corrections together, we obtain the final completeness correction, or sample selection function $p(M,z)$. This function gives the probability that a quasar with a given absolute magnitude $M_{1450}$ and redshift $z$ can be selected by our method.
We adopt the method described in \citet{2006AJ....131.2766R} to compute the absolute magnitude $M_{1450}$ from the observed $i$-band apparent magnitude. We choose $M_{1450}$ as the reference magnitude because the rest-frame 1450 \AA \ lies in a relatively line-free region of the quasar ultraviolet spectrum, and thus provides a robust tracer of the intrinsic AGN continuum with minimal contamination from strong emission lines.
Figure \ref{fig:Selection_function} shows the quasar selection function in the deep region and the wide region. The low-completeness regions at the faint end is mainly caused by the apparent-magnitude limit. In these regions, the sources are too faint to be detected, rather than being missed by our selection.

\subsection{QLFs}

We derive binned QLFs first. We set $p(M,z)$ as the quasar selection function, and the available volume \citep{1980ApJ...235..694A} for a quasar  with absolute magnitude $M$ and redshift $z$ in a magnitude bin $\Delta M$ and redshift bin $\Delta z$ is given by
\begin{equation}\label{V_a} 
  V_a = \iint_{\Delta M \Delta z} p(M,z) \frac{dV}{dz} dz dM.
\end{equation}
The binned QLF and its statistical uncertainties can be expressed as follows:
\begin{equation}\label{Phi(M,z)} 
  \Phi(M,z) = \sum \frac{1}{V_a^i},
\end{equation}

\begin{equation}\label{sigma_Phi(M,z)} 
  \sigma(\Phi) = \left[ \sum \left( \frac{1}{V_a^i} \right) ^2 \right] ^{\frac{1}{2}},
\end{equation}
where $V_a^i$ is the available volume in each bin, and the sum is over all quasars in each bin. 

We divide the redshift range of $1.0 < z < 3.5$ into 5 bins with a bin size of 0.5. In each redshift bin, we further divide the luminosity into several magnitude bins with a bin size of 0.5 magnitudes.

The $K$-correction is applied following the prescription given in \citet{2006AJ....131.2766R}. The calculated results are shown in Table \ref{tab: Binned QLFs}. The final errors $\Delta{\Phi}$ are obtained by adding the errors from each step of the completeness corrections to the statistical uncertainties $\sigma(\Phi)$. We plot the binned QLFs in Figure \ref{fig: QLFs}.

\subsection{Model fitting}

In order to characterize the QLFs at $1.0 < z < 3.5$, we adopt a double power-law model \citep{2000MNRAS.317.1014B} as the parametric form for the QLFs,

\begin{equation}\label{Phi_par} 
  \Phi_{par}(M,z) = \frac{\Phi^*}{10^{0.4(\alpha+1)(M-M^*)}+10^{0.4(\beta+1)(M-M^*)}},
\end{equation}
where $\alpha$ and $\beta$ represent the slopes of the faint and bright ends of the function, respectively, $M^*$ is the characteristic magnitude, and $\Phi^*$ is the normalization of the density.

We apply the maximum likelihood method \citep{1983ApJ...269...35M} to fit the derived luminosity function. The likelihood function is expressed as follows:
\begin{equation}  
    \begin{aligned}\label{S} 
      S = & -2 \sum \ln [\Phi(M_i,z_i)p(M_i,z_i)] \\ 
          & +2\int_{\Delta M} \int_{\Delta z} \Phi(M,z)p(M,z) \frac{dV}{dz} dz dM,
    \end{aligned}
\end{equation}

\newpage
\renewcommand{\arraystretch}{0.91}
\begin{longtable}{ccccc} 

\caption{Binned QLFs}
    \\
    \hline
    \hline
    $M_{1450}$ & $N$ & $\log(\Phi)$ & $\Delta{\Phi} /10^{-9}$ & Regions\\
    \hline
    \endfirsthead
    \hline
    \hline
    $M_{1450}$ & $N$ & $\log(\Phi)$ & $\Delta{\Phi} /10^9$ & Regions\\
    \hline
    \endhead
    \hline
    $1.0 < z < 1.5$ \\
    \hline
    $ -26.75 $ & $ 4 $ & $ -8.02 $ & $ 6.52 $ & Deep region \\
    $ -26.25 $ & $ 11 $ & $ -7.54 $ & $ 15.94 $ & Deep region \\
    $ -25.75 $ & $ 24 $ & $ -7.26 $ & $ 20.52 $ & Deep region \\
    $ -25.25 $ & $ 67 $ & $ -6.9 $ & $ 26.97 $ & Deep region \\
    $ -24.75 $ & $ 162 $ & $ -6.54 $ & $ 39.49 $ & Deep region \\
    $ -24.25 $ & $ 386 $ & $ -6.17 $ & $ 68.49 $ & Deep region \\
    $ -23.75 $ & $ 563 $ & $ -5.98 $ & $ 133.55 $ & Deep region \\
    $ -23.25 $ & $ 765 $ & $ -5.8 $ & $ 264.14 $ & Deep region \\
    $ -22.75 $ & $ 951 $ & $ -5.64 $ & $ 543.0 $ & Deep region \\
    $ -22.25 $ & $ 894 $ & $ -5.54 $ & $ 1047.3 $ & Deep region \\
    $ -27.75 $ & $ 3 $ & $ -8.99 $ & $ 0.74 $ & Wide region \\
    $ -27.25 $ & $ 5 $ & $ -8.74 $ & $ 1.1 $ & Wide region \\
    $ -26.75 $ & $ 20 $ & $ -8.21 $ & $ 2.04 $ & Wide region \\
    $ -26.25 $ & $ 57 $ & $ -7.74 $ & $ 4.53 $ & Wide region \\
    $ -25.75 $ & $ 172 $ & $ -7.29 $ & $ 8.49 $ & Wide region \\
    $ -25.25 $ & $ 435 $ & $ -6.9 $ & $ 15.56 $ & Wide region \\
    $ -24.75 $ & $ 1002 $ & $ -6.54 $ & $ 29.04 $ & Wide region \\
    $ -24.25 $ & $ 2084 $ & $ -6.23 $ & $ 48.91 $ & Wide region \\
    $ -23.75 $ & $ 3488 $ & $ -6.0 $ & $ 93.0 $ & Wide region \\
    $ -23.25 $ & $ 4838 $ & $ -5.8 $ & $ 199.84 $ & Wide region \\
    $ -22.75 $ & $ 4382 $ & $ -5.65 $ & $ 453.17 $ & Wide region \\
    $ -22.25 $ & $ 1715 $ & $ -5.55 $ & $ 953.36 $ & Wide region \\
    
    \hline
    $1.5 < z < 2.0$ \\
    \hline
    $ -27.75 $ & $ 2 $ & $ -8.57 $ & $ 1.94 $ & Deep region \\
    $ -27.25 $ & $ 4 $ & $ -8.27 $ & $ 2.75 $ & Deep region \\
    $ -26.75 $ & $ 6 $ & $ -8.02 $ & $ 4.63 $ & Deep region \\
    $ -26.25 $ & $ 32 $ & $ -7.36 $ & $ 8.0 $ & Deep region \\
    $ -25.75 $ & $ 91 $ & $ -6.87 $ & $ 20.76 $ & Deep region \\
    $ -25.25 $ & $ 222 $ & $ -6.49 $ & $ 37.11 $ & Deep region \\
    $ -24.75 $ & $ 410 $ & $ -6.21 $ & $ 70.63 $ & Deep region \\
    $ -24.25 $ & $ 709 $ & $ -5.96 $ & $ 128.96 $ & Deep region \\
    $ -23.75 $ & $ 875 $ & $ -5.77 $ & $ 363.15 $ & Deep region \\
    $ -23.25 $ & $ 777 $ & $ -5.62 $ & $ 923.19 $ & Deep region \\
    $ -22.75 $ & $ 544 $ & $ -5.52 $ & $ 1646.65 $ & Deep region \\
    $ -22.25 $ & $ 235 $ & $ -5.45 $ & $ 2756.83 $ & Deep region \\
    $ -27.75 $ & $ 4 $ & $ -8.82 $ & $ 1.11 $ & Wide region \\
    $ -27.25 $ & $ 17 $ & $ -8.41 $ & $ 1.12 $ & Wide region \\
    $ -26.75 $ & $ 49 $ & $ -7.93 $ & $ 2.45 $ & Wide region \\
    $ -26.25 $ & $ 203 $ & $ -7.32 $ & $ 6.21 $ & Wide region \\
    $ -25.75 $ & $ 542 $ & $ -6.9 $ & $ 11.61 $ & Wide region \\
    $ -25.25 $ & $ 1345 $ & $ -6.49 $ & $ 30.12 $ & Wide region \\
    $ -24.75 $ & $ 2604 $ & $ -6.2 $ & $ 57.14 $ & Wide region \\
    $ -24.25 $ & $ 4047 $ & $ -5.96 $ & $ 146.19 $ & Wide region \\
    $ -23.75 $ & $ 3848 $ & $ -5.77 $ & $ 392.25 $ & Wide region \\
    $ -23.25 $ & $ 1405 $ & $ -5.63 $ & $ 882.37 $ & Wide region \\
    
    \hline
    $2.0 < z < 2.5$ \\
    \hline
    $ -27.75 $ & $ 1 $ & $ -8.89 $ & $ 1.29 $ & Deep region \\
    $ -27.25 $ & $ 5 $ & $ -8.11 $ & $ 4.15 $ & Deep region \\
    $ -26.75 $ & $ 21 $ & $ -7.53 $ & $ 7.97 $ & Deep region \\
    $ -26.25 $ & $ 61 $ & $ -7.07 $ & $ 14.88 $ & Deep region \\
    $ -25.75 $ & $ 160 $ & $ -6.66 $ & $ 27.22 $ & Deep region \\
    $ -25.25 $ & $ 329 $ & $ -6.32 $ & $ 57.11 $ & Deep region \\
    $ -24.75 $ & $ 491 $ & $ -6.11 $ & $ 120.95 $ & Deep region \\
    $ -24.25 $ & $ 682 $ & $ -5.91 $ & $ 253.19 $ & Deep region \\
    $ -23.75 $ & $ 654 $ & $ -5.77 $ & $ 620.74 $ & Deep region \\
    $ -23.25 $ & $ 303 $ & $ -5.66 $ & $ 1456.5 $ & Deep region \\
    $ -27.75 $ & $ 5 $ & $ -8.99 $ & $ 0.48 $ & Wide region \\
    $ -27.25 $ & $ 38 $ & $ -8.09 $ & $ 1.69 $ & Wide region \\
    $ -26.75 $ & $ 138 $ & $ -7.5 $ & $ 4.84 $ & Wide region \\
    $ -26.25 $ & $ 386 $ & $ -7.05 $ & $ 11.72 $ & Wide region \\
    $ -25.75 $ & $ 890 $ & $ -6.69 $ & $ 22.8 $ & Wide region \\
    $ -25.25 $ & $ 1930 $ & $ -6.35 $ & $ 46.54 $ & Wide region \\
    $ -24.75 $ & $ 3212 $ & $ -6.1 $ & $ 99.41 $ & Wide region \\
    $ -24.25 $ & $ 2416 $ & $ -5.93 $ & $ 283.37 $ & Wide region \\
    
    \hline
    $2.5 < z < 3.0$ \\
    \hline
    $ -27.75 $ & $ 4 $ & $ -8.25 $ & $ 3.03 $ & Deep region \\
    $ -27.25 $ & $ 19 $ & $ -7.53 $ & $ 9.59 $ & Deep region \\
    $ -26.75 $ & $ 26 $ & $ -7.39 $ & $ 11.98 $ & Deep region \\
    $ -26.25 $ & $ 81 $ & $ -6.93 $ & $ 20.46 $ & Deep region \\
    $ -25.75 $ & $ 163 $ & $ -6.6 $ & $ 44.01 $ & Deep region \\
    $ -25.25 $ & $ 229 $ & $ -6.38 $ & $ 100.33 $ & Deep region \\
    $ -24.75 $ & $ 263 $ & $ -6.19 $ & $ 237.74 $ & Deep region \\
    $ -24.25 $ & $ 227 $ & $ -6.01 $ & $ 503.53 $ & Deep region \\
    $ -23.75 $ & $ 111 $ & $ -5.92 $ & $ 950.11 $ & Deep region \\
    $ -27.75 $ & $ 18 $ & $ -8.28 $ & $ 2.29 $ & Wide region \\
    $ -27.25 $ & $ 69 $ & $ -7.74 $ & $ 4.59 $ & Wide region \\
    $ -26.75 $ & $ 191 $ & $ -7.28 $ & $ 12.47 $ & Wide region \\
    $ -26.25 $ & $ 434 $ & $ -6.94 $ & $ 20.87 $ & Wide region \\
    $ -25.75 $ & $ 912 $ & $ -6.62 $ & $ 41.33 $ & Wide region \\
    $ -25.25 $ & $ 1483 $ & $ -6.38 $ & $ 83.24 $ & Wide region \\
    $ -24.75 $ & $ 1469 $ & $ -6.18 $ & $ 219.73 $ & Wide region \\
    $ -24.25 $ & $ 347 $ & $ -6.04 $ & $ 468.58 $ & Wide region \\
    
    \hline
    $3.0 < z < 3.5$ \\
    \hline
    $ -27.75 $ & $ 4 $ & $ -8.12 $ & $ 5.16 $ & Deep region \\
    $ -27.25 $ & $ 9 $ & $ -7.89 $ & $ 4.87 $ & Deep region \\
    $ -26.75 $ & $ 26 $ & $ -7.39 $ & $ 11.34 $ & Deep region \\
    $ -26.25 $ & $ 52 $ & $ -7.03 $ & $ 28.46 $ & Deep region \\
    $ -25.75 $ & $ 94 $ & $ -6.78 $ & $ 42.02 $ & Deep region \\
    $ -25.25 $ & $ 169 $ & $ -6.5 $ & $ 81.08 $ & Deep region \\
    $ -24.75 $ & $ 140 $ & $ -6.38 $ & $ 198.06 $ & Deep region \\
    $ -24.25 $ & $ 101 $ & $ -6.24 $ & $ 364.78 $ & Deep region \\
    $ -27.75 $ & $ 10 $ & $ -8.55 $ & $ 1.32 $ & Wide region \\
    $ -27.25 $ & $ 37 $ & $ -7.98 $ & $ 3.33 $ & Wide region \\
    $ -26.75 $ & $ 149 $ & $ -7.38 $ & $ 9.7 $ & Wide region \\
    $ -26.25 $ & $ 277 $ & $ -7.1 $ & $ 18.16 $ & Wide region \\
    $ -25.75 $ & $ 546 $ & $ -6.81 $ & $ 32.97 $ & Wide region \\
    $ -25.25 $ & $ 782 $ & $ -6.56 $ & $ 81.57 $ & Wide region \\
    $ -24.75 $ & $ 249 $ & $ -6.41 $ & $ 203.65 $ & Wide region \\
    \hline
    \multicolumn{5}{l}{\textup{Note: $\Phi$ and $\Delta{\Phi}$ is in units of $\mathrm{Mpc^{-3}mag^{-1}}$.}} \\
\label{tab: Binned QLFs}
\end{longtable}

\begin{figure*}[htbp]
    \centering
    \subfigure[$1.0<z<1.5$]{
        \includegraphics[width=0.31\textwidth]{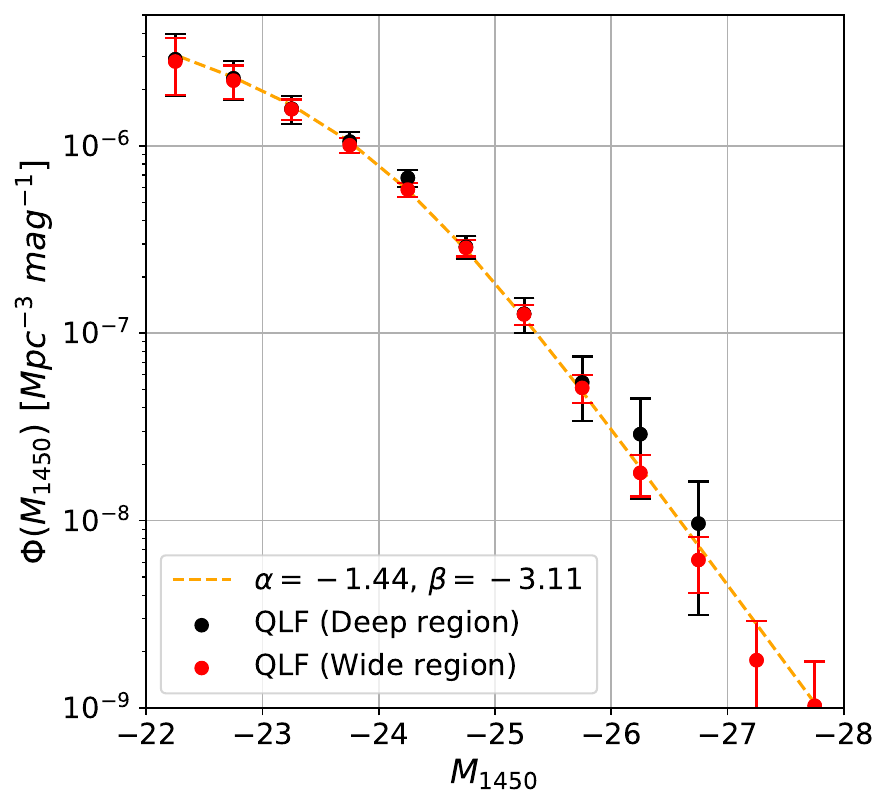}
        \label{fig:QLF1}
    }
    \hfill
    \subfigure[$1.5<z<2.0$]{
        \includegraphics[width=0.31\textwidth]{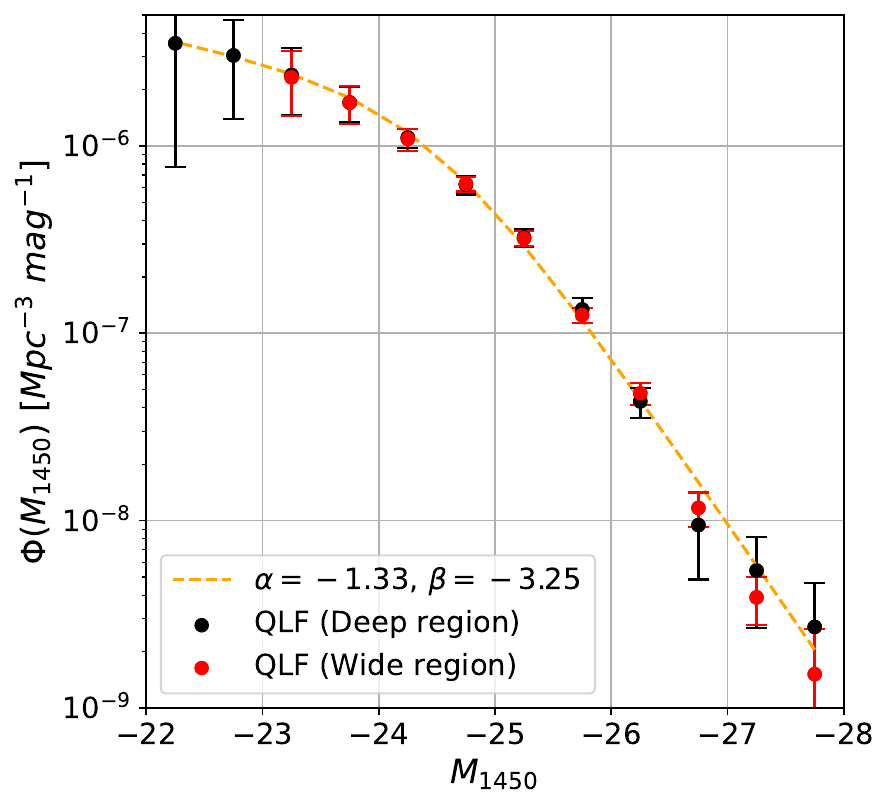}
        \label{fig:QLF2}
    }
    \hfill
    \subfigure[$2.0<z<2.5$]{
        \includegraphics[width=0.31\textwidth]{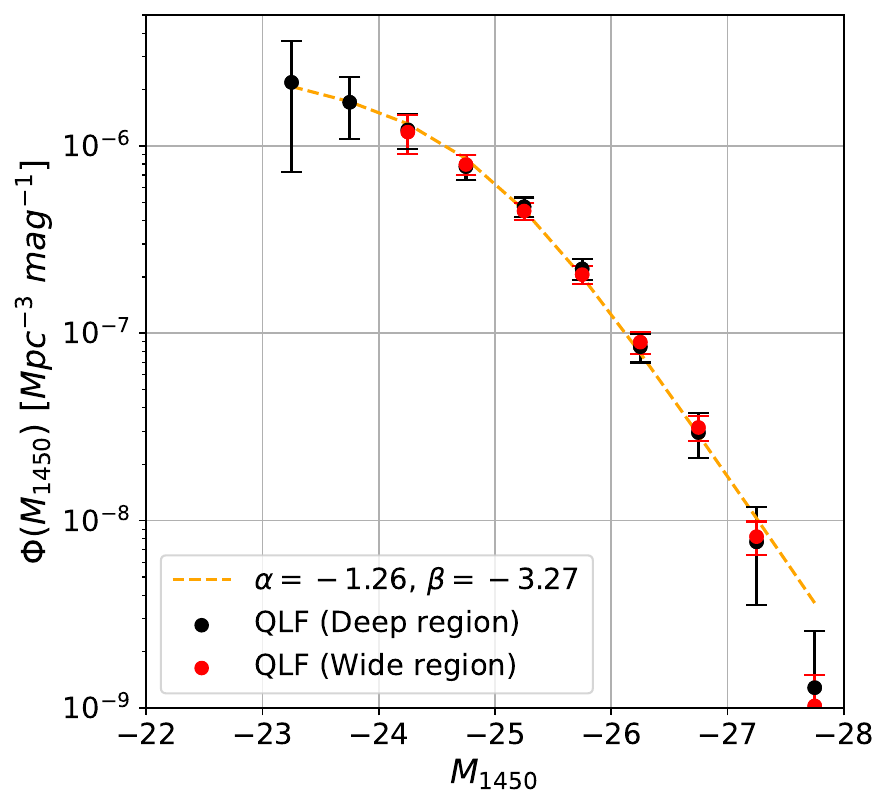}
        \label{fig:QLF3}
    }

    \subfigure[$2.0<z<3.0$]{
        \includegraphics[width=0.31\textwidth]{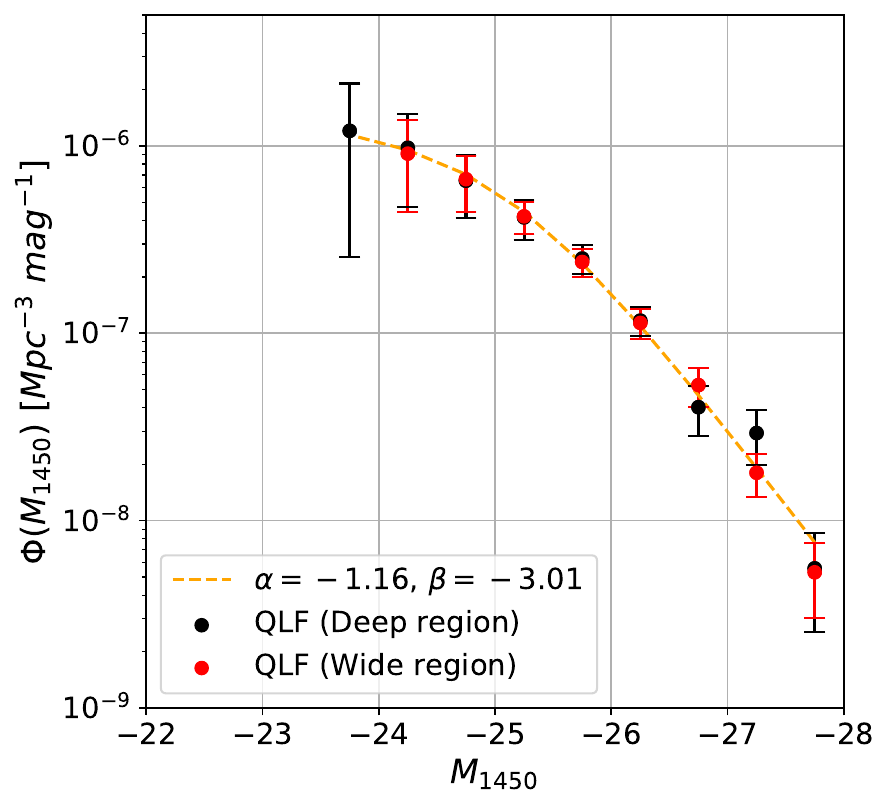}
        \label{fig:QLF4}
    }
    \hfill
    \subfigure[$3.0<z<3.5$]{
        \includegraphics[width=0.31\textwidth]{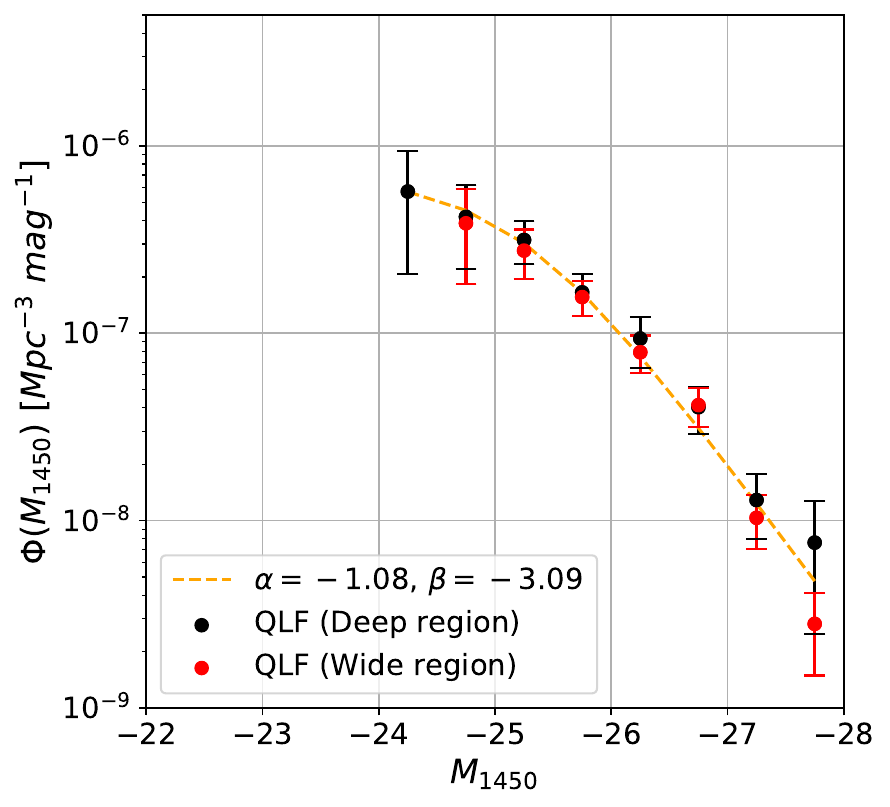}
        \label{fig:QLF5}
    }
    \hfill
    \subfigure{
        \includegraphics[width=0.31\textwidth]{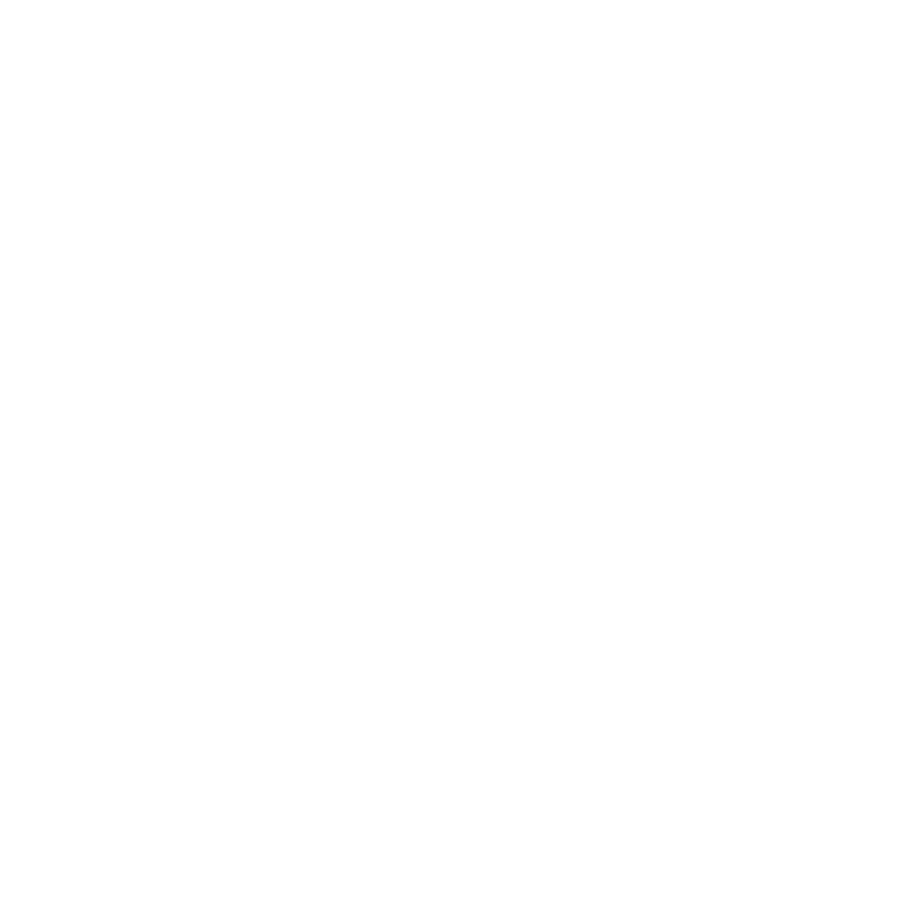}
        \label{fig:blank}
    }

    \caption{QLFs at $1.0<z<3.5$. The black solid dots represent the binned QLFs in the deep region, and the red solid dots represent the binned QLFs in the wide region. The orange dashed line represents the result of fitting the parameterized luminosity function using the maximum likelihood method, derived by combining the deep region data with the wide region data.}
    
    \label{fig: QLFs}
\end{figure*}

\newpage

where $p(M,z)$ is the selection function of quasars. We apply the MCMC method to obtain the best-fit results and estimate the associated uncertainties. The best-fit results are presented in Table \ref{tab: Parameters of the Best Fits}. We also plot the parametric QLFs with the orange dashed line in Figure \ref{fig: QLFs}. The results show that as the universe evolves, there is no clear evolutionary trend in the bright- and faint-end slopes of the QLFs, while the characteristic luminosity continues to decrease and the density normalization increases.

\section{Discussion}
\label{sec: Discussion}

\begin{table*}
\renewcommand{\arraystretch}{1.2}
\centering
\caption{Parameters of the Best Fits}
\begin{tabular}{ccccc}
\hline
\hline
redshift & $\alpha$ & $\beta$ & $M^*$ & $\log \Phi^*$\\
\hline
$1.0 < z < 1.5$ & $-1.44_{-0.05}^{+0.04}$ & $-3.11_{-0.06}^{+0.05}$ & $-23.98_{-0.10}^{+0.09}$ & $-5.79_{-0.05}^{+0.04}$\\
$1.5 < z < 2.0$ & $-1.33_{-0.04}^{+0.04}$ & $-3.25_{-0.04}^{+0.04}$ & $-24.46_{-0.06}^{+0.06}$ & $-5.73_{-0.03}^{+0.03}$\\
$2.0 < z < 2.5$ & $-1.26_{-0.07}^{+0.07}$ & $-3.27_{-0.06}^{+0.06}$ & $-24.89_{-0.09}^{+0.10}$ & $-5.83_{-0.05}^{+0.05}$\\
$2.5 < z < 3.0$ & $-1.16_{-0.14}^{+0.15}$ & $-3.01_{-0.09}^{+0.08}$ & $-25.12_{-0.19}^{+0.18}$ & $-5.99_{-0.09}^{+0.08}$\\
$3.0 < z < 3.5$ & $-1.08_{-0.13}^{+0.06}$ & $-3.09_{-0.07}^{+0.07}$ & $-25.23_{-0.11}^{+0.10}$ & $-6.21_{-0.05}^{+0.04}$\\
\hline
\end{tabular}
\label{tab: Parameters of the Best Fits}
\end{table*}

\subsection{Evolution models}

Most previous works suggested that QLFs follow the pure luminosity evolution (PLE) model at $z \lesssim 2.2$ and the luminosity evolution density evolution (LEDE) model at $z \gtrsim 2.2$ \citep{2009MNRAS.399.1755C, 2013ApJ...773...14R, 2013A&A...551A..29P, 2016A&A...587A..41P, 2018AJ....155..110Y}. However, some studies suggest that for QLFs with $z \gtrsim 2.0$, the pure density evolution (PDE) model is sufficient to describe the quasar evolution  \citep{2021ApJ...910L..11K, 2022ApJ...928..172P}. We test the PLE model, PDE model, and LEDE model separately. For the PLE model, we use the following definition \citep{2009MNRAS.399.1755C}:
\begin{equation}\label{PLE} 
  M^{*}(z) = M^{*}(0) - 2.5(k_1 z + k_2 z^{2}).
\end{equation}
The PDE model follows the formula by \cite{2021ApJ...910L..11K}:
\begin{equation}\label{PDE} 
  \log [\Phi^{*}(z)] = \log [\Phi^{*}(z=z_p)] + k_1(z-z_p),
\end{equation}
we set $z_p = 1.0$  and 2.5 for the two redshift ranges $1.0 < z < 2.5$ and $z > 2.5$, respectively. The LEDE model follows the formula by \cite{2013ApJ...773...14R}:
\begin{equation}\label{LEDE_1} 
  \log [\Phi^{*}(z)] = \log [\Phi^{*}(z=z_p)] + k_1(z-z_p),
\end{equation}
\begin{equation}\label{LEDE_2} 
  M^{*}(z) = M^{*}(z = z_p) + k_2(z - z_p).
\end{equation}
In addition to the three models mentioned above, we also explore the model in which all parameters evolve, referred to as a ``free case":
\begin{equation}\label{Free_case} 
  X(z) = X_0 + k_i(z - z_p),
\end{equation}
where $X \in \{\log \Phi^*, M^*, \alpha, \beta \}$, and the coefficient $k_i \in \{k_{\log \Phi^*}, k_{M^*}, k_{\alpha}, k_{\beta} \}$.

We use the four models to fit the QLFs, and then use the reduced $\chi ^2_{\nu}$ to evaluate the goodness of fit:
\begin{equation}\label{chi} 
  \chi ^2_{\nu} = \frac{\sum{[(\Phi_{obs}-\Phi_{th})/{\sigma}]^2}}{n- \nu},
\end{equation}
where $\Phi_{obs}$ is the binned QLF obtained from the calculation, $\Phi_{th}$ is the binned QLF derived using the quasar evolution model, $n$ is the number of data points, and $\nu$ is the number of free parameters.

The results are listed in Table \ref{tab: Evolution models}. The limited number of data points leads to overfitting, so we consider both the $\chi ^2$ values and the degrees of freedom (dof) when evaluating the goodness of fit. In the redshift range of $1.0 < z < 2.5$, the PLE model has the best fitting performance, which is consistent with previous results \citep{2013ApJ...773...14R, 2016A&A...587A..41P, 2018AJ....155..110Y}. This indicates that during this stage, the number density of quasars remains approximately constant, while their average luminosity evolves systematically over time. Therefore, the PLE model provides a better description of the observational data. In the redshift range of $2.5 < z < 3.5$, the $\chi ^2_{\nu}$ values of the three models show no significant differences, mainly because only two redshift bins are included. However, since the PDE model has one fewer parameter compared to the PLE and LEDE models, we consider that the PDE model is enough to describe quasar evolution within this redshift range. This conclusion is consistent with the results of \cite{2021ApJ...910L..11K}.

\subsection{Comparison with previous results}

\begin{figure*}[htbp]
    \centering
    \subfigure[$1.0<z<1.5$]{
        \includegraphics[width=0.31\textwidth]{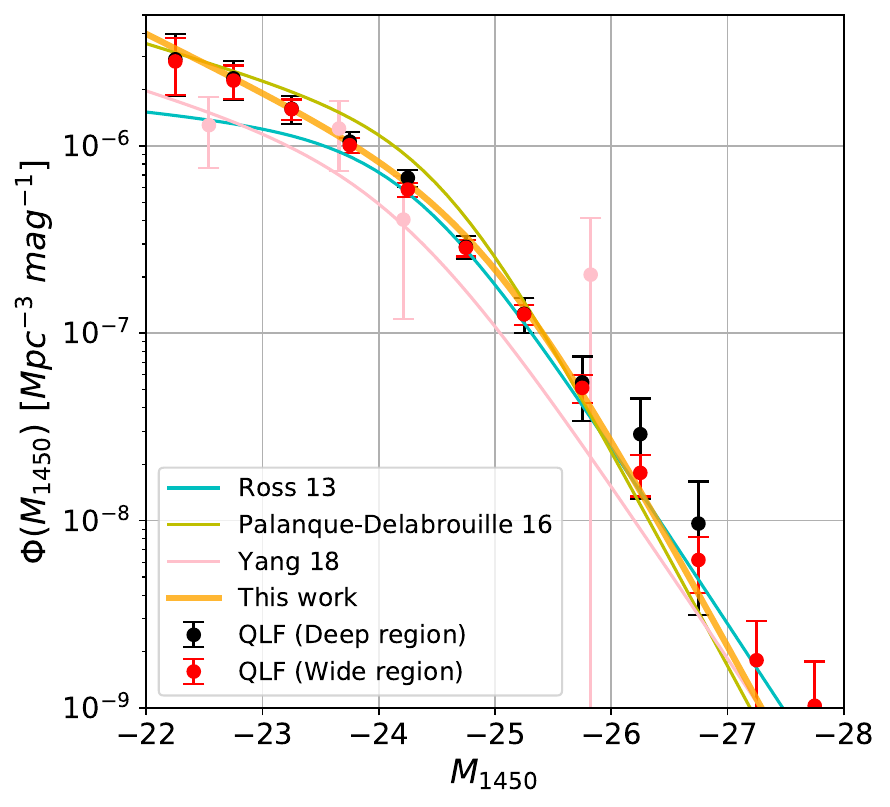}
        \label{fig:QLF1_Comparison}
    }
    \hfill
    \subfigure[$1.5<z<2.0$]{
        \includegraphics[width=0.31\textwidth]{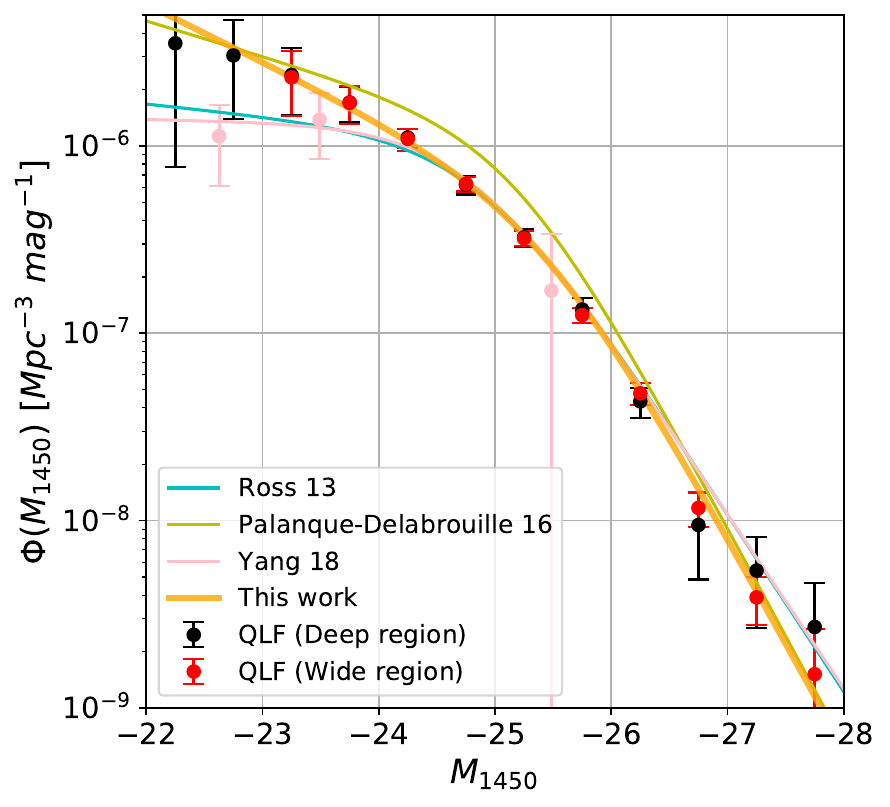}
        \label{fig:QLF2_Comparison}
    }
    \hfill
    \subfigure[$2.0<z<2.5$]{
        \includegraphics[width=0.31\textwidth]{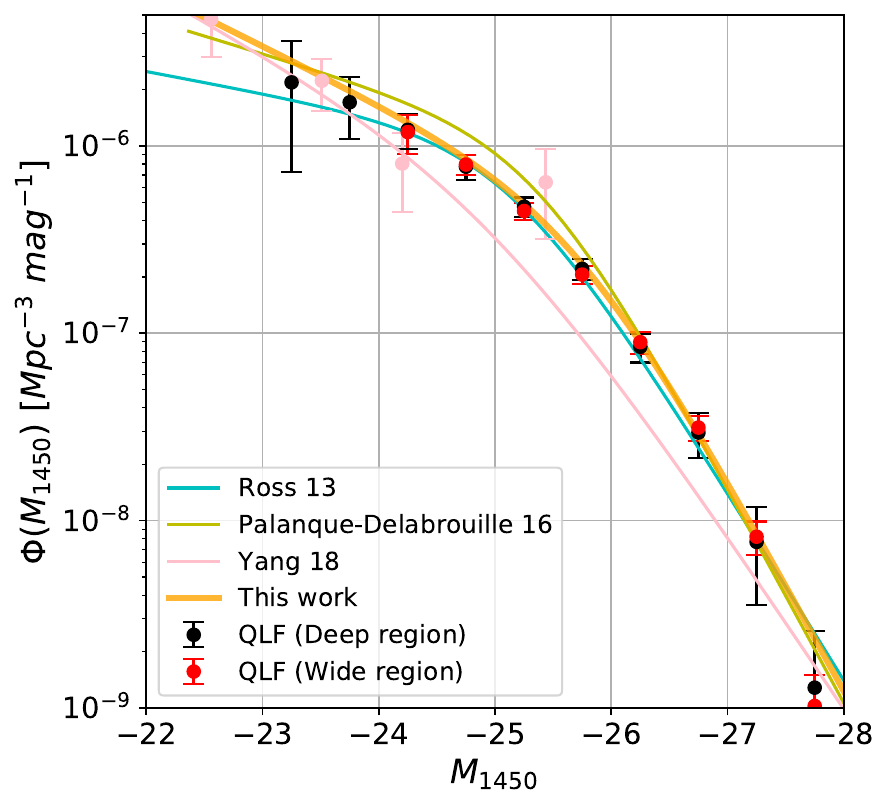}
        \label{fig:QLF3_Comparison}
    }

    \subfigure[$2.5<z<3.0$]{
        \includegraphics[width=0.31\textwidth]{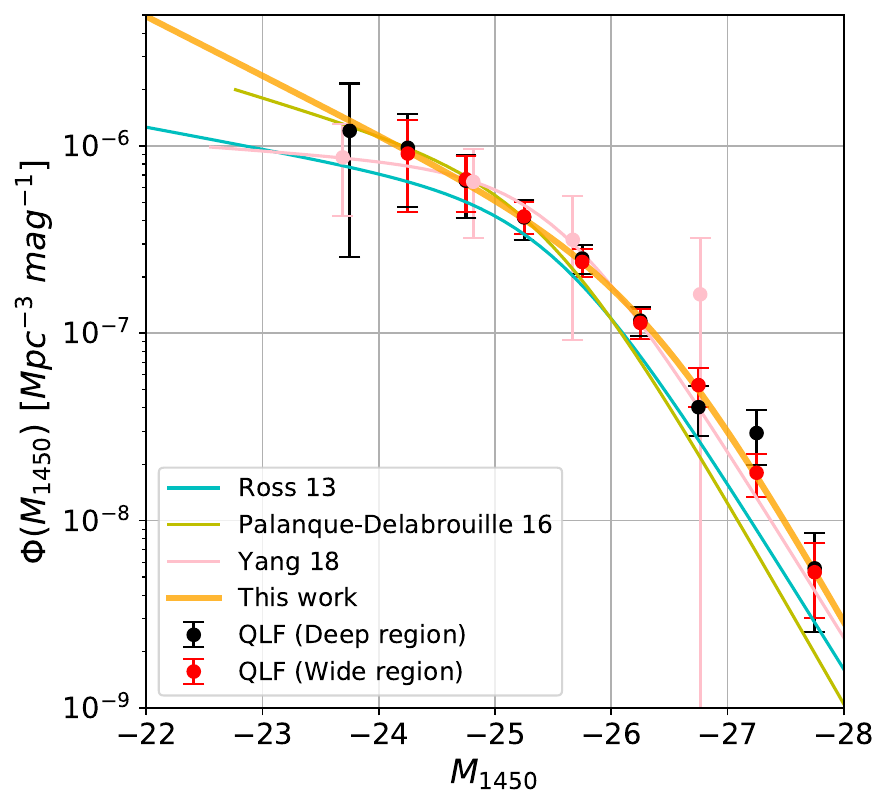}
        \label{fig:QLF4_Comparison}
    }
    \hfill
    \subfigure[$3.0<z<3.5$]{
        \includegraphics[width=0.31\textwidth]{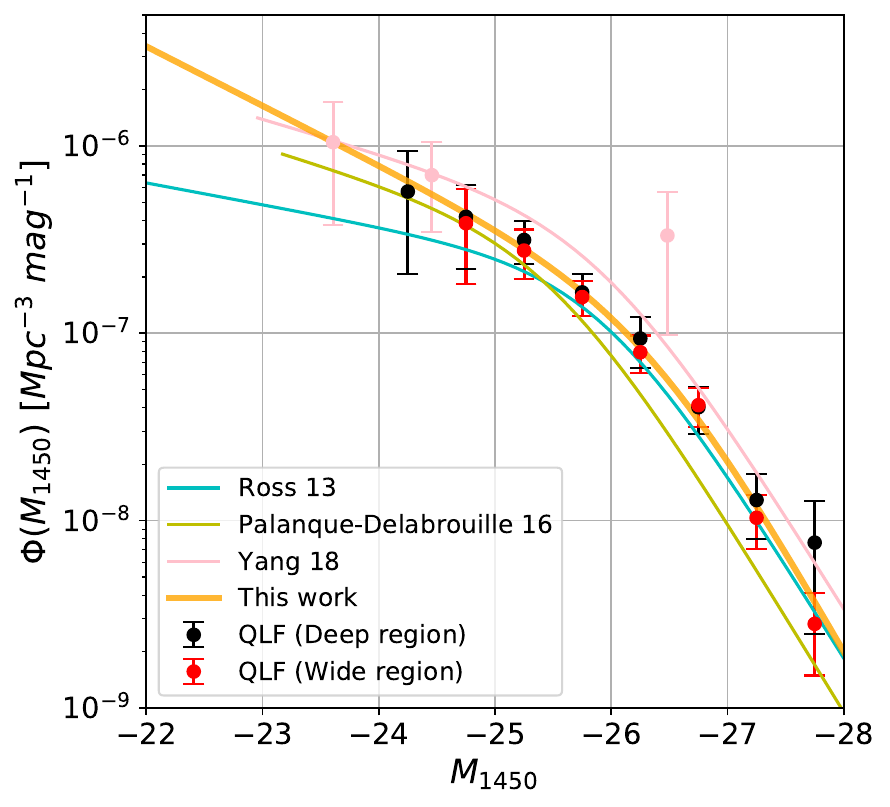}
        \label{fig:QLF5_Comparison}
    }
    \hfill
    \subfigure{
        \includegraphics[width=0.31\textwidth]{QLFblank.pdf}
    }

    \caption{Comparison of QLFs with others. The solid orange line represents the QLFs using the PLE model for the redshift range $1.0 < z < 2.5$ and the PDE model for the redshift range $2.5 < z < 3.5$. For comparison, the solid cyan line represents the results from \cite{2013ApJ...773...14R}, the solid yellow line shows the results from \cite{2016A&A...587A..41P}, and the pink line indicates the results from \cite{2018AJ....155..110Y}. The faint end of the results from \cite{2016A&A...587A..41P} and \cite{2018AJ....155..110Y} are truncated based on the limiting apparent magnitude of their quasar selection.}
    \label{fig:QLFs_Comparison}
\end{figure*}

In Figure \ref{fig:QLFs_Comparison}, we compare our QLFs with those from previous studies \citep{2013ApJ...773...14R, 2016A&A...587A..41P, 2018AJ....155..110Y}. For the results from \cite{2013ApJ...773...14R}, we use their PLE model for $z < 2$ and the LEDE model for the deep region when $z > 2$. For the results from \cite{2016A&A...587A..41P}, we use their PLE+LEDE model for comparison. For the results from \cite{2018AJ....155..110Y}, we apply the PLE model for $z < 2.5$ and the LEDE model for $z > 2.5$. To facilitate a direct comparison with previous studies, we convert the magnitudes used in their QLFs into $M_{1450}$.

Compared with \cite{2013ApJ...773...14R}, our QLFs show higher number densities at the faint end in most redshift bins. At $z\sim2.7$, our bright-end QLF is also higher. These differences are mainly related to the different quasar selections and completeness estimates. To make a direct comparison, we select a common sky region and focus on a redshift range $2.5<z<3.0$. In this region, nearly all quasars in the statistical sample of \cite{2013ApJ...773...14R} are recovered by our optical-infrared color selection. In addition, our selection identifies about 1.5 times more spectroscopically confirmed quasars than those included in the sample of \cite{2013ApJ...773...14R}. These additional quasars are systematically redder than those in the sample of \cite{2013ApJ...773...14R}. This is because the quasar selection in \cite{2013ApJ...773...14R} was based on the BOSS XDQSO color selection without infrared data. As a result, the completeness is relatively low. Their Figure 7 shows that the selection efficiency is usually about $30\%$--$75\%$ in most bins, and the completeness is only $\sim 50\%$ even for bright quasars at $z\sim2.7$. At this redshift, quasar colors are very similar to those of A/F stars in the optical. In contrast, our optical-infrared color selection reaches a higher completeness of $\sim 80\%$ for bright quasars at the same redshift. This allows us to recover quasars more effectively, and explains why our bright-end QLF is higher near $z\sim2.7$.

Compared with \cite{2016A&A...587A..41P}, our QLF is higher at the bright end at $2.5<z<3.5$. This difference is related to the survey area and the limited number of very bright quasars in their sample. Their analysis is mainly based on the Stripe~82 region, which covers only about $94.5,\mathrm{deg}^2$. Indeed, their sample contains no quasars brighter than $g<18$ mag. We further test this effect using our deep-field data. For quasars at $2.5<z<3.5$ and $-28<M_{1450}<-27$, about half of them have apparent magnitudes brighter than $g=18$ mag. Therefore, a non-negligible fraction of the most luminous quasars in this redshift and luminosity range would not be represented in their sample. The absence of these very bright quasars explains why their bright-end QLF is lower than our result.

The \cite{2018AJ....155..110Y} QLF was derived from a deep survey covering only $\sim2\,\mathrm{deg}^2$, which is much smaller than the area used in our study. This leads to large statistical uncertainties, and thus a direct comparison is not reasonable.


\begin{figure}[ht!]
\centering
\includegraphics[width=0.45\textwidth]{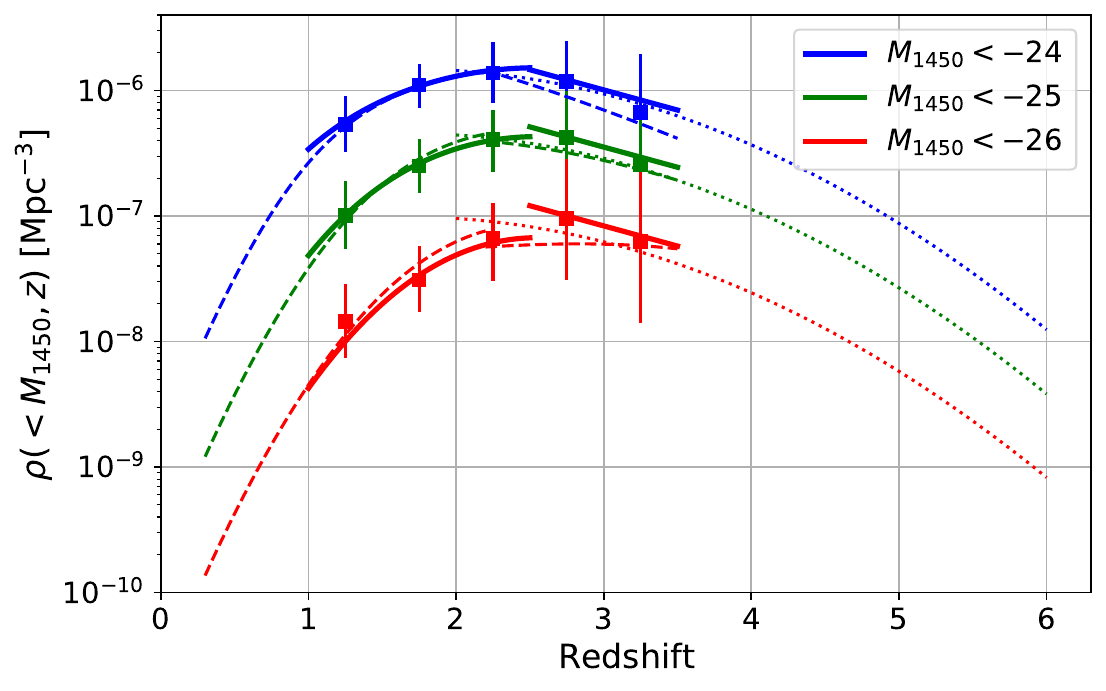}
\caption{Cumulative density evolution of quasars at $1.0 < z < 3.5$. The blue, green, and red lines represent magnitude ranges $M_{1450} < - 25$, $ - 26$, and $ - 27$ mag, respectively. Square points and solid lines are our results. For comparison, the dashed line represents the results from \cite{2013ApJ...773...14R}, and the dotted line shows the results from \cite{2021ApJ...910L..11K}.}
\label{fig: Cumulative_Spatial_Density_Evolution}
\end{figure}

We also investigated the cumulative spatial density evolution of quasars within the redshift range $1.0<z<3.5$. The spatial density of quasars brighter than a specified magnitude $M$ is determined by integrating the QLF,
\begin{equation}\label{rho} 
  \rho(<M, z) = \int_{- \infty}^{M} \Phi(M,z) dM,
\end{equation}
where we apply the PLE model for the redshift range $1.0 < z < 2.5$ and the PDE model for $2.5<z<3.5$. We set the specified magnitude $M$ to $-24,-25$, and $-26$, respectively. Figure \ref{fig: Cumulative_Spatial_Density_Evolution} shows the cumulative spatial density evolution based on our model. Overall, our results are consistent with those from previous studies. However, a slight difference is observed in the range of $1.0<z<2.5$, where our evolution is shallower compared with results from \cite{2013ApJ...773...14R}. 

Since our results are based on the DESI Jura dataset, which will not be publicly released, we further examined the differences between the Jura and Loa datasets. The Jura dataset is derived from a version of Year 3 data that closely corresponds to, but is not identical with, DESI Data Release 2, due to differences in both the number of observed targets and the data processing pipeline \citep{2023AJ....165..144G}. The Loa dataset will be included with DR2. By cross-matching the spectra from Loa and Jura, we found that approximately $3.5 \%$ of the spectra are included in Loa but absent in Jura. Given that the quasar luminosity function is calculated in logarithmic space, this small difference has a negligible impact and does not affect our final results.

\begin{table*}[htbp]
\renewcommand{\arraystretch}{1.2}
\centering
\caption{Evolution Models}
\begin{tabular}{cccccccccc}
\hline
\hline
{} & ${\alpha}_0$ & ${\beta}_0$ & $M_s$ & ${\Phi}^*(z=z_p)$ & $k_1$ & $k_2$ & $k_3$ & $k_4$ &$\chi ^2_{\nu} (\chi ^2 / dof)$ \\
\hline
$1.0 < z < 2.5$ \\
\hline
PLE & $-1.77_{-0.07}^{+0.08}$ & $-3.84_{-0.14}^{+0.12}$ & $-22.60_{-0.03}^{+0.03}$ & $-6.30_{-0.02}^{+0.02}$ & $0.98_{-0.01}^{+0.01}$ & $-0.19_{-0.01}^{+0.01}$ & $\cdots$ & $\cdots$ & 0.41(22.73/56) \\ 
PDE & $-1.96_{-0.04}^{+0.05}$ & $-3.82_{-0.14}^{+0.12}$ & $-25.57_{-0.03}^{+0.03}$ & $-6.84_{-0.02}^{+0.02}$ & $0.47_{-0.02}^{+0.02}$ & $\cdots$ & $\cdots$ & $\cdots$ & 1.12(63.59/57) \\
LEDE & $-1.83_{-0.06}^{+0.07}$ & $-3.89_{-0.15}^{+0.13}$ & $-24.96_{-0.03}^{+0.03}$ & $-6.44_{-0.02}^{+0.02}$ & $0.08_{-0.02}^{+0.02}$ & $-0.64_{-0.03}^{+0.04}$ & $\cdots$& $\cdots$ & 0.70(39.34/56) \\ 
Free case & $-1.77_{-0.08}^{+0.09}$ & $-3.39_{-0.17}^{+0.14}$ & $-24.61_{-0.04}^{+0.04}$ & $-6.23_{-0.02}^{+0.02}$ & $-0.11_{-0.03}^{+0.03}$ & $-0.96_{-0.04}^{+0.04}$ & $0.00_{-0.11}^{+0.13}$ & $-0.51_{-0.19}^{+0.15}$ & 0.60(32.64/54) \\
\hline

\hline
$2.5 < z < 3.5$ \\
\hline
PLE & $-1.87_{-0.16}^{+0.20}$ & $-3.82_{-0.49}^{+0.36}$ & $-21.37_{-0.08}^{+0.09}$ & $-6.83_{-0.06}^{+0.05}$ & $1.55_{-0.01}^{+0.01}$ & $-0.29_{-0.01}^{+0.01}$ & $\cdots$ & $\cdots$ & 0.25(6.61/26) \\ 
PDE & $-1.79_{-0.16}^{+0.21}$ & $-3.77_{-0.39}^{+0.30}$ & $-26.34_{-0.07}^{+0.08}$ & $-6.60_{-0.05}^{+0.04}$ & $-0.32_{-0.11}^{+0.09}$ & $\cdots$ & $\cdots$ & $\cdots$ & 0.19(5.25/27) \\ 
LEDE & $-1.79_{-0.18}^{+0.23}$ & $-3.77_{-0.44}^{+0.32}$ & $-26.32_{-0.08}^{+0.09}$ & $-6.59_{-0.06}^{+0.05}$ & $-0.34_{-0.12}^{+0.10}$ & $-0.03_{-0.15}^{+0.18}$ & $\cdots$ & $\cdots$ & 0.20(5.24/26) \\ 
Free case & $-1.49_{-0.20}^{+0.27}$ & $-2.99_{-0.46}^{+0.31}$ & $-25.59_{-0.09}^{+0.10}$ & $-6.11_{-0.07}^{+0.06}$ & $-1.34_{-0.14}^{+0.11}$ & $-1.54_{-0.16}^{+0.20}$ & $-0.59_{-0.31}^{+0.45}$ & $-1.87_{-1.37}^{+0.93}$ & 0.18(4.33/24) \\
\hline

\hline
\end{tabular}
\label{tab: Evolution models}
\end{table*}

\section{Summary}
\label{sec: Summary}

In this paper, we have used the quasar catalogs from SDSS and DESI as a library to study the evolution of type 1 quasars at $1.0<z<3.5$. We defined two sky regions, including a deep region of $\sim$265 deg$^2$ with a depth reaching 21.5 mag, and a wide region of $\sim$1700 deg$^2$ with a depth of 20.5 mag. Using photometric data from SDSS, UKIDSS, VISTA, and WISE, and employing a straightforward color selection method, we built a quasar sample with high completeness and high purity.

We then measured the QLFs at $1.0 < z < 3.5$. We first assessed the completeness due to the morphological classification, cross-matching, color selection methods, and incomplete spectra. Subsequently, we derived binned QLFs and fitted them using a double power-law model. Compared with previous studies, our QLF shows some differences. In most redshift bins, it has higher number densities at the faint end, and it is also higher at the bright end at $2.5<z<3.5$. These differences are mainly due to different quasar selection methods, completeness corrections, survey areas, and fitting methods. In our analysis, we apply a uniform color selection to a large spectroscopic quasar sample from SDSS and DESI, together with a consistent completeness correction. With a higher completeness and a larger survey area, our result gives a more uniform and self-consistent picture of quasar evolution over the redshift range of $1.0 < z < 3.5$.

We further examined the evolution of QLFs by evaluating the fit of three models: PLE, PDE, and LEDE. Our analysis found that the PLE model effectively describes the quasar evolution in the redshift range of $1.0 < z < 2.5$. It implies that the changes in the luminosity distribution of these quasars likely stem from systematic evolution in black hole accretion efficiency, rather than significant variations in their population size. For the higher redshift range of $2.5 < z < 3.5$, the PDE model is enough to characterize the quasar evolution in this redshift range. It suggests a continuous increase in the spatial number density of this quasar population during this epoch. This trend reflects the cosmological characteristics of their early formation and active phase. The transition between these two evolutionary modes reveals fundamental differences in the dominant physical mechanisms governing quasars across different cosmic epochs.

Looking ahead, with the upcoming deployment of a new generation of survey projects such as the Euclid Space Telescope \citep{2025A&A...697A...1E} and the China Space Station Telescope (CSST) \citep{2025arXiv250704618C}, we will gain access to deeper and more systematic near-infrared data. These datasets will substantially improve constraints on the shape and evolution of QLFs at high redshifts, offering critical insights into the physical connection between quasar activity and the growth of SMBHs.

\section{acknowledgments}
We acknowledge support from the National Key R\&D Program of China (2021YFA1600404) and the National Science Foundation of China (12225301).

This research used data obtained with the Sloan Digital Sky Survey (SDSS). Funding for SDSS IV has been provided by the Alfred P. Sloan Foundation, the U.S. Department of Energy Office of Science, and the Participating Institutions. SDSS-IV acknowledges support and resources from the Center for High Performance Computing at the University of Utah. The SDSS website is www.sdss4.org.

This research used data obtained with the Dark Energy Spectroscopic Instrument (DESI). This material is based upon work supported by the U.S. Department of Energy (DOE), Office of Science, Office of High-Energy Physics, under Contract No. DE–AC02–05CH11231, and by the National Energy Research Scientific Computing Center, a DOE Office of Science User Facility under the same contract. Additional support for DESI was provided by the U.S. National Science Foundation (NSF), Division of Astronomical Sciences under Contract No. AST-0950945 to the NSF’s National Optical-Infrared Astronomy Research Laboratory; the Science and Technology Facilities Council of the United Kingdom; the Gordon and Betty Moore Foundation; the Heising-Simons Foundation; the French Alternative Energies and Atomic Energy Commission (CEA); the National Council of Humanities, Science and Technology of Mexico (CONAHCYT); the Ministry of Science, Innovation and Universities of Spain (MICIU/AEI/10.13039/501100011033), and by the DESI Member Institutions: \url{https://www.desi.lbl.gov/collaborating-institutions}. Any opinions, findings, and conclusions or recommendations expressed in this material are those of the author(s) and do not necessarily reflect the views of the U. S. National Science Foundation, the U. S. Department of Energy, or any of the listed funding agencies. The authors are honored to be permitted to conduct scientific research on I'oligam Du'ag (Kitt Peak), a mountain with particular significance to the Tohono O'odham Nation.

S.P. is supported by the international Gemini Observatory, a program of NSF NOIRLab, which is managed by the Association of Universities for Research in Astronomy (AURA) under a cooperative agreement with the U.S. National Science Foundation, on behalf of the Gemini partnership of Argentina, Brazil, Canada, Chile, the Republic of Korea, and the United States of America.

$Facilities$: SDSS, DESI, UKIDSS, VISTA, WISE

$Software$: Astropy \citep{2013A&A...558A..33A, 2018AJ....156..123A, 2022ApJ...935..167A}, Topcat \citep{2005ASPC..347...29T}, emcee \citep{2013PASP..125..306F}.

All data records needed to reproduce the figures of this publication are available on Zenodo at \href{https://doi.org/10.5281/zenodo.21442096}{doi:10.5281/zenodo.21442096}.

\bibliography{MyRef}{}

@ARTICLE{2011ApJ...739...56D,
       author = {{De Rosa}, G. and {Decarli}, R. and {Walter}, F. and {Fan}, X. and {Jiang}, L. and {Kurk}, J. and {Pasquali}, A. and {Rix}, H.~W.},
        title = "{Evidence for Non-evolving Fe II/Mg II Ratios in Rapidly Accreting z \raisebox{-0.5ex}\textasciitilde 6 QSOs}",
      journal = {\apj},
         year = 2011,
        month = oct,
       volume = {739},
       number = {2},
          eid = {56},
        pages = {56},
          doi = {10.1088/0004-637X/739/2/56},
archivePrefix = {arXiv},
       eprint = {1106.5501},
 primaryClass = {astro-ph.CO},
       adsurl = {https://ui.adsabs.harvard.edu/abs/2011ApJ...739...56D}
}

@ARTICLE{2000ApJ...533..631K,
       author = {{Kaspi}, Shai and {Smith}, Paul S. and {Netzer}, Hagai and {Maoz}, Dan and {Jannuzi}, Buell T. and {Giveon}, Uriel},
        title = "{Reverberation Measurements for 17 Quasars and the Size-Mass-Luminosity Relations in Active Galactic Nuclei}",
      journal = {\apj},
         year = 2000,
        month = apr,
       volume = {533},
       number = {2},
        pages = {631-649},
          doi = {10.1086/308704},
archivePrefix = {arXiv},
       eprint = {astro-ph/9911476},
 primaryClass = {astro-ph},
       adsurl = {https://ui.adsabs.harvard.edu/abs/2000ApJ...533..631K}
}

@ARTICLE{2005Natur.433..604D,
       author = {{Di Matteo}, Tiziana and {Springel}, Volker and {Hernquist}, Lars},
        title = "{Energy input from quasars regulates the growth and activity of black holes and their host galaxies}",
      journal = {\nat},
         year = 2005,
        month = feb,
       volume = {433},
       number = {7026},
        pages = {604-607},
          doi = {10.1038/nature03335},
archivePrefix = {arXiv},
       eprint = {astro-ph/0502199},
 primaryClass = {astro-ph},
       adsurl = {https://ui.adsabs.harvard.edu/abs/2005Natur.433..604D}
}

@ARTICLE{2005MNRAS.361..776S,
       author = {{Springel}, Volker and {Di Matteo}, Tiziana and {Hernquist}, Lars},
        title = "{Modelling feedback from stars and black holes in galaxy mergers}",
      journal = {\mnras},
         year = 2005,
        month = aug,
       volume = {361},
       number = {3},
        pages = {776-794},
          doi = {10.1111/j.1365-2966.2005.09238.x},
archivePrefix = {arXiv},
       eprint = {astro-ph/0411108},
 primaryClass = {astro-ph},
       adsurl = {https://ui.adsabs.harvard.edu/abs/2005MNRAS.361..776S}
}

@ARTICLE{2013ARA&A..51..511K,
       author = {{Kormendy}, John and {Ho}, Luis C.},
        title = "{Coevolution (Or Not) of Supermassive Black Holes and Host Galaxies}",
      journal = {\araa},
         year = 2013,
        month = aug,
       volume = {51},
       number = {1},
        pages = {511-653},
          doi = {10.1146/annurev-astro-082708-101811},
archivePrefix = {arXiv},
       eprint = {1304.7762},
 primaryClass = {astro-ph.CO},
       adsurl = {https://ui.adsabs.harvard.edu/abs/2013ARA&A..51..511K}
}

@ARTICLE{2021MNRAS.507....1V,
       author = {{Valentini}, Milena and {Gallerani}, Simona and {Ferrara}, Andrea},
        title = "{Host galaxies of high-redshift quasars: SMBH growth and feedback}",
      journal = {\mnras},
         year = 2021,
        month = oct,
       volume = {507},
       number = {1},
        pages = {1-26},
          doi = {10.1093/mnras/stab1992},
archivePrefix = {arXiv},
       eprint = {2107.05638},
 primaryClass = {astro-ph.GA},
       adsurl = {https://ui.adsabs.harvard.edu/abs/2021MNRAS.507....1V}
}

@ARTICLE{2006AJ....132..117F,
       author = {{Fan}, Xiaohui and {Strauss}, Michael A. and {Becker}, Robert H. and {White}, Richard L. and {Gunn}, James E. and {Knapp}, Gillian R. and {Richards}, Gordon T. and {Schneider}, Donald P. and {Brinkmann}, J. and {Fukugita}, Masataka},
        title = "{Constraining the Evolution of the Ionizing Background and the Epoch of Reionization with z\raisebox{-0.5ex}\textasciitilde6 Quasars. II. A Sample of 19 Quasars}",
      journal = {\aj},
         year = 2006,
        month = jul,
       volume = {132},
       number = {1},
        pages = {117-136},
          doi = {10.1086/504836},
archivePrefix = {arXiv},
       eprint = {astro-ph/0512082},
 primaryClass = {astro-ph},
       adsurl = {https://ui.adsabs.harvard.edu/abs/2006AJ....132..117F}
}

@ARTICLE{2010ApJ...714..834C,
       author = {{Carilli}, C.~L. and {Wang}, Ran and {Fan}, X. and {Walter}, F. and {Kurk}, J. and {Riechers}, D. and {Wagg}, J. and {Hennawi}, J. and {Jiang}, L. and {Menten}, K.~M. and {Bertoldi}, F. and {Strauss}, Michael A. and {Cox}, P.},
        title = "{Ionization Near Zones Associated with Quasars at z \raisebox{-0.5ex}\textasciitilde 6}",
      journal = {\apj},
         year = 2010,
        month = may,
       volume = {714},
       number = {1},
        pages = {834-839},
          doi = {10.1088/0004-637X/714/1/834},
archivePrefix = {arXiv},
       eprint = {1003.0016},
 primaryClass = {astro-ph.CO},
       adsurl = {https://ui.adsabs.harvard.edu/abs/2010ApJ...714..834C}
}

@ARTICLE{2015MNRAS.447..499M,
       author = {{McGreer}, Ian D. and {Mesinger}, Andrei and {D'Odorico}, Valentina},
        title = "{Model-independent evidence in favour of an end to reionization by z {\ensuremath{\approx}} 6}",
      journal = {\mnras},
         year = 2015,
        month = feb,
       volume = {447},
       number = {1},
        pages = {499-505},
          doi = {10.1093/mnras/stu2449},
archivePrefix = {arXiv},
       eprint = {1411.5375},
 primaryClass = {astro-ph.CO},
       adsurl = {https://ui.adsabs.harvard.edu/abs/2015MNRAS.447..499M}
}

@ARTICLE{2013AJ....145...10D,
       author = {{Dawson}, Kyle S. and {Schlegel}, David J. and {Ahn}, Christopher P. and {Anderson}, Scott F. and {Aubourg}, {\'E}ric and {Bailey}, Stephen and {Barkhouser}, Robert H. and {Bautista}, Julian E. and {Beifiori}, Alessandra and {Berlind}, Andreas A. and {Bhardwaj}, Vaishali and {Bizyaev}, Dmitry and {Blake}, Cullen H. and {Blanton}, Michael R. and {Blomqvist}, Michael and {Bolton}, Adam S. and {Borde}, Arnaud and {Bovy}, Jo and {Brandt}, W.~N. and {Brewington}, Howard and {Brinkmann}, Jon and {Brown}, Peter J. and {Brownstein}, Joel R. and {Bundy}, Kevin and {Busca}, N.~G. and {Carithers}, William and {Carnero}, Aurelio R. and {Carr}, Michael A. and {Chen}, Yanmei and {Comparat}, Johan and {Connolly}, Natalia and {Cope}, Frances and {Croft}, Rupert A.~C. and {Cuesta}, Antonio J. and {da Costa}, Luiz N. and {Davenport}, James R.~A. and {Delubac}, Timoth{\'e}e and {de Putter}, Roland and {Dhital}, Saurav and {Ealet}, Anne and {Ebelke}, Garrett L. and {Eisenstein}, Daniel J. and {Escoffier}, S. and {Fan}, Xiaohui and {Filiz Ak}, N. and {Finley}, Hayley and {Font-Ribera}, Andreu and {G{\'e}nova-Santos}, R. and {Gunn}, James E. and {Guo}, Hong and {Haggard}, Daryl and {Hall}, Patrick B. and {Hamilton}, Jean-Christophe and {Harris}, Ben and {Harris}, David W. and {Ho}, Shirley and {Hogg}, David W. and {Holder}, Diana and {Honscheid}, Klaus and {Huehnerhoff}, Joe and {Jordan}, Beatrice and {Jordan}, Wendell P. and {Kauffmann}, Guinevere and {Kazin}, Eyal A. and {Kirkby}, David and {Klaene}, Mark A. and {Kneib}, Jean-Paul and {Le Goff}, Jean-Marc and {Lee}, Khee-Gan and {Long}, Daniel C. and {Loomis}, Craig P. and {Lundgren}, Britt and {Lupton}, Robert H. and {Maia}, Marcio A.~G. and {Makler}, Martin and {Malanushenko}, Elena and {Malanushenko}, Viktor and {Mandelbaum}, Rachel and {Manera}, Marc and {Maraston}, Claudia and {Margala}, Daniel and {Masters}, Karen L. and {McBride}, Cameron K. and {McDonald}, Patrick and {McGreer}, Ian D. and {McMahon}, Richard G. and {Mena}, Olga and {Miralda-Escud{\'e}}, Jordi and {Montero-Dorta}, Antonio D. and {Montesano}, Francesco and {Muna}, Demitri and {Myers}, Adam D. and {Naugle}, Tracy and {Nichol}, Robert C. and {Noterdaeme}, Pasquier and {Nuza}, Sebasti{\'a}n E. and {Olmstead}, Matthew D. and {Oravetz}, Audrey and {Oravetz}, Daniel J. and {Owen}, Russell and {Padmanabhan}, Nikhil and {Palanque-Delabrouille}, Nathalie and {Pan}, Kaike and {Parejko}, John K. and {P{\^a}ris}, Isabelle and {Percival}, Will J. and {P{\'e}rez-Fournon}, Ismael and {P{\'e}rez-R{\`a}fols}, Ignasi and {Petitjean}, Patrick and {Pfaffenberger}, Robert and {Pforr}, Janine and {Pieri}, Matthew M. and {Prada}, Francisco and {Price-Whelan}, Adrian M. and {Raddick}, M. Jordan and {Rebolo}, Rafael and {Rich}, James and {Richards}, Gordon T. and {Rockosi}, Constance M. and {Roe}, Natalie A. and {Ross}, Ashley J. and {Ross}, Nicholas P. and {Rossi}, Graziano and {Rubi{\~n}o-Martin}, J.~A. and {Samushia}, Lado and {S{\'a}nchez}, Ariel G. and {Sayres}, Conor and {Schmidt}, Sarah J. and {Schneider}, Donald P. and {Sc{\'o}ccola}, C.~G. and {Seo}, Hee-Jong and {Shelden}, Alaina and {Sheldon}, Erin and {Shen}, Yue and {Shu}, Yiping and {Slosar}, An{\v{z}}e and {Smee}, Stephen A. and {Snedden}, Stephanie A. and {Stauffer}, Fritz and {Steele}, Oliver and {Strauss}, Michael A. and {Streblyanska}, Alina and {Suzuki}, Nao and {Swanson}, Molly E.~C. and {Tal}, Tomer and {Tanaka}, Masayuki and {Thomas}, Daniel and {Tinker}, Jeremy L. and {Tojeiro}, Rita and {Tremonti}, Christy A. and {Vargas Maga{\~n}a}, M. and {Verde}, Licia and {Viel}, Matteo and {Wake}, David A. and {Watson}, Mike and {Weaver}, Benjamin A. and {Weinberg}, David H. and {Weiner}, Benjamin J. and {West}, Andrew A. and {White}, Martin and {Wood-Vasey}, W.~M. and {Yeche}, Christophe and {Zehavi}, Idit and {Zhao}, Gong-Bo and {Zheng}, Zheng},
        title = "{The Baryon Oscillation Spectroscopic Survey of SDSS-III}",
      journal = {\aj},
         year = 2013,
        month = jan,
       volume = {145},
       number = {1},
          eid = {10},
        pages = {10},
          doi = {10.1088/0004-6256/145/1/10},
archivePrefix = {arXiv},
       eprint = {1208.0022},
 primaryClass = {astro-ph.CO},
       adsurl = {https://ui.adsabs.harvard.edu/abs/2013AJ....145...10D}
}

@ARTICLE{2014JCAP...05..027F,
       author = {{Font-Ribera}, Andreu and {Kirkby}, David and {Busca}, Nicolas and {Miralda-Escud{\'e}}, Jordi and {Ross}, Nicholas P. and {Slosar}, An{\v{z}}e and {Rich}, James and {Aubourg}, {\'E}ric and {Bailey}, Stephen and {Bhardwaj}, Vaishali and {Bautista}, Julian and {Beutler}, Florian and {Bizyaev}, Dmitry and {Blomqvist}, Michael and {Brewington}, Howard and {Brinkmann}, Jon and {Brownstein}, Joel R. and {Carithers}, Bill and {Dawson}, Kyle S. and {Delubac}, Timoth{\'e}e and {Ebelke}, Garrett and {Eisenstein}, Daniel J. and {Ge}, Jian and {Kinemuchi}, Karen and {Lee}, Khee-Gan and {Malanushenko}, Viktor and {Malanushenko}, Elena and {Marchante}, Moses and {Margala}, Daniel and {Muna}, Demitri and {Myers}, Adam D. and {Noterdaeme}, Pasquier and {Oravetz}, Daniel and {Palanque-Delabrouille}, Nathalie and {P{\^a}ris}, Isabelle and {Petitjean}, Patrick and {Pieri}, Matthew M. and {Rossi}, Graziano and {Schneider}, Donald P. and {Simmons}, Audrey and {Viel}, Matteo and {Yeche}, Christophe and {York}, Donald G.},
        title = "{Quasar-Lyman {\ensuremath{\alpha}} forest cross-correlation from BOSS DR11: Baryon Acoustic Oscillations}",
      journal = {\jcap},
         year = 2014,
        month = may,
       volume = {2014},
       number = {5},
          eid = {027},
        pages = {027},
          doi = {10.1088/1475-7516/2014/05/027},
archivePrefix = {arXiv},
       eprint = {1311.1767},
 primaryClass = {astro-ph.CO},
       adsurl = {https://ui.adsabs.harvard.edu/abs/2014JCAP...05..027F}
}

@ARTICLE{2015A&A...574A..59D,
       author = {{Delubac}, Timoth{\'e}e and {Bautista}, Julian E. and {Busca}, Nicol{\'a}s G. and {Rich}, James and {Kirkby}, David and {Bailey}, Stephen and {Font-Ribera}, Andreu and {Slosar}, An{\v{z}}e and {Lee}, Khee-Gan and {Pieri}, Matthew M. and {Hamilton}, Jean-Christophe and {Aubourg}, {\'E}ric and {Blomqvist}, Michael and {Bovy}, Jo and {Brinkmann}, Jon and {Carithers}, William and {Dawson}, Kyle S. and {Eisenstein}, Daniel J. and {Gontcho}, Satya Gontcho A. and {Kneib}, Jean-Paul and {Le Goff}, Jean-Marc and {Margala}, Daniel and {Miralda-Escud{\'e}}, Jordi and {Myers}, Adam D. and {Nichol}, Robert C. and {Noterdaeme}, Pasquier and {O'Connell}, Ross and {Olmstead}, Matthew D. and {Palanque-Delabrouille}, Nathalie and {P{\^a}ris}, Isabelle and {Petitjean}, Patrick and {Ross}, Nicholas P. and {Rossi}, Graziano and {Schlegel}, David J. and {Schneider}, Donald P. and {Weinberg}, David H. and {Y{\`e}che}, Christophe and {York}, Donald G.},
        title = "{Baryon acoustic oscillations in the Ly{\ensuremath{\alpha}} forest of BOSS DR11 quasars}",
      journal = {\aap},
         year = 2015,
        month = feb,
       volume = {574},
          eid = {A59},
        pages = {A59},
          doi = {10.1051/0004-6361/201423969},
archivePrefix = {arXiv},
       eprint = {1404.1801},
 primaryClass = {astro-ph.CO},
       adsurl = {https://ui.adsabs.harvard.edu/abs/2015A&A...574A..59D}
}

@ARTICLE{2021PhRvD.103h3533A,
       author = {{Alam}, Shadab and {Aubert}, Marie and {Avila}, Santiago and {Balland}, Christophe and {Bautista}, Julian E. and {Bershady}, Matthew A. and {Bizyaev}, Dmitry and {Blanton}, Michael R. and {Bolton}, Adam S. and {Bovy}, Jo and {Brinkmann}, Jonathan and {Brownstein}, Joel R. and {Burtin}, Etienne and {Chabanier}, Sol{\`e}ne and {Chapman}, Michael J. and {Choi}, Peter Doohyun and {Chuang}, Chia-Hsun and {Comparat}, Johan and {Cousinou}, Marie-Claude and {Cuceu}, Andrei and {Dawson}, Kyle S. and {de la Torre}, Sylvain and {de Mattia}, Arnaud and {Agathe}, Victoria de Sainte and {des Bourboux}, H{\'e}lion du Mas and {Escoffier}, Stephanie and {Etourneau}, Thomas and {Farr}, James and {Font-Ribera}, Andreu and {Frinchaboy}, Peter M. and {Fromenteau}, Sebastien and {Gil-Mar{\'\i}n}, H{\'e}ctor and {Le Goff}, Jean-Marc and {Gonzalez-Morales}, Alma X. and {Gonzalez-Perez}, Violeta and {Grabowski}, Kathleen and {Guy}, Julien and {Hawken}, Adam J. and {Hou}, Jiamin and {Kong}, Hui and {Parker}, James and {Klaene}, Mark and {Kneib}, Jean-Paul and {Lin}, Sicheng and {Long}, Daniel and {Lyke}, Brad W. and {de la Macorra}, Axel and {Martini}, Paul and {Masters}, Karen and {Mohammad}, Faizan G. and {Moon}, Jeongin and {Mueller}, Eva-Maria and {Mu{\~n}oz-Guti{\'e}rrez}, Andrea and {Myers}, Adam D. and {Nadathur}, Seshadri and {Neveux}, Richard and {Newman}, Jeffrey A. and {Noterdaeme}, Pasquier and {Oravetz}, Audrey and {Oravetz}, Daniel and {Palanque-Delabrouille}, Nathalie and {Pan}, Kaike and {Paviot}, Romain and {Percival}, Will J. and {P{\'e}rez-R{\`a}fols}, Ignasi and {Petitjean}, Patrick and {Pieri}, Matthew M. and {Prakash}, Abhishek and {Raichoor}, Anand and {Ravoux}, Corentin and {Rezaie}, Mehdi and {Rich}, James and {Ross}, Ashley J. and {Rossi}, Graziano and {Ruggeri}, Rossana and {Ruhlmann-Kleider}, Vanina and {S{\'a}nchez}, Ariel G. and {S{\'a}nchez}, F. Javier and {S{\'a}nchez-Gallego}, Jos{\'e} R. and {Sayres}, Conor and {Schneider}, Donald P. and {Seo}, Hee-Jong and {Shafieloo}, Arman and {Slosar}, An{\v{z}}e and {Smith}, Alex and {Stermer}, Julianna and {Tamone}, Amelie and {Tinker}, Jeremy L. and {Tojeiro}, Rita and {Vargas-Maga{\~n}a}, Mariana and {Variu}, Andrei and {Wang}, Yuting and {Weaver}, Benjamin A. and {Weijmans}, Anne-Marie and {Y{\`e}che}, Christophe and {Zarrouk}, Pauline and {Zhao}, Cheng and {Zhao}, Gong-Bo and {Zheng}, Zheng},
        title = "{Completed SDSS-IV extended Baryon Oscillation Spectroscopic Survey: Cosmological implications from two decades of spectroscopic surveys at the Apache Point Observatory}",
      journal = {\prd},
         year = 2021,
        month = apr,
       volume = {103},
       number = {8},
          eid = {083533},
        pages = {083533},
          doi = {10.1103/PhysRevD.103.083533},
archivePrefix = {arXiv},
       eprint = {2007.08991},
 primaryClass = {astro-ph.CO},
       adsurl = {https://ui.adsabs.harvard.edu/abs/2021PhRvD.103h3533A}
}

@ARTICLE{2002AJ....123.2945R,
       author = {{Richards}, Gordon T. and {Fan}, Xiaohui and {Newberg}, Heidi Jo and {Strauss}, Michael A. and {Vanden Berk}, Daniel E. and {Schneider}, Donald P. and {Yanny}, Brian and {Boucher}, Adam and {Burles}, Scott and {Frieman}, Joshua A. and {Gunn}, James E. and {Hall}, Patrick B. and {Ivezi{\'c}}, {\v{Z}}eljko and {Kent}, Stephen and {Loveday}, Jon and {Lupton}, Robert H. and {Rockosi}, Constance M. and {Schlegel}, David J. and {Stoughton}, Chris and {SubbaRao}, Mark and {York}, Donald G.},
        title = "{Spectroscopic Target Selection in the Sloan Digital Sky Survey: The Quasar Sample}",
      journal = {\aj},
         year = 2002,
        month = jun,
       volume = {123},
       number = {6},
        pages = {2945-2975},
          doi = {10.1086/340187},
archivePrefix = {arXiv},
       eprint = {astro-ph/0202251},
 primaryClass = {astro-ph},
       adsurl = {https://ui.adsabs.harvard.edu/abs/2002AJ....123.2945R}
}

@ARTICLE{2006AJ....131.2766R,
       author = {{Richards}, Gordon T. and {Strauss}, Michael A. and {Fan}, Xiaohui and {Hall}, Patrick B. and {Jester}, Sebastian and {Schneider}, Donald P. and {Vanden Berk}, Daniel E. and {Stoughton}, Chris and {Anderson}, Scott F. and {Brunner}, Robert J. and {Gray}, Jim and {Gunn}, James E. and {Ivezi{\'c}}, {\v{Z}}eljko and {Kirkland}, Margaret K. and {Knapp}, G.~R. and {Loveday}, Jon and {Meiksin}, Avery and {Pope}, Adrian and {Szalay}, Alexander S. and {Thakar}, Anirudda R. and {Yanny}, Brian and {York}, Donald G. and {Barentine}, J.~C. and {Brewington}, Howard J. and {Brinkmann}, J. and {Fukugita}, Masataka and {Harvanek}, Michael and {Kent}, Stephen M. and {Kleinman}, S.~J. and {Krzesi{\'n}ski}, Jurek and {Long}, Daniel C. and {Lupton}, Robert H. and {Nash}, Thomas and {Neilsen}, Eric H., Jr. and {Nitta}, Atsuko and {Schlegel}, David J. and {Snedden}, Stephanie A.},
        title = "{The Sloan Digital Sky Survey Quasar Survey: Quasar Luminosity Function from Data Release 3}",
      journal = {\aj},
         year = 2006,
        month = jun,
       volume = {131},
       number = {6},
        pages = {2766-2787},
          doi = {10.1086/503559},
archivePrefix = {arXiv},
       eprint = {astro-ph/0601434},
 primaryClass = {astro-ph},
       adsurl = {https://ui.adsabs.harvard.edu/abs/2006AJ....131.2766R}
}

@ARTICLE{2015ApJS..221...27M,
       author = {{Myers}, Adam D. and {Palanque-Delabrouille}, Nathalie and {Prakash}, Abhishek and {P{\^a}ris}, Isabelle and {Yeche}, Christophe and {Dawson}, Kyle S. and {Bovy}, Jo and {Lang}, Dustin and {Schlegel}, David J. and {Newman}, Jeffrey A. and {Petitjean}, Patrick and {Kneib}, Jean-Paul and {Laurent}, Pierre and {Percival}, Will J. and {Ross}, Ashley J. and {Seo}, Hee-Jong and {Tinker}, Jeremy L. and {Armengaud}, Eric and {Brownstein}, Joel and {Burtin}, Etienne and {Cai}, Zheng and {Comparat}, Johan and {Kasliwal}, Mansi and {Kulkarni}, Shrinivas R. and {Laher}, Russ and {Levitan}, David and {McBride}, Cameron K. and {McGreer}, Ian D. and {Miller}, Adam A. and {Nugent}, Peter and {Ofek}, Eran and {Rossi}, Graziano and {Ruan}, John and {Schneider}, Donald P. and {Sesar}, Branimir and {Streblyanska}, Alina and {Surace}, Jason},
        title = "{The SDSS-IV Extended Baryon Oscillation Spectroscopic Survey: Quasar Target Selection}",
      journal = {\apjs},
         year = 2015,
        month = dec,
       volume = {221},
       number = {2},
          eid = {27},
        pages = {27},
          doi = {10.1088/0067-0049/221/2/27},
archivePrefix = {arXiv},
       eprint = {1508.04472},
 primaryClass = {astro-ph.CO},
       adsurl = {https://ui.adsabs.harvard.edu/abs/2015ApJS..221...27M}
}

@ARTICLE{2023ApJ...944..107C,
       author = {{Chaussidon}, Edmond and {Y{\`e}che}, Christophe and {Palanque-Delabrouille}, Nathalie and {Alexander}, David M. and {Yang}, Jinyi and {Ahlen}, Steven and {Bailey}, Stephen and {Brooks}, David and {Cai}, Zheng and {Chabanier}, Sol{\`e}ne and {Davis}, Tamara M. and {Dawson}, Kyle and {de laMacorra}, Axel and {Dey}, Arjun and {Dey}, Biprateep and {Eftekharzadeh}, Sarah and {Eisenstein}, Daniel J. and {Fanning}, Kevin and {Font-Ribera}, Andreu and {Gazta{\~n}aga}, Enrique and {A Gontcho}, Satya Gontcho and {Gonzalez-Morales}, Alma X. and {Guy}, Julien and {Herrera-Alcantar}, Hiram K. and {Honscheid}, Klaus and {Ishak}, Mustapha and {Jiang}, Linhua and {Juneau}, Stephanie and {Kehoe}, Robert and {Kisner}, Theodore and {Kov{\'a}cs}, Andras and {Kremin}, Anthony and {Lan}, Ting-Wen and {Landriau}, Martin and {Le Guillou}, Laurent and {Levi}, Michael E. and {Magneville}, Christophe and {Martini}, Paul and {Meisner}, Aaron M. and {Moustakas}, John and {Mu{\~n}oz-Guti{\'e}rrez}, Andrea and {Myers}, Adam D. and {Newman}, Jeffrey A. and {Nie}, Jundan and {Percival}, Will J. and {Poppett}, Claire and {Prada}, Francisco and {Raichoor}, Anand and {Ravoux}, Corentin and {Ross}, Ashley J. and {Schlafly}, Edward and {Schlegel}, David and {Tan}, Ting and {Tarl{\'e}}, Gregory and {Zhou}, Rongpu and {Zhou}, Zhimin and {Zou}, Hu},
        title = "{Target Selection and Validation of DESI Quasars}",
      journal = {\apj},
         year = 2023,
        month = feb,
       volume = {944},
       number = {1},
          eid = {107},
        pages = {107},
          doi = {10.3847/1538-4357/acb3c2},
archivePrefix = {arXiv},
       eprint = {2208.08511},
 primaryClass = {astro-ph.CO},
       adsurl = {https://ui.adsabs.harvard.edu/abs/2023ApJ...944..107C}
}

@ARTICLE{2011ApJ...728...26M,
       author = {{MacLeod}, C.~L. and {Brooks}, K. and {Ivezi{\'c}}, {\v{Z}}. and {Kochanek}, C.~S. and {Gibson}, R. and {Meisner}, A. and {Koz{\l}owski}, S. and {Sesar}, B. and {Becker}, A.~C. and {de Vries}, W.~H.},
        title = "{Quasar Selection Based on Photometric Variability}",
      journal = {\apj},
         year = 2011,
        month = feb,
       volume = {728},
       number = {1},
          eid = {26},
        pages = {26},
          doi = {10.1088/0004-637X/728/1/26},
archivePrefix = {arXiv},
       eprint = {1009.2081},
 primaryClass = {astro-ph.CO},
       adsurl = {https://ui.adsabs.harvard.edu/abs/2011ApJ...728...26M}
}

@ARTICLE{2016A&A...587A..41P,
       author = {{Palanque-Delabrouille}, N. and {Magneville}, Ch. and {Y{\`e}che}, Ch. and {P{\^a}ris}, I. and {Petitjean}, P. and {Burtin}, E. and {Dawson}, K. and {McGreer}, I. and {Myers}, A.~D. and {Rossi}, G. and {Schlegel}, D. and {Schneider}, D. and {Streblyanska}, A. and {Tinker}, J.},
        title = "{The extended Baryon Oscillation Spectroscopic Survey: Variability selection and quasar luminosity function}",
      journal = {\aap},
         year = 2016,
        month = mar,
       volume = {587},
          eid = {A41},
        pages = {A41},
          doi = {10.1051/0004-6361/201527392},
archivePrefix = {arXiv},
       eprint = {1509.05607},
 primaryClass = {astro-ph.CO},
       adsurl = {https://ui.adsabs.harvard.edu/abs/2016A&A...587A..41P}
}

@ARTICLE{2005A&A...441..417H,
       author = {{Hasinger}, G. and {Miyaji}, T. and {Schmidt}, M.},
        title = "{Luminosity-dependent evolution of soft X-ray selected AGN. New Chandra and XMM-Newton surveys}",
      journal = {\aap},
         year = 2005,
        month = oct,
       volume = {441},
       number = {2},
        pages = {417-434},
          doi = {10.1051/0004-6361:20042134},
archivePrefix = {arXiv},
       eprint = {astro-ph/0506118},
 primaryClass = {astro-ph},
       adsurl = {https://ui.adsabs.harvard.edu/abs/2005A&A...441..417H}
}

@ARTICLE{2014MNRAS.445.3557V,
       author = {{Vito}, F. and {Gilli}, R. and {Vignali}, C. and {Comastri}, A. and {Brusa}, M. and {Cappelluti}, N. and {Iwasawa}, K.},
        title = "{The hard X-ray luminosity function of high-redshift (3 < z {\ensuremath{\lesssim}} 5) active galactic nuclei}",
      journal = {\mnras},
         year = 2014,
        month = dec,
       volume = {445},
       number = {4},
        pages = {3557-3574},
          doi = {10.1093/mnras/stu2004},
archivePrefix = {arXiv},
       eprint = {1409.6918},
 primaryClass = {astro-ph.GA},
       adsurl = {https://ui.adsabs.harvard.edu/abs/2014MNRAS.445.3557V}
}

@ARTICLE{2015MNRAS.453.1946G,
       author = {{Georgakakis}, A. and {Aird}, J. and {Buchner}, J. and {Salvato}, M. and {Menzel}, M. -L. and {Brandt}, W.~N. and {McGreer}, I.~D. and {Dwelly}, T. and {Mountrichas}, G. and {Koki}, C. and {Georgantopoulos}, I. and {Hsu}, L. -T. and {Merloni}, A. and {Liu}, Z. and {Nandra}, K. and {Ross}, N.~P.},
        title = "{The X-ray luminosity function of active galactic nuclei in the redshift interval z=3-5}",
      journal = {\mnras},
         year = 2015,
        month = oct,
       volume = {453},
       number = {2},
        pages = {1946-1964},
          doi = {10.1093/mnras/stv1703},
archivePrefix = {arXiv},
       eprint = {1507.07558},
 primaryClass = {astro-ph.HE},
       adsurl = {https://ui.adsabs.harvard.edu/abs/2015MNRAS.453.1946G}
}

@ARTICLE{2009AJ....138.1925M,
       author = {{McGreer}, Ian D. and {Helfand}, David J. and {White}, Richard L.},
        title = "{Radio-Selected Quasars in the Sloan Digital Sky Survey}",
      journal = {\aj},
         year = 2009,
        month = dec,
       volume = {138},
       number = {6},
        pages = {1925-1937},
          doi = {10.1088/0004-6256/138/6/1925},
archivePrefix = {arXiv},
       eprint = {0909.4091},
 primaryClass = {astro-ph.CO},
       adsurl = {https://ui.adsabs.harvard.edu/abs/2009AJ....138.1925M}
}

@ARTICLE{2011ApJ...736...57Z,
       author = {{Zeimann}, Gregory R. and {White}, Richard L. and {Becker}, Robert H. and {Hodge}, Jacqueline A. and {Stanford}, Spencer A. and {Richards}, Gordon T.},
        title = "{Discovery of a Radio-selected z \raisebox{-0.5ex}\textasciitilde 6 Quasar}",
      journal = {\apj},
         year = 2011,
        month = jul,
       volume = {736},
       number = {1},
          eid = {57},
        pages = {57},
          doi = {10.1088/0004-637X/736/1/57},
archivePrefix = {arXiv},
       eprint = {1105.2047},
 primaryClass = {astro-ph.CO},
       adsurl = {https://ui.adsabs.harvard.edu/abs/2011ApJ...736...57Z}
}

@ARTICLE{2015ApJ...804..118B,
       author = {{Ba{\~n}ados}, E. and {Venemans}, B.~P. and {Morganson}, E. and {Hodge}, J. and {Decarli}, R. and {Walter}, F. and {Stern}, D. and {Schlafly}, E. and {Farina}, E.~P. and {Greiner}, J. and {Chambers}, K.~C. and {Fan}, X. and {Rix}, H. -W. and {Burgett}, W.~S. and {Draper}, P.~W. and {Flewelling}, J. and {Kaiser}, N. and {Metcalfe}, N. and {Morgan}, J.~S. and {Tonry}, J.~L. and {Wainscoat}, R.~J.},
        title = "{Constraining the Radio-loud Fraction of Quasars at z > 5.5}",
      journal = {\apj},
         year = 2015,
        month = may,
       volume = {804},
       number = {2},
          eid = {118},
        pages = {118},
          doi = {10.1088/0004-637X/804/2/118},
archivePrefix = {arXiv},
       eprint = {1503.04214},
 primaryClass = {astro-ph.GA},
       adsurl = {https://ui.adsabs.harvard.edu/abs/2015ApJ...804..118B}
}

@ARTICLE{2020A&A...644A..17H,
       author = {{Heintz}, K.~E. and {Fynbo}, J.~P.~U. and {Geier}, S.~J. and {M{\o}ller}, P. and {Krogager}, J. -K. and {Konstantopoulou}, C. and {de Burgos}, A. and {Christensen}, L. and {Steinhardt}, C.~L. and {Milvang-Jensen}, B. and {Jakobsson}, P. and {H{\o}g}, E. and {Arvedlund}, B.~E.~H.~K. and {Christiansen}, C.~R. and {Hansen}, T.~B. and {Henriksen}, P.~D. and {Kuszon}, K.~B. and {McKenzie}, I.~B. and {Mosekj{\ae}r}, K.~A. and {Paulsen}, M.~F.~K. and {Sukstorf}, M.~N. and {Wilson}, S.~N. and {{\O}rgaard}, S.~K.~K.},
        title = "{Spectroscopic classification of a complete sample of astrometrically-selected quasar candidates using Gaia DR2}",
      journal = {\aap},
         year = 2020,
        month = dec,
       volume = {644},
          eid = {A17},
        pages = {A17},
          doi = {10.1051/0004-6361/202039262},
archivePrefix = {arXiv},
       eprint = {2010.05934},
 primaryClass = {astro-ph.GA},
       adsurl = {https://ui.adsabs.harvard.edu/abs/2020A&A...644A..17H}
}

@ARTICLE{2021ApJS..254....6F,
       author = {{Fu}, Yuming and {Wu}, Xue-Bing and {Yang}, Qian and {Brown}, Anthony G.~A. and {Feng}, Xiaotong and {Ma}, Qinchun and {Li}, Shuyan},
        title = "{Finding Quasars behind the Galactic Plane. I. Candidate Selections with Transfer Learning}",
      journal = {\apjs},
         year = 2021,
        month = may,
       volume = {254},
       number = {1},
          eid = {6},
        pages = {6},
          doi = {10.3847/1538-4365/abe85e},
archivePrefix = {arXiv},
       eprint = {2102.09770},
 primaryClass = {astro-ph.GA},
       adsurl = {https://ui.adsabs.harvard.edu/abs/2021ApJS..254....6F}
}

@ARTICLE{2023arXiv230617749S,
       author = {{Storey-Fisher}, Kate and {Hogg}, David W. and {Rix}, Hans-Walter and {Eilers}, Anna-Christina and {Fabbian}, Giulio and {Blanton}, Michael and {Alonso}, David},
        title = "{Quaia, the Gaia-unWISE Quasar Catalog: An All-Sky Spectroscopic Quasar Sample}",
      journal = {arXiv e-prints},
         year = 2023,
        month = jun,
          eid = {arXiv:2306.17749},
        pages = {arXiv:2306.17749},
          doi = {10.48550/arXiv.2306.17749},
archivePrefix = {arXiv},
       eprint = {2306.17749},
 primaryClass = {astro-ph.GA},
       adsurl = {https://ui.adsabs.harvard.edu/abs/2023arXiv230617749S}
}

@ARTICLE{1999AJ....117.2528F,
       author = {{Fan}, Xiaohui},
        title = "{Simulation of Stellar Objects in SDSS Color Space}",
      journal = {\aj},
         year = 1999,
        month = may,
       volume = {117},
       number = {5},
        pages = {2528-2551},
          doi = {10.1086/300848},
archivePrefix = {arXiv},
       eprint = {astro-ph/9902063},
 primaryClass = {astro-ph},
       adsurl = {https://ui.adsabs.harvard.edu/abs/1999AJ....117.2528F}
}

@ARTICLE{2013ApJ...773...14R,
       author = {{Ross}, Nicholas P. and {McGreer}, Ian D. and {White}, Martin and {Richards}, Gordon T. and {Myers}, Adam D. and {Palanque-Delabrouille}, Nathalie and {Strauss}, Michael A. and {Anderson}, Scott F. and {Shen}, Yue and {Brandt}, W.~N. and {Y{\`e}che}, Christophe and {Swanson}, Molly E.~C. and {Aubourg}, {\'E}ric and {Bailey}, Stephen and {Bizyaev}, Dmitry and {Bovy}, Jo and {Brewington}, Howard and {Brinkmann}, J. and {DeGraf}, Colin and {Di Matteo}, Tiziana and {Ebelke}, Garrett and {Fan}, Xiaohui and {Ge}, Jian and {Malanushenko}, Elena and {Malanushenko}, Viktor and {Mandelbaum}, Rachel and {Maraston}, Claudia and {Muna}, Demitri and {Oravetz}, Daniel and {Pan}, Kaike and {P{\^a}ris}, Isabelle and {Petitjean}, Patrick and {Schawinski}, Kevin and {Schlegel}, David J. and {Schneider}, Donald P. and {Silverman}, John D. and {Simmons}, Audrey and {Snedden}, Stephanie and {Streblyanska}, Alina and {Suzuki}, Nao and {Weinberg}, David H. and {York}, Donald},
        title = "{The SDSS-III Baryon Oscillation Spectroscopic Survey: The Quasar Luminosity Function from Data Release Nine}",
      journal = {\apj},
         year = 2013,
        month = aug,
       volume = {773},
       number = {1},
          eid = {14},
        pages = {14},
          doi = {10.1088/0004-637X/773/1/14},
archivePrefix = {arXiv},
       eprint = {1210.6389},
 primaryClass = {astro-ph.CO},
       adsurl = {https://ui.adsabs.harvard.edu/abs/2013ApJ...773...14R}
}

@ARTICLE{2022ApJ...928..172P,
       author = {{Pan}, Zhiwei and {Jiang}, Linhua and {Fan}, Xiaohui and {Wu}, Jin and {Yang}, Jinyi},
        title = "{Quasar UV Luminosity Function at 3.5 < z < 5.0 from SDSS Deep Imaging Data}",
      journal = {\apj},
         year = 2022,
        month = apr,
       volume = {928},
       number = {2},
          eid = {172},
        pages = {172},
          doi = {10.3847/1538-4357/ac5aab},
archivePrefix = {arXiv},
       eprint = {2112.07801},
 primaryClass = {astro-ph.GA},
       adsurl = {https://ui.adsabs.harvard.edu/abs/2022ApJ...928..172P}
}

@ARTICLE{1980ApJ...235..694A,
       author = {{Avni}, Y. and {Bahcall}, J.~N.},
        title = "{On the simultaneous analysis of several complete samples. The V/Vmax and Ve/Va variables, with applications to quasars.}",
      journal = {\apj},
         year = 1980,
        month = feb,
       volume = {235},
        pages = {694-716},
          doi = {10.1086/157673},
       adsurl = {https://ui.adsabs.harvard.edu/abs/1980ApJ...235..694A}
}

@ARTICLE{1983ApJ...269...35M,
       author = {{Marshall}, H.~L. and {Tananbaum}, H. and {Avni}, Y. and {Zamorani}, G.},
        title = "{Analysis of complete quasar samples to obtain parameters of luminosity and evolution functions}",
      journal = {\apj},
         year = 1983,
        month = jun,
       volume = {269},
        pages = {35-41},
          doi = {10.1086/161016},
       adsurl = {https://ui.adsabs.harvard.edu/abs/1983ApJ...269...35M}
}

@ARTICLE{2000MNRAS.317.1014B,
       author = {{Boyle}, B.~J. and {Shanks}, T. and {Croom}, S.~M. and {Smith}, R.~J. and {Miller}, L. and {Loaring}, N. and {Heymans}, C.},
        title = "{The 2dF QSO Redshift Survey - I. The optical luminosity function of quasi-stellar objects}",
      journal = {\mnras},
         year = 2000,
        month = oct,
       volume = {317},
       number = {4},
        pages = {1014-1022},
          doi = {10.1046/j.1365-8711.2000.03730.x},
archivePrefix = {arXiv},
       eprint = {astro-ph/0005368},
 primaryClass = {astro-ph},
       adsurl = {https://ui.adsabs.harvard.edu/abs/2000MNRAS.317.1014B}
}

@ARTICLE{2018AJ....155..110Y,
       author = {{Yang}, Jinyi and {Wu}, Xue-Bing and {Liu}, Dezi and {Fan}, Xiaohui and {Yang}, Qian and {Wang}, Feige and {McGreer}, Ian D. and {Fan}, Zuhui and {Yuan}, Shuo and {Shan}, Huanyuan},
        title = "{Deep CFHT Y-band Imaging of VVDS-F22 Field. II. Quasar Selection and Quasar Luminosity Function}",
      journal = {\aj},
         year = 2018,
        month = mar,
       volume = {155},
       number = {3},
          eid = {110},
        pages = {110},
          doi = {10.3847/1538-3881/aaa543},
archivePrefix = {arXiv},
       eprint = {1801.01245},
 primaryClass = {astro-ph.GA},
       adsurl = {https://ui.adsabs.harvard.edu/abs/2018AJ....155..110Y}
}

@ARTICLE{2009MNRAS.399.1755C,
       author = {{Croom}, Scott M. and {Richards}, Gordon T. and {Shanks}, Tom and {Boyle}, Brian J. and {Strauss}, Michael A. and {Myers}, Adam D. and {Nichol}, Robert C. and {Pimbblet}, Kevin A. and {Ross}, Nicholas P. and {Schneider}, Donald P. and {Sharp}, Robert G. and {Wake}, David A.},
        title = "{The 2dF-SDSS LRG and QSO survey: the QSO luminosity function at 0.4 < z < 2.6}",
      journal = {\mnras},
         year = 2009,
        month = nov,
       volume = {399},
       number = {4},
        pages = {1755-1772},
          doi = {10.1111/j.1365-2966.2009.15398.x},
archivePrefix = {arXiv},
       eprint = {0907.2727},
 primaryClass = {astro-ph.CO},
       adsurl = {https://ui.adsabs.harvard.edu/abs/2009MNRAS.399.1755C}
}

@ARTICLE{2013A&A...551A..29P,
       author = {{Palanque-Delabrouille}, N. and {Magneville}, Ch. and {Y{\`e}che}, Ch. and {Eftekharzadeh}, S. and {Myers}, A.~D. and {Petitjean}, P. and {P{\^a}ris}, I. and {Aubourg}, E. and {McGreer}, I. and {Fan}, X. and {Dey}, A. and {Schlegel}, D. and {Bailey}, S. and {Bizayev}, D. and {Bolton}, A. and {Dawson}, K. and {Ebelke}, G. and {Ge}, J. and {Malanushenko}, E. and {Malanushenko}, V. and {Oravetz}, D. and {Pan}, K. and {Ross}, N.~P. and {Schneider}, D.~P. and {Sheldon}, E. and {Simmons}, A. and {Tinker}, J. and {White}, M. and {Willmer}, Ch.},
        title = "{Luminosity function from dedicated SDSS-III and MMT data of quasars in 0.7 < z < 4.0 selected with a new approach}",
      journal = {\aap},
         year = 2013,
        month = mar,
       volume = {551},
          eid = {A29},
        pages = {A29},
          doi = {10.1051/0004-6361/201220379},
archivePrefix = {arXiv},
       eprint = {1209.3968},
 primaryClass = {astro-ph.CO},
       adsurl = {https://ui.adsabs.harvard.edu/abs/2013A&A...551A..29P}
}

@ARTICLE{2021ApJ...910L..11K,
       author = {{Kim}, Yongjung and {Im}, Myungshin},
        title = "{Pure Density Evolution of the Ultraviolet Quasar Luminosity Function at 2 {\ensuremath{\lesssim}} z {\ensuremath{\lesssim}} 6}",
      journal = {\apjl},
         year = 2021,
        month = mar,
       volume = {910},
       number = {1},
          eid = {L11},
        pages = {L11},
          doi = {10.3847/2041-8213/abed58},
archivePrefix = {arXiv},
       eprint = {2103.06265},
 primaryClass = {astro-ph.GA},
       adsurl = {https://ui.adsabs.harvard.edu/abs/2021ApJ...910L..11K}
}

@ARTICLE{2019ApJS..240...30S,
       author = {{Schlafly}, Edward F. and {Meisner}, Aaron M. and {Green}, Gregory M.},
        title = "{The unWISE Catalog: Two Billion Infrared Sources from Five Years of WISE Imaging}",
      journal = {\apjs},
         year = 2019,
        month = feb,
       volume = {240},
       number = {2},
          eid = {30},
        pages = {30},
          doi = {10.3847/1538-4365/aafbea},
archivePrefix = {arXiv},
       eprint = {1901.03337},
 primaryClass = {astro-ph.IM},
       adsurl = {https://ui.adsabs.harvard.edu/abs/2019ApJS..240...30S}
}

@ARTICLE{2000AJ....120.1579Y,
       author = {{York}, Donald G. and {Adelman}, J. and {Anderson}, John E., Jr. and {Anderson}, Scott F. and {Annis}, James and {Bahcall}, Neta A. and {Bakken}, J.~A. and {Barkhouser}, Robert and {Bastian}, Steven and {Berman}, Eileen and {Boroski}, William N. and {Bracker}, Steve and {Briegel}, Charlie and {Briggs}, John W. and {Brinkmann}, J. and {Brunner}, Robert and {Burles}, Scott and {Carey}, Larry and {Carr}, Michael A. and {Castander}, Francisco J. and {Chen}, Bing and {Colestock}, Patrick L. and {Connolly}, A.~J. and {Crocker}, J.~H. and {Csabai}, Istv{\'a}n and {Czarapata}, Paul C. and {Davis}, John Eric and {Doi}, Mamoru and {Dombeck}, Tom and {Eisenstein}, Daniel and {Ellman}, Nancy and {Elms}, Brian R. and {Evans}, Michael L. and {Fan}, Xiaohui and {Federwitz}, Glenn R. and {Fiscelli}, Larry and {Friedman}, Scott and {Frieman}, Joshua A. and {Fukugita}, Masataka and {Gillespie}, Bruce and {Gunn}, James E. and {Gurbani}, Vijay K. and {de Haas}, Ernst and {Haldeman}, Merle and {Harris}, Frederick H. and {Hayes}, J. and {Heckman}, Timothy M. and {Hennessy}, G.~S. and {Hindsley}, Robert B. and {Holm}, Scott and {Holmgren}, Donald J. and {Huang}, Chi-hao and {Hull}, Charles and {Husby}, Don and {Ichikawa}, Shin-Ichi and {Ichikawa}, Takashi and {Ivezi{\'c}}, {\v{Z}}eljko and {Kent}, Stephen and {Kim}, Rita S.~J. and {Kinney}, E. and {Klaene}, Mark and {Kleinman}, A.~N. and {Kleinman}, S. and {Knapp}, G.~R. and {Korienek}, John and {Kron}, Richard G. and {Kunszt}, Peter Z. and {Lamb}, D.~Q. and {Lee}, B. and {Leger}, R. French and {Limmongkol}, Siriluk and {Lindenmeyer}, Carl and {Long}, Daniel C. and {Loomis}, Craig and {Loveday}, Jon and {Lucinio}, Rich and {Lupton}, Robert H. and {MacKinnon}, Bryan and {Mannery}, Edward J. and {Mantsch}, P.~M. and {Margon}, Bruce and {McGehee}, Peregrine and {McKay}, Timothy A. and {Meiksin}, Avery and {Merelli}, Aronne and {Monet}, David G. and {Munn}, Jeffrey A. and {Narayanan}, Vijay K. and {Nash}, Thomas and {Neilsen}, Eric and {Neswold}, Rich and {Newberg}, Heidi Jo and {Nichol}, R.~C. and {Nicinski}, Tom and {Nonino}, Mario and {Okada}, Norio and {Okamura}, Sadanori and {Ostriker}, Jeremiah P. and {Owen}, Russell and {Pauls}, A. George and {Peoples}, John and {Peterson}, R.~L. and {Petravick}, Donald and {Pier}, Jeffrey R. and {Pope}, Adrian and {Pordes}, Ruth and {Prosapio}, Angela and {Rechenmacher}, Ron and {Quinn}, Thomas R. and {Richards}, Gordon T. and {Richmond}, Michael W. and {Rivetta}, Claudio H. and {Rockosi}, Constance M. and {Ruthmansdorfer}, Kurt and {Sandford}, Dale and {Schlegel}, David J. and {Schneider}, Donald P. and {Sekiguchi}, Maki and {Sergey}, Gary and {Shimasaku}, Kazuhiro and {Siegmund}, Walter A. and {Smee}, Stephen and {Smith}, J. Allyn and {Snedden}, S. and {Stone}, R. and {Stoughton}, Chris and {Strauss}, Michael A. and {Stubbs}, Christopher and {SubbaRao}, Mark and {Szalay}, Alexander S. and {Szapudi}, Istvan and {Szokoly}, Gyula P. and {Thakar}, Anirudda R. and {Tremonti}, Christy and {Tucker}, Douglas L. and {Uomoto}, Alan and {Vanden Berk}, Dan and {Vogeley}, Michael S. and {Waddell}, Patrick and {Wang}, Shu-i. and {Watanabe}, Masaru and {Weinberg}, David H. and {Yanny}, Brian and {Yasuda}, Naoki and {SDSS Collaboration}},
        title = "{The Sloan Digital Sky Survey: Technical Summary}",
      journal = {\aj},
         year = 2000,
        month = sep,
       volume = {120},
       number = {3},
        pages = {1579-1587},
          doi = {10.1086/301513},
archivePrefix = {arXiv},
       eprint = {astro-ph/0006396},
 primaryClass = {astro-ph},
       adsurl = {https://ui.adsabs.harvard.edu/abs/2000AJ....120.1579Y}
}

@ARTICLE{2020ApJS..250....8L,
       author = {{Lyke}, Brad W. and {Higley}, Alexandra N. and {McLane}, J.~N. and {Schurhammer}, Danielle P. and {Myers}, Adam D. and {Ross}, Ashley J. and {Dawson}, Kyle and {Chabanier}, Sol{\`e}ne and {Martini}, Paul and {Busca}, Nicol{\'a}s G. and {Mas des Bourboux}, H{\'e}lion du and {Salvato}, Mara and {Streblyanska}, Alina and {Zarrouk}, Pauline and {Burtin}, Etienne and {Anderson}, Scott F. and {Bautista}, Julian and {Bizyaev}, Dmitry and {Brandt}, W.~N. and {Brinkmann}, Jonathan and {Brownstein}, Joel R. and {Comparat}, Johan and {Green}, Paul and {de la Macorra}, Axel and {Mu{\~n}oz Guti{\'e}rrez}, Andrea and {Hou}, Jiamin and {Newman}, Jeffrey A. and {Palanque-Delabrouille}, Nathalie and {P{\^a}ris}, Isabelle and {Percival}, Will J. and {Petitjean}, Patrick and {Rich}, James and {Rossi}, Graziano and {Schneider}, Donald P. and {Smith}, Alexander and {Vivek}, M. and {Weaver}, Benjamin Alan},
        title = "{The Sloan Digital Sky Survey Quasar Catalog: Sixteenth Data Release}",
      journal = {\apjs},
         year = 2020,
        month = sep,
       volume = {250},
       number = {1},
          eid = {8},
        pages = {8},
          doi = {10.3847/1538-4365/aba623},
archivePrefix = {arXiv},
       eprint = {2007.09001},
 primaryClass = {astro-ph.GA},
       adsurl = {https://ui.adsabs.harvard.edu/abs/2020ApJS..250....8L}
}

@INPROCEEDINGS{1998ASPC..152...71L,
       author = {{Lewis}, I.~J. and {Glazebrook}, K. and {Taylor}, K.},
        title = "{The Anglo-Australian Observatory 2dF Project: Current Status and the First Year of Science}",
    booktitle = {Fiber Optics in Astronomy III},
         year = 1998,
       editor = {{Arribas}, S. and {Mediavilla}, E. and {Watson}, F.},
       series = {Astronomical Society of the Pacific Conference Series},
       volume = {152},
        month = jan,
        pages = {71},
       adsurl = {https://ui.adsabs.harvard.edu/abs/1998ASPC..152...71L}
}

@ARTICLE{2004MNRAS.349.1397C,
       author = {{Croom}, S.~M. and {Smith}, R.~J. and {Boyle}, B.~J. and {Shanks}, T. and {Miller}, L. and {Outram}, P.~J. and {Loaring}, N.~S.},
        title = "{The 2dF QSO Redshift Survey - XII. The spectroscopic catalogue and luminosity function}",
      journal = {\mnras},
         year = 2004,
        month = apr,
       volume = {349},
       number = {4},
        pages = {1397-1418},
          doi = {10.1111/j.1365-2966.2004.07619.x},
archivePrefix = {arXiv},
       eprint = {astro-ph/0403040},
 primaryClass = {astro-ph},
       adsurl = {https://ui.adsabs.harvard.edu/abs/2004MNRAS.349.1397C}
}

@ARTICLE{2019ApJS..240....6Y,
       author = {{Yao}, Su and {Wu}, Xue-Bing and {Ai}, Y.~L. and {Yang}, Jinyi and {Yang}, Qian and {Dong}, Xiaoyi and {Joshi}, Ravi and {Wang}, Feige and {Feng}, Xiaotong and {Fu}, Yuming and {Hou}, Wen and {Luo}, A. -L. and {Kong}, Xiao and {Liu}, Yuanqi and {Zhao}, Y. -H. and {Zhang}, Y. -X. and {Yuan}, H. -L. and {Shen}, Shiyin},
        title = "{The Large Sky Area Multi-object Fiber Spectroscopic Telescope (LAMOST) Quasar Survey: The Fourth and Fifth Data Releases}",
      journal = {\apjs},
         year = 2019,
        month = jan,
       volume = {240},
       number = {1},
          eid = {6},
        pages = {6},
          doi = {10.3847/1538-4365/aaef88},
archivePrefix = {arXiv},
       eprint = {1811.01570},
 primaryClass = {astro-ph.GA},
       adsurl = {https://ui.adsabs.harvard.edu/abs/2019ApJS..240....6Y}
}

@ARTICLE{2003MNRAS.346.1055K,
       author = {{Kauffmann}, Guinevere and {Heckman}, Timothy M. and {Tremonti}, Christy and {Brinchmann}, Jarle and {Charlot}, St{\'e}phane and {White}, Simon D.~M. and {Ridgway}, Susan E. and {Brinkmann}, Jon and {Fukugita}, Masataka and {Hall}, Patrick B. and {Ivezi{\'c}}, {\v{Z}}eljko and {Richards}, Gordon T. and {Schneider}, Donald P.},
        title = "{The host galaxies of active galactic nuclei}",
      journal = {\mnras},
         year = 2003,
        month = dec,
       volume = {346},
       number = {4},
        pages = {1055-1077},
          doi = {10.1111/j.1365-2966.2003.07154.x},
archivePrefix = {arXiv},
       eprint = {astro-ph/0304239},
 primaryClass = {astro-ph},
       adsurl = {https://ui.adsabs.harvard.edu/abs/2003MNRAS.346.1055K}
}

@ARTICLE{2024ApJS..271...54F,
       author = {{Fu}, Yuming and {Wu}, Xue-Bing and {Li}, Yifan and {Pang}, Yuxuan and {Joshi}, Ravi and {Zhang}, Shuo and {Wang}, Qiyue and {Yang}, Jing and {Ng}, FanLam and {Liu}, Xingjian and et al.},
        title = "{CatNorth: An Improved Gaia DR3 Quasar Candidate Catalog with Pan-STARRS1 and CatWISE}",
      journal = {\apjs},
         year = 2024,
        month = apr,
       volume = {271},
       number = {2},
          eid = {54},
        pages = {54},
          doi = {10.3847/1538-4365/ad2ae6},
archivePrefix = {arXiv},
       eprint = {2310.12704},
 primaryClass = {astro-ph.GA},
       adsurl = {https://ui.adsabs.harvard.edu/abs/2024ApJS..271...54F}
}

@ARTICLE{2016ApJ...819...24W,
       author = {{Wang}, Feige and {Wu}, Xue-Bing and {Fan}, Xiaohui and {Yang}, Jinyi and {Yi}, Weimin and {Bian}, Fuyan and {McGreer}, Ian D. and {Yang}, Qian and {Ai}, Yanli and {Dong}, Xiaoyi and {Zuo}, Wenwen and {Jiang}, Linhua and {Green}, Richard and {Wang}, Shu and {Cai}, Zheng and {Wang}, Ran and {Yue}, Minghao},
        title = "{A Survey of Luminous High-redshift Quasars with SDSS and WISE. I. Target Selection and Optical Spectroscopy}",
      journal = {\apj},
         year = 2016,
        month = mar,
       volume = {819},
       number = {1},
          eid = {24},
        pages = {24},
          doi = {10.3847/0004-637X/819/1/24},
archivePrefix = {arXiv},
       eprint = {1602.04659},
 primaryClass = {astro-ph.GA},
       adsurl = {https://ui.adsabs.harvard.edu/abs/2016ApJ...819...24W}
}

@ARTICLE{2007MNRAS.379.1599L,
       author = {{Lawrence}, A. and {Warren}, S.~J. and {Almaini}, O. and {Edge}, A.~C. and {Hambly}, N.~C. and {Jameson}, R.~F. and {Lucas}, P. and {Casali}, M. and {Adamson}, A. and {Dye}, S. and {Emerson}, J.~P. and {Foucaud}, S. and {Hewett}, P. and {Hirst}, P. and {Hodgkin}, S.~T. and {Irwin}, M.~J. and {Lodieu}, N. and {McMahon}, R.~G. and {Simpson}, C. and {Smail}, I. and {Mortlock}, D. and {Folger}, M.},
        title = "{The UKIRT Infrared Deep Sky Survey (UKIDSS)}",
      journal = {\mnras},
         year = 2007,
        month = aug,
       volume = {379},
       number = {4},
        pages = {1599-1617},
          doi = {10.1111/j.1365-2966.2007.12040.x},
archivePrefix = {arXiv},
       eprint = {astro-ph/0604426},
 primaryClass = {astro-ph},
       adsurl = {https://ui.adsabs.harvard.edu/abs/2007MNRAS.379.1599L}
}

@ARTICLE{2010AJ....140.1868W,
       author = {{Wright}, Edward L. and {Eisenhardt}, Peter R.~M. and {Mainzer}, Amy K. and {Ressler}, Michael E. and {Cutri}, Roc M. and {Jarrett}, Thomas and {Kirkpatrick}, J. Davy and {Padgett}, Deborah and {McMillan}, Robert S. and {Skrutskie}, Michael and {Stanford}, S.~A. and {Cohen}, Martin and {Walker}, Russell G. and {Mather}, John C. and {Leisawitz}, David and {Gautier}, III, Thomas N. and {McLean}, Ian and {Benford}, Dominic and {Lonsdale}, Carol J. and {Blain}, Andrew and {Mendez}, Bryan and {Irace}, William R. and {Duval}, Valerie and {Liu}, Fengchuan and {Royer}, Don and {Heinrichsen}, Ingolf and {Howard}, Joan and {Shannon}, Mark and {Kendall}, Martha and {Walsh}, Amy L. and {Larsen}, Mark and {Cardon}, Joel G. and {Schick}, Scott and {Schwalm}, Mark and {Abid}, Mohamed and {Fabinsky}, Beth and {Naes}, Larry and {Tsai}, Chao-Wei},
        title = "{The Wide-field Infrared Survey Explorer (WISE): Mission Description and Initial On-orbit Performance}",
      journal = {\aj},
         year = 2010,
        month = dec,
       volume = {140},
       number = {6},
        pages = {1868-1881},
          doi = {10.1088/0004-6256/140/6/1868},
archivePrefix = {arXiv},
       eprint = {1008.0031},
 primaryClass = {astro-ph.IM},
       adsurl = {https://ui.adsabs.harvard.edu/abs/2010AJ....140.1868W}
}

@ARTICLE{2013A&A...558A..33A,
       author = {{Astropy Collaboration} and {Robitaille}, Thomas P. and {Tollerud}, Erik J. and {Greenfield}, Perry and {Droettboom}, Michael and {Bray}, Erik and {Aldcroft}, Tom and {Davis}, Matt and {Ginsburg}, Adam and {Price-Whelan}, Adrian M. and {Kerzendorf}, Wolfgang E. and {Conley}, Alexander and {Crighton}, Neil and {Barbary}, Kyle and {Muna}, Demitri and {Ferguson}, Henry and {Grollier}, Fr{\'e}d{\'e}ric and {Parikh}, Madhura M. and {Nair}, Prasanth H. and {Unther}, Hans M. and {Deil}, Christoph and {Woillez}, Julien and {Conseil}, Simon and {Kramer}, Roban and {Turner}, James E.~H. and {Singer}, Leo and {Fox}, Ryan and {Weaver}, Benjamin A. and {Zabalza}, Victor and {Edwards}, Zachary I. and {Azalee Bostroem}, K. and {Burke}, D.~J. and {Casey}, Andrew R. and {Crawford}, Steven M. and {Dencheva}, Nadia and {Ely}, Justin and {Jenness}, Tim and {Labrie}, Kathleen and {Lim}, Pey Lian and {Pierfederici}, Francesco and {Pontzen}, Andrew and {Ptak}, Andy and {Refsdal}, Brian and {Servillat}, Mathieu and {Streicher}, Ole},
        title = "{Astropy: A community Python package for astronomy}",
      journal = {\aap},
         year = 2013,
        month = oct,
       volume = {558},
          eid = {A33},
        pages = {A33},
          doi = {10.1051/0004-6361/201322068},
archivePrefix = {arXiv},
       eprint = {1307.6212},
 primaryClass = {astro-ph.IM},
       adsurl = {https://ui.adsabs.harvard.edu/abs/2013A&A...558A..33A}
}

@ARTICLE{2018AJ....156..123A,
       author = {{Astropy Collaboration} and {Price-Whelan}, A.~M. and {Sip{\H{o}}cz}, B.~M. and {G{\"u}nther}, H.~M. and {Lim}, P.~L. and {Crawford}, S.~M. and {Conseil}, S. and {Shupe}, D.~L. and {Craig}, M.~W. and {Dencheva}, N. and {Ginsburg}, A. and {VanderPlas}, J.~T. and {Bradley}, L.~D. and {P{\'e}rez-Su{\'a}rez}, D. and {de Val-Borro}, M. and {Aldcroft}, T.~L. and {Cruz}, K.~L. and {Robitaille}, T.~P. and {Tollerud}, E.~J. and {Ardelean}, C. and {Babej}, T. and {Bach}, Y.~P. and {Bachetti}, M. and {Bakanov}, A.~V. and {Bamford}, S.~P. and {Barentsen}, G. and {Barmby}, P. and {Baumbach}, A. and {Berry}, K.~L. and {Biscani}, F. and {Boquien}, M. and {Bostroem}, K.~A. and {Bouma}, L.~G. and {Brammer}, G.~B. and {Bray}, E.~M. and {Breytenbach}, H. and {Buddelmeijer}, H. and {Burke}, D.~J. and {Calderone}, G. and {Cano Rodr{\'\i}guez}, J.~L. and {Cara}, M. and {Cardoso}, J.~V.~M. and {Cheedella}, S. and {Copin}, Y. and {Corrales}, L. and {Crichton}, D. and {D'Avella}, D. and {Deil}, C. and {Depagne}, {\'E}. and {Dietrich}, J.~P. and {Donath}, A. and {Droettboom}, M. and {Earl}, N. and {Erben}, T. and {Fabbro}, S. and {Ferreira}, L.~A. and {Finethy}, T. and {Fox}, R.~T. and {Garrison}, L.~H. and {Gibbons}, S.~L.~J. and {Goldstein}, D.~A. and {Gommers}, R. and {Greco}, J.~P. and {Greenfield}, P. and {Groener}, A.~M. and {Grollier}, F. and {Hagen}, A. and {Hirst}, P. and {Homeier}, D. and {Horton}, A.~J. and {Hosseinzadeh}, G. and {Hu}, L. and {Hunkeler}, J.~S. and {Ivezi{\'c}}, {\v{Z}}. and {Jain}, A. and {Jenness}, T. and {Kanarek}, G. and {Kendrew}, S. and {Kern}, N.~S. and {Kerzendorf}, W.~E. and {Khvalko}, A. and {King}, J. and {Kirkby}, D. and {Kulkarni}, A.~M. and {Kumar}, A. and {Lee}, A. and {Lenz}, D. and {Littlefair}, S.~P. and {Ma}, Z. and {Macleod}, D.~M. and {Mastropietro}, M. and {McCully}, C. and {Montagnac}, S. and {Morris}, B.~M. and {Mueller}, M. and {Mumford}, S.~J. and {Muna}, D. and {Murphy}, N.~A. and {Nelson}, S. and {Nguyen}, G.~H. and {Ninan}, J.~P. and {N{\"o}the}, M. and {Ogaz}, S. and {Oh}, S. and {Parejko}, J.~K. and {Parley}, N. and {Pascual}, S. and {Patil}, R. and {Patil}, A.~A. and {Plunkett}, A.~L. and {Prochaska}, J.~X. and {Rastogi}, T. and {Reddy Janga}, V. and {Sabater}, J. and {Sakurikar}, P. and {Seifert}, M. and {Sherbert}, L.~E. and {Sherwood-Taylor}, H. and {Shih}, A.~Y. and {Sick}, J. and {Silbiger}, M.~T. and {Singanamalla}, S. and {Singer}, L.~P. and {Sladen}, P.~H. and {Sooley}, K.~A. and {Sornarajah}, S. and {Streicher}, O. and {Teuben}, P. and {Thomas}, S.~W. and {Tremblay}, G.~R. and {Turner}, J.~E.~H. and {Terr{\'o}n}, V. and {van Kerkwijk}, M.~H. and {de la Vega}, A. and {Watkins}, L.~L. and {Weaver}, B.~A. and {Whitmore}, J.~B. and {Woillez}, J. and {Zabalza}, V. and {Astropy Contributors}},
        title = "{The Astropy Project: Building an Open-science Project and Status of the v2.0 Core Package}",
      journal = {\aj},
         year = 2018,
        month = sep,
       volume = {156},
       number = {3},
          eid = {123},
        pages = {123},
          doi = {10.3847/1538-3881/aabc4f},
archivePrefix = {arXiv},
       eprint = {1801.02634},
 primaryClass = {astro-ph.IM},
       adsurl = {https://ui.adsabs.harvard.edu/abs/2018AJ....156..123A}
}

@ARTICLE{2013PASP..125..306F,
       author = {{Foreman-Mackey}, Daniel and {Hogg}, David W. and {Lang}, Dustin and {Goodman}, Jonathan},
        title = "{emcee: The MCMC Hammer}",
      journal = {\pasp},
         year = 2013,
        month = mar,
       volume = {125},
       number = {925},
        pages = {306},
          doi = {10.1086/670067},
archivePrefix = {arXiv},
       eprint = {1202.3665},
 primaryClass = {astro-ph.IM},
       adsurl = {https://ui.adsabs.harvard.edu/abs/2013PASP..125..306F}
}

@INPROCEEDINGS{2005ASPC..347...29T,
       author = {{Taylor}, M.~B.},
        title = "{TOPCAT \& STIL: Starlink Table/VOTable Processing Software}",
    booktitle = {Astronomical Data Analysis Software and Systems XIV},
         year = 2005,
       editor = {{Shopbell}, P. and {Britton}, M. and {Ebert}, R.},
       series = {Astronomical Society of the Pacific Conference Series},
       volume = {347},
        month = dec,
        pages = {29},
       adsurl = {https://ui.adsabs.harvard.edu/abs/2005ASPC..347...29T}
}

@ARTICLE{2022NatAs...6..850J,
       author = {{Jiang}, Linhua and {Ning}, Yuanhang and {Fan}, Xiaohui and {Ho}, Luis C. and {Luo}, Bin and {Wang}, Feige and {Wu}, Jin and {Wu}, Xue-Bing and {Yang}, Jinyi and {Zheng}, Zhen-Ya},
        title = "{Definitive upper bound on the negligible contribution of quasars to cosmic reionization}",
      journal = {Nature Astronomy},
         year = 2022,
        month = jun,
       volume = {6},
        pages = {850-856},
          doi = {10.1038/s41550-022-01708-w},
archivePrefix = {arXiv},
       eprint = {2206.07825},
 primaryClass = {astro-ph.GA},
       adsurl = {https://ui.adsabs.harvard.edu/abs/2022NatAs...6..850J}
}

@ARTICLE{2004A&A...424..793W,
       author = {{Wu}, X. -B. and {Wang}, R. and {Kong}, M.~Z. and {Liu}, F.~K. and {Han}, J.~L.},
        title = "{Black hole mass estimation using a relation between the BLR size and emission line luminosity of AGN}",
      journal = {\aap},
         year = 2004,
        month = sep,
       volume = {424},
        pages = {793-798},
          doi = {10.1051/0004-6361:20035845},
archivePrefix = {arXiv},
       eprint = {astro-ph/0403243},
 primaryClass = {astro-ph},
       adsurl = {https://ui.adsabs.harvard.edu/abs/2004A&A...424..793W}
}

@ARTICLE{2011ApJ...735...68K,
       author = {{Kim}, Dae-Won and {Protopapas}, Pavlos and {Byun}, Yong-Ik and {Alcock}, Charles and {Khardon}, Roni and {Trichas}, Markos},
        title = "{Quasi-stellar Object Selection Algorithm Using Time Variability and Machine Learning: Selection of 1620 Quasi-stellar Object Candidates from MACHO Large Magellanic Cloud Database}",
      journal = {\apj},
         year = 2011,
        month = jul,
       volume = {735},
       number = {2},
          eid = {68},
        pages = {68},
          doi = {10.1088/0004-637X/735/2/68},
archivePrefix = {arXiv},
       eprint = {1101.3316},
 primaryClass = {astro-ph.IM},
       adsurl = {https://ui.adsabs.harvard.edu/abs/2011ApJ...735...68K}
}

@ARTICLE{2019MNRAS.485.4539J,
       author = {{Jin}, Xin and {Zhang}, Yanxia and {Zhang}, Jingyi and {Zhao}, Yongheng and {Wu}, Xue-bing and {Fan}, Dongwei},
        title = "{Efficient selection of quasar candidates based on optical and infrared photometric data using machine learning}",
      journal = {\mnras},
         year = 2019,
        month = jun,
       volume = {485},
       number = {4},
        pages = {4539-4549},
          doi = {10.1093/mnras/stz680},
archivePrefix = {arXiv},
       eprint = {1903.03335},
 primaryClass = {astro-ph.IM},
       adsurl = {https://ui.adsabs.harvard.edu/abs/2019MNRAS.485.4539J}
}

@software{2021ascl.soft06008M,
       author = {{McGreer}, Ian and {Moustakas}, John and {Schindler}, JT},
        title = "{simqso: Simulated quasar spectra generator}",
 howpublished = {Astrophysics Source Code Library, record ascl:2106.008},
         year = 2021,
        month = jun,
          eid = {ascl:2106.008},
       adsurl = {https://ui.adsabs.harvard.edu/abs/2021ascl.soft06008M}
}

@ARTICLE{2013Msngr.154...35M,
       author = {{McMahon}, R.~G. and {Banerji}, M. and {Gonzalez}, E. and {Koposov}, S.~E. and {Bejar}, V.~J. and {Lodieu}, N. and {Rebolo}, R. and {VHS Collaboration}},
        title = "{First Scientific Results from the VISTA Hemisphere Survey (VHS)}",
      journal = {The Messenger},
         year = 2013,
        month = dec,
       volume = {154},
        pages = {35-37},
       adsurl = {https://ui.adsabs.harvard.edu/abs/2013Msngr.154...35M}
}

@ARTICLE{2016arXiv161100037D,
       author = {{DESI Collaboration} and {Aghamousa}, Amir and {Aguilar}, Jessica and {Ahlen}, Steve and {Alam}, Shadab and {Allen}, Lori E. and {Allende Prieto}, Carlos and {Annis}, James and {Bailey}, Stephen and {Balland}, Christophe and {Ballester}, Otger and {Baltay}, Charles and {Beaufore}, Lucas and {Bebek}, Chris and {Beers}, Timothy C. and {Bell}, Eric F. and {Bernal}, Jos{\'e} Luis and {Besuner}, Robert and {Beutler}, Florian and {Blake}, Chris and {Bleuler}, Hannes and {Blomqvist}, Michael and {Blum}, Robert and {Bolton}, Adam S. and {Briceno}, Cesar and {Brooks}, David and {Brownstein}, Joel R. and {Buckley-Geer}, Elizabeth and {Burden}, Angela and {Burtin}, Etienne and {Busca}, Nicolas G. and {Cahn}, Robert N. and {Cai}, Yan-Chuan and {Cardiel-Sas}, Laia and {Carlberg}, Raymond G. and {Carton}, Pierre-Henri and {Casas}, Ricard and {Castander}, Francisco J. and {Cervantes-Cota}, Jorge L. and {Claybaugh}, Todd M. and {Close}, Madeline and {Coker}, Carl T. and {Cole}, Shaun and {Comparat}, Johan and {Cooper}, Andrew P. and {Cousinou}, M. -C. and {Crocce}, Martin and {Cuby}, Jean-Gabriel and {Cunningham}, Daniel P. and {Davis}, Tamara M. and {Dawson}, Kyle S. and {de la Macorra}, Axel and {De Vicente}, Juan and {Delubac}, Timoth{\'e}e and {Derwent}, Mark and {Dey}, Arjun and {Dhungana}, Govinda and {Ding}, Zhejie and {Doel}, Peter and {Duan}, Yutong T. and {Ealet}, Anne and {Edelstein}, Jerry and {Eftekharzadeh}, Sarah and {Eisenstein}, Daniel J. and {Elliott}, Ann and {Escoffier}, St{\'e}phanie and {Evatt}, Matthew and {Fagrelius}, Parker and {Fan}, Xiaohui and {Fanning}, Kevin and {Farahi}, Arya and {Farihi}, Jay and {Favole}, Ginevra and {Feng}, Yu and {Fernandez}, Enrique and {Findlay}, Joseph R. and {Finkbeiner}, Douglas P. and {Fitzpatrick}, Michael J. and {Flaugher}, Brenna and {Flender}, Samuel and {Font-Ribera}, Andreu and {Forero-Romero}, Jaime E. and {Fosalba}, Pablo and {Frenk}, Carlos S. and {Fumagalli}, Michele and {Gaensicke}, Boris T. and {Gallo}, Giuseppe and {Garcia-Bellido}, Juan and {Gaztanaga}, Enrique and {Pietro Gentile Fusillo}, Nicola and {Gerard}, Terry and {Gershkovich}, Irena and {Giannantonio}, Tommaso and {Gillet}, Denis and {Gonzalez-de-Rivera}, Guillermo and {Gonzalez-Perez}, Violeta and {Gott}, Shelby and {Graur}, Or and {Gutierrez}, Gaston and {Guy}, Julien and {Habib}, Salman and {Heetderks}, Henry and {Heetderks}, Ian and {Heitmann}, Katrin and {Hellwing}, Wojciech A. and {Herrera}, David A. and {Ho}, Shirley and {Holland}, Stephen and {Honscheid}, Klaus and {Huff}, Eric and {Hutchinson}, Timothy A. and {Huterer}, Dragan and {Hwang}, Ho Seong and {Illa Laguna}, Joseph Maria and {Ishikawa}, Yuzo and {Jacobs}, Dianna and {Jeffrey}, Niall and {Jelinsky}, Patrick and {Jennings}, Elise and {Jiang}, Linhua and {Jimenez}, Jorge and {Johnson}, Jennifer and {Joyce}, Richard and {Jullo}, Eric and {Juneau}, St{\'e}phanie and {Kama}, Sami and {Karcher}, Armin and {Karkar}, Sonia and {Kehoe}, Robert and {Kennamer}, Noble and {Kent}, Stephen and {Kilbinger}, Martin and {Kim}, Alex G. and {Kirkby}, David and {Kisner}, Theodore and {Kitanidis}, Ellie and {Kneib}, Jean-Paul and {Koposov}, Sergey and {Kovacs}, Eve and {Koyama}, Kazuya and {Kremin}, Anthony and {Kron}, Richard and {Kronig}, Luzius and {Kueter-Young}, Andrea and {Lacey}, Cedric G. and {Lafever}, Robin and {Lahav}, Ofer and {Lambert}, Andrew and {Lampton}, Michael and {Landriau}, Martin and {Lang}, Dustin and {Lauer}, Tod R. and {Le Goff}, Jean-Marc and {Le Guillou}, Laurent and {Le Van Suu}, Auguste and {Lee}, Jae Hyeon and {Lee}, Su-Jeong and {Leitner}, Daniela and {Lesser}, Michael and {Levi}, Michael E. and {L'Huillier}, Benjamin and {Li}, Baojiu and {Liang}, Ming and {Lin}, Huan and {Linder}, Eric and {Loebman}, Sarah R. and {Luki{\'c}}, Zarija and {Ma}, Jun and {MacCrann}, Niall and {Magneville}, Christophe and {Makarem}, Laleh and {Manera}, Marc and {Manser}, Christopher J. and {Marshall}, Robert and {Martini}, Paul and {Massey}, Richard and {Matheson}, Thomas and {McCauley}, Jeremy and {McDonald}, Patrick and {McGreer}, Ian D. and {Meisner}, Aaron and {Metcalfe}, Nigel and {Miller}, Timothy N. and {Miquel}, Ramon and {Moustakas}, John and {Myers}, Adam and {Naik}, Milind and {Newman}, Jeffrey A. and {Nichol}, Robert C. and {Nicola}, Andrina and {Nicolati da Costa}, Luiz and {Nie}, Jundan and {Niz}, Gustavo and {Norberg}, Peder and {Nord}, Brian and {Norman}, Dara and {Nugent}, Peter and {O'Brien}, Thomas and {Oh}, Minji and {Olsen}, Knut A.~G.},
        title = "{The DESI Experiment Part II: Instrument Design}",
      journal = {arXiv e-prints},
         year = 2016,
        month = oct,
          eid = {arXiv:1611.00037},
        pages = {arXiv:1611.00037},
          doi = {10.48550/arXiv.1611.00037},
archivePrefix = {arXiv},
       eprint = {1611.00037},
 primaryClass = {astro-ph.IM},
       adsurl = {https://ui.adsabs.harvard.edu/abs/2016arXiv161100037D}
}

@ARTICLE{2022AJ....164..207D,
       author = {{DESI Collaboration} and {Abareshi}, B. and {Aguilar}, J. and {Ahlen}, S. and {Alam}, Shadab and {Alexander}, David M. and {Alfarsy}, R. and {Allen}, L. and {Allende Prieto}, C. and {Alves}, O. and {Ameel}, J. and {Armengaud}, E. and {Asorey}, J. and {Aviles}, Alejandro and {Bailey}, S. and {Balaguera-Antol{\'\i}nez}, A. and {Ballester}, O. and {Baltay}, C. and {Bault}, A. and {Beltran}, S.~F. and {Benavides}, B. and {BenZvi}, S. and {Berti}, A. and {Besuner}, R. and {Beutler}, Florian and {Bianchi}, D. and {Blake}, C. and {Blanc}, P. and {Blum}, R. and {Bolton}, A. and {Bose}, S. and {Bramall}, D. and {Brieden}, S. and {Brodzeller}, A. and {Brooks}, D. and {Brownewell}, C. and {Buckley-Geer}, E. and {Cahn}, R.~N. and {Cai}, Z. and {Canning}, R. and {Capasso}, R. and {Carnero Rosell}, A. and {Carton}, P. and {Casas}, R. and {Castander}, F.~J. and {Cervantes-Cota}, J.~L. and {Chabanier}, S. and {Chaussidon}, E. and {Chuang}, C. and {Circosta}, C. and {Cole}, S. and {Cooper}, A.~P. and {da Costa}, L. and {Cousinou}, M. -C. and {Cuceu}, A. and {Davis}, T.~M. and {Dawson}, K. and {de la Cruz-Noriega}, R. and {de la Macorra}, A. and {de Mattia}, A. and {Della Costa}, J. and {Demmer}, P. and {Derwent}, M. and {Dey}, A. and {Dey}, B. and {Dhungana}, G. and {Ding}, Z. and {Dobson}, C. and {Doel}, P. and {Donald-McCann}, J. and {Donaldson}, J. and {Douglass}, K. and {Duan}, Y. and {Dunlop}, P. and {Edelstein}, J. and {Eftekharzadeh}, S. and {Eisenstein}, D.~J. and {Enriquez-Vargas}, M. and {Escoffier}, S. and {Evatt}, M. and {Fagrelius}, P. and {Fan}, X. and {Fanning}, K. and {Fawcett}, V.~A. and {Ferraro}, S. and {Ereza}, J. and {Flaugher}, B. and {Font-Ribera}, A. and {Forero-Romero}, J.~E. and {Frenk}, C.~S. and {Fromenteau}, S. and {G{\"a}nsicke}, B.~T. and {Garcia-Quintero}, C. and {Garrison}, L. and {Gazta{\~n}aga}, E. and {Gerardi}, F. and {Gil-Mar{\'\i}n}, H. and {Gontcho A Gontcho}, S. and {Gonzalez-Morales}, Alma X. and {Gonzalez-de-Rivera}, G. and {Gonzalez-Perez}, V. and {Gordon}, C. and {Graur}, O. and {Green}, D. and {Grove}, C. and {Gruen}, D. and {Gutierrez}, G. and {Guy}, J. and {Hahn}, C. and {Harris}, S. and {Herrera}, D. and {Herrera-Alcantar}, Hiram K. and {Honscheid}, K. and {Howlett}, C. and {Huterer}, D. and {Ir{\v{s}}i{\v{c}}}, V. and {Ishak}, M. and {Jelinsky}, P. and {Jiang}, L. and {Jimenez}, J. and {Jing}, Y.~P. and {Joyce}, R. and {Jullo}, E. and {Juneau}, S. and {Kara{\c{c}}ayl{\i}}, N.~G. and {Karamanis}, M. and {Karcher}, A. and {Karim}, T. and {Kehoe}, R. and {Kent}, S. and {Kirkby}, D. and {Kisner}, T. and {Kitaura}, F. and {Koposov}, S.~E. and {Kov{\'a}cs}, A. and {Kremin}, A. and {Krolewski}, Alex and {L'Huillier}, B. and {Lahav}, O. and {Lambert}, A. and {Lamman}, C. and {Lan}, Ting-Wen and {Landriau}, M. and {Lane}, S. and {Lang}, D. and {Lange}, J.~U. and {Lasker}, J. and {Le Guillou}, L. and {Leauthaud}, A. and {Le Van Suu}, A. and {Levi}, Michael E. and {Li}, T.~S. and {Magneville}, C. and {Manera}, M. and {Manser}, Christopher J. and {Marshall}, B. and {Martini}, Paul and {McCollam}, W. and {McDonald}, P. and {Meisner}, Aaron M. and {Mena-Fern{\'a}ndez}, J. and {Meneses-Rizo}, J. and {Mezcua}, M. and {Miller}, T. and {Miquel}, R. and {Montero-Camacho}, P. and {Moon}, J. and {Moustakas}, J. and {Mueller}, E. and {Mu{\~n}oz-Guti{\'e}rrez}, Andrea and {Myers}, Adam D. and {Nadathur}, S. and {Najita}, J. and {Napolitano}, L. and {Neilsen}, E. and {Newman}, Jeffrey A. and {Nie}, J.~D. and {Ning}, Y. and {Niz}, G. and {Norberg}, P. and {Noriega}, Hern{\'a}n E. and {O'Brien}, T. and {Obuljen}, A. and {Palanque-Delabrouille}, N. and {Palmese}, A. and {Zhiwei}, P. and {Pappalardo}, D. and {PENG}, X. and {Percival}, W.~J. and {Perruchot}, S. and {Pogge}, R. and {Poppett}, C. and {Porredon}, A. and {Prada}, F. and {Prochaska}, J. and {Pucha}, R. and {P{\'e}rez-Fern{\'a}ndez}, A. and {P{\'e}rez-R{\`a}fols}, I. and {Rabinowitz}, D. and {Raichoor}, A.},
        title = "{Overview of the Instrumentation for the Dark Energy Spectroscopic Instrument}",
      journal = {\aj},
         year = 2022,
        month = nov,
       volume = {164},
       number = {5},
          eid = {207},
        pages = {207},
          doi = {10.3847/1538-3881/ac882b},
archivePrefix = {arXiv},
       eprint = {2205.10939},
 primaryClass = {astro-ph.IM},
       adsurl = {https://ui.adsabs.harvard.edu/abs/2022AJ....164..207D}
}

@ARTICLE{2014AJ....147..108L,
       author = {{Lang}, Dustin},
        title = "{unWISE: Unblurred Coadds of the WISE Imaging}",
      journal = {\aj},
         year = 2014,
        month = may,
       volume = {147},
       number = {5},
          eid = {108},
        pages = {108},
          doi = {10.1088/0004-6256/147/5/108},
archivePrefix = {arXiv},
       eprint = {1405.0308},
 primaryClass = {astro-ph.IM},
       adsurl = {https://ui.adsabs.harvard.edu/abs/2014AJ....147..108L}
}

@ARTICLE{2016AJ....151...36L,
       author = {{Lang}, Dustin and {Hogg}, David W. and {Schlegel}, David J.},
        title = "{WISE Photometry for 400 Million SDSS Sources}",
      journal = {\aj},
         year = 2016,
        month = feb,
       volume = {151},
       number = {2},
          eid = {36},
        pages = {36},
          doi = {10.3847/0004-6256/151/2/36},
archivePrefix = {arXiv},
       eprint = {1410.7397},
 primaryClass = {astro-ph.IM},
       adsurl = {https://ui.adsabs.harvard.edu/abs/2016AJ....151...36L}
}
\bibliographystyle{aasjournal}

\end{document}